\documentclass[10pt]{book}
\usepackage{authblk}
\usepackage[utf8]{inputenc}
\usepackage[english]{babel}
\usepackage{blindtext}
\usepackage{siunitx}
\usepackage[a5paper]{geometry}
\usepackage{scrextend}
\changefontsizes{10.0pt}
\usepackage{lettrine}
\usepackage{fancyhdr}
\usepackage{amsmath}
\usepackage{amssymb}
\usepackage{amsthm}
\usepackage{wasysym}
\usepackage[colorlinks,bookmarks=false,citecolor=blue,linkcolor=red,urlcolor=blue,hypertexnames=true]{hyperref}
\usepackage{colortbl}
\usepackage{verbatim}
\usepackage{graphicx}
\usepackage{epsfig}
\usepackage{dcolumn}
\usepackage{bm}

\makeatletter

\newenvironment{chapabstract}{%
\begin{center}
\bfseries {Abstract}
\end{center}}

\renewcommand*\env@matrix[1][*\c@MaxMatrixCols c]{%
  \hskip -\arraycolsep
  \let\@ifnextchar\new@ifnextchar
  \array{#1}}
\makeatother

\usepackage{appendix}

\title{Imbalanced Fermi systems and exotic Superconducting Phases}
\author{Madhuparna Karmakar}
\affil{Department of Physics, Indian Institute of 
technology, Madras, Chennai-600036, India.}
\begin{document}
\maketitle
\begin{chapabstract}
{Quantum many body phases, in competition or coexistence has always been a 
fascinating area of research in condensed matter physics. One of the most 
widely investigated scenario of phase competition is that between superconductivity 
and magnetism, which in spite of several decades of research continues to bring up 
new surprises till date. While superconductivity and magnetism are naively considered 
to be antagonists, there are several circumstances where these many body phases can be 
found to be cooperating, coexisting and even promoting each other. The results of such 
cooperative behavior are found to show up as exotic quantum phases.

In this chapter we discuss about such an exotic superconducting phase, which comes to life 
when a particular class of superconducting material viz. ``Pauli limited superconductors'' 
are subjected to an imbalance between the population of the fermionic species undergoing pairing. 
The consequence of such an imbalance is the emergence of a non-trivial spatially modulated superconducting 
phase, popularly known as the Fulde-Ferrell-Larkin-Ovchinnikov (FFLO) phase. Characterized by a 
finite momentum pairing between the fermionic species the FFLO phase hosts a non zero magnetization, 
leading to an unconventional coexistent phase of superconductivity and magnetism.

Starting with an introduction to the Pauli limited superconducting systems, in this article we 
will review the solid state and ultra cold atomic gas setups which hosts the FFLO superconducting 
phase, along with the relevant experimental diagnostics and their reported observations. This would 
be followed up by the theoretical attempts made to understand the physics of these novel superconducting 
systems, and the progress made therein. Finally, a non perturbative numerical technique would be discussed, 
which is suitable to capture the behavior of such many body systems, and edges over the existing ones in 
several aspects. Being a vibrant area of strong correlation physics, the Pauli limited superconductors 
hold several exciting promises for future investigations. We will touch upon them towards the end of 
the chapter.}
\end{chapabstract}

\section{Introduction}

Superconductivity is one of the most fascinating phenomena of quantum many body systems. 
Right from its discovery in mercury (Hg) by Kamerlingh Onnes in 1911 \cite{tinkham_book}, superconductivity 
has continued to intrigue the condensed matter physics community over the past more than a century. 
The basic idea that underlies any superconducting material is the formation of a macroscopic 
condensate with long range order, which manifests itself via properties like perfect diamagnetism, 
absence of electrical resistivity etc. below a transition temperature (T$_{c}$). It was established by Bardeen, 
Cooper and Schrieffer (BCS) in their pioneering work that the microscopic mechanism that gives rise to 
the superconducting behavior in a system is the formation of ``Cooper pairs'',  wherein, two 
fermions with equal and opposite momenta (${\bf K}_{\uparrow}$, ${-\bf K}_{\downarrow}$) pairs 
up by overcoming the repulsion between them \cite{bcs}. 

The BCS theory was originally devised to capture the physics of ``conventional'' superconductors, 
with low transition temperature such as, lead, mercury etc., and it was shown that in these 
materials the lattice phonons served as the attractive ``glue'' to pair up the fermions 
\cite{tinkham_book,bcs}. While the BCS theory is still the most celebrated microscopic theory 
in the history of superconductivity, the discovery of high temperature superconducting cuprates 
brought forth the limitations of the BCS 
theory \cite{muller_hightc}. The transition temperatures reported for the cuprates and subsequently for 
several other categories of superconducting materials were well above what could possibly be  
achieved by a phonon mediated coupling between the fermions \cite{lee_rmp2006,pfleiderer_rmp2009,stewart_rmp2011,johnson_adv2010}. 
In other words, there was some alternate mechanism which served as the attractive ``glue'' 
between the fermions in these ``unconventional'' superconductors with high transition temperature. 

Inspite of the change in the mechanism of superconductivity, some of the aspects however remained 
unaltered between the conventional 
and unconventional superconductors. For example the superconductivity was still dictated by the pairing 
between the fermions having equal and opposite momenta, i. e. the net momentum of the Cooper pair was zero. 
Keeping it consistent with the existing literature, we would refer to such superconducting pairing as 
the ``zero-momentum'' superconductivity, so as to distinguish them from the ``finite-momentum'' 
superconductivity, which would be introduced in the later sections of this chapter. 

It is obvious that once this pairing between the fermions is broken by some perturbation, superconductivity 
is destroyed. One 
of the most prominent superconducting pair breaking perturbation is the magnetic field. It is now well 
established that an orbital magnetic field when applied to a type-II superconductor leads to the formation of 
superconducting vortices beyond a lower critical field, H$\ge$H$_{c1}^{orb}$ \cite{de_gennes_book,saint_james_book}.
 As the 
magnetic field is increased the number of vortices penetrating the superconductor progressively increases 
and gives rise to the Abrikosov vortex lattice \cite{abrikosov1957}. As large number of vortices crowd the 
superconducting system the vortex cores begin to overlap each other at an upper critical field, 
H$_{c2}^{orb}=\Phi_{0}/2\pi\xi^{2}$, where 
$\Phi_{0}=\pi \hbar c/\vert e\vert$ is the flux quantum, and $\xi$ is the superconducting coherence length. 
The upper critical field (H$_{c2}^{orb}$) thus sets the strength of the orbital magnetic field required to 
destroy the superconductivity of a material \cite{tinkham_book,de_gennes_book,saint_james_book}. 

Superconductivity can also be destroyed by magnetic field via an alternate mechanism known as the 
Pauli paramagnetic pair breaking effect, originating from the Zeeman splitting of the single electron
energy levels. When a Zeeman field is applied to a normal metal the electrons undergo polarization because 
the Zeeman effect splits the Fermi surfaces corresponding to the up ($\uparrow$) and down ($\downarrow$) 
fermion species. This is known as Pauli paramagnetism.
On the other hand, a superconductor comprising of Cooper pairs is not readily polarized and one needs to 
break the pair in order to polarize the system. This pair breaking takes place when the Pauli paramagnetic 
energy, $E_{p}=(1/2)\chi_{n}H^{2}$ equals the superconducting condensation energy $E_{c}=(1/2)N(0)\Delta^{2}$. 
Here, $\chi_{n}=(1/2)(g\mu_{B})^{2}N(0)$ is the spin susceptibility of the normal state, $g$ is the spectroscopic 
splitting factor of an electron, $\mu_{B}$ is the Bohr magneton, $\Delta$ is the superconducting energy 
gap and $N(0)$ is the density of states at the Fermi level. The upper critical field at which the Pauli 
paramagnetic pair breaking effect takes place in a superconductor is estimated to be 
H$_{c2}^{P} = \sqrt{2}\Delta/g\mu_{B}$ and is known as the Chandrasekhar-Clogston 
limit \cite{chandrasekhar1962,clogston1962}.      

In general, the upper critical magnetic field of a superconductor is determined based on both the orbital and 
Pauli paramagnetic pair breaking effects. The relative importance of these effects is determined 
in terms of the Maki parameter $\alpha$, defined as \cite{saint_james_book}, 

\begin{eqnarray}
\alpha & = & \sqrt{2}\frac{H_{c2}^{orb}}{H_{c2}^{P}} \sim \Delta/\epsilon_{F} 
\end{eqnarray}
which is the ratio of the two upper critical magnetic fields at zero temperature (T=0). Here, $\epsilon_{F}$ 
is the Fermi energy. 
For most of the superconductors $\alpha < 1$, and thus the system loses superconductivity via the 
overlap of superconducting vortices much before the applied magnetic field reaches H$_{c2}^{P}$. 
There are however situations when $\alpha > 1$ and thus H$_{c2}^{P} < $ H$_{c2}^{orb}$, viz,  
(i) the Fermi energy is strongly suppressed, 
(ii) the superconducting system is two dimensional and is subjected to an in-plane magnetic field, 
such that the orbital effects can be neglected, and  
(iii) the system is charge neutral. 
The class of superconductors with $\alpha>1$ and which loses its superconductivity by Pauli paramagnetic 
pair breaking effect are known as the Pauli limited superconductors. Prominent examples include 
heavy fermion superconductor CeCoIn$_{5}$ \cite{bianchi2003}, layered two-dimensional organic superconductor 
$\kappa$-(BEDT-TTF)$_{2}$Cu(NCS)$_{2}$ \cite{lortz2007}, iron superconductor KFe$_{2}$As$_{2}$ \cite{zocco2013} 
and ultracold Fermi gas with imbalance in the populations of fermionic species \cite{ketterle_nature2008}.   

\subsection{Pauli limited superconductivity}

\begin{figure}
\centering
\includegraphics[width=0.96\linewidth]{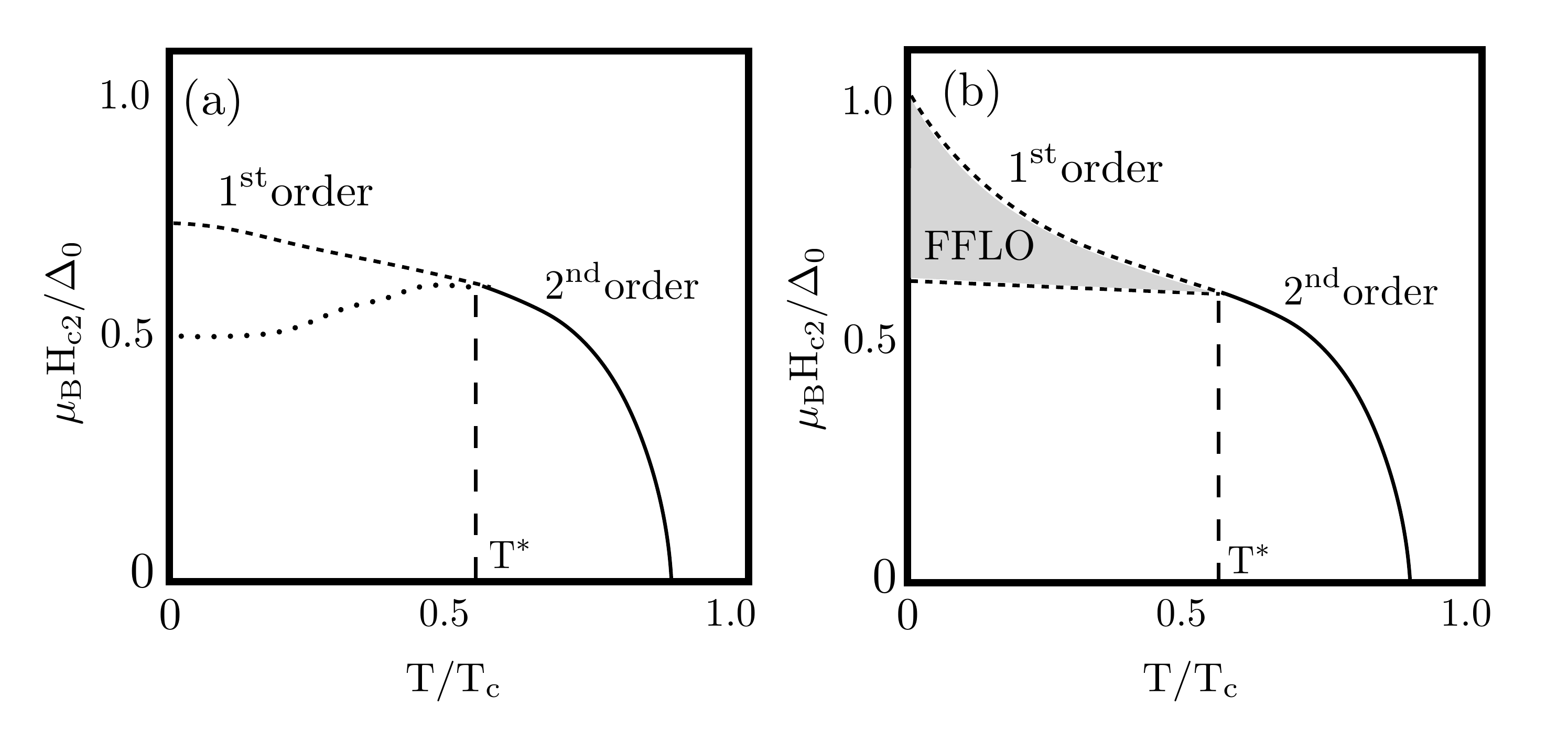}
\caption{Color online: Schematic phase diagram showing (a) the Chandrashekhar-Clogston limit of Pauli 
limited superconductor, (b) spatially modulated superconducting (FFLO) phase at high magnetic 
fields.}
\end{figure}

A Pauli limited superconductor in the absence of any orbital effect is known to demonstrate 
non trivial behavior in the in-plane magnetic field-temperature ($H-T$) phase diagram 
\cite{saint_james_book,ketterson_song_book},  as shown by the schematic in Figure 1(a). 
In the vicinity of T$_{c}$,  dH$_{c2}$/dT $\propto$ 
$\sqrt{T_{c}/(T_{c}-T)}$ and the initial slope of H$_{c2}^{P}$ is infinite, unlike the orbital 
limiting case for which the slope is finite. In the absence of Pauli limiting effect the loss 
of superconductivity at H$_{c2}^{orb}$ is via a second order phase transition. For a Pauli limited 
superconductor, however, the order of thermal phase transition changes below a tricritical 
temperature, T$^{*}$, say. While the transition is of second order for T$>$T$^{*}$, a first order 
transition is realized for T$<$T$^{*}$. In Figure 1(a) we represent the second and first order 
phase transitions by solid and dashed curves, respectively. The second order transition line 
(dotted) at T$<$T$^{*}$ correspond to a metastable state. The low temperature high magnetic field regime of
this phase diagram thus correspond to a first order thermal transition regime, and the associated 
field at T=0 is H$_{c2}^{P}$.

This Chandrasekhar-Clogston limit H$_{c2}^{P}$ was considered to be the upper critical field of Pauli 
limited superconductivity until Fulde-Ferrell (FF) \cite{ff1964} and Larkin-Ovchinnikov (LO) \cite{lo1964} 
pointed out that this critical field can be enhanced further, by taking into account the possibility of 
an inhomogeneous superconducting state in the high magnetic field regime. The Pauli paramagnetic pair 
breaking effect is reduced by a new 
pairing state (${\bf K}_{\uparrow}$, ${-{\bf K}+{\bf Q}}_{\downarrow}$) between the Zeeman split Fermi surfaces,  
where ${\bf Q}\neq 0$ is the centre of mass momentum of the Cooper pair. Thus, rather than a zero momentum 
superconducting state, the system hosts a finite momentum superconducting state in the regime of high 
magnetic field and low temperatures. We represent this possibility of finite momentum pairing by the 
schematic shown in Figure 1(b). Below the tricritical point (T$^{*}$) the transition is always first order, as 
was demonstrated later for continuum two and three-dimensional systems \cite{mora2004,mora2005}. 
This exotic finite momentum superconducting phase is known as the Fulde-Ferrell-
Larkin-Ovchinnikov (FFLO) state. The schematic presented in Figure 2 shows the finite momentum pairing between 
the fermions in the FFLO state in comparison with the zero momentum pairing in the BCS state. As is 
evident from the figure an FFLO state comprises of a mismatch in the size of the Fermi surfaces 
corresponding to the up and down fermion species. Such a Fermi surface mismatch gives rise to unequal 
populations of the fermionic species i. e. a population imbalance.
\begin{figure}
\centering
\includegraphics[width=0.86\linewidth]{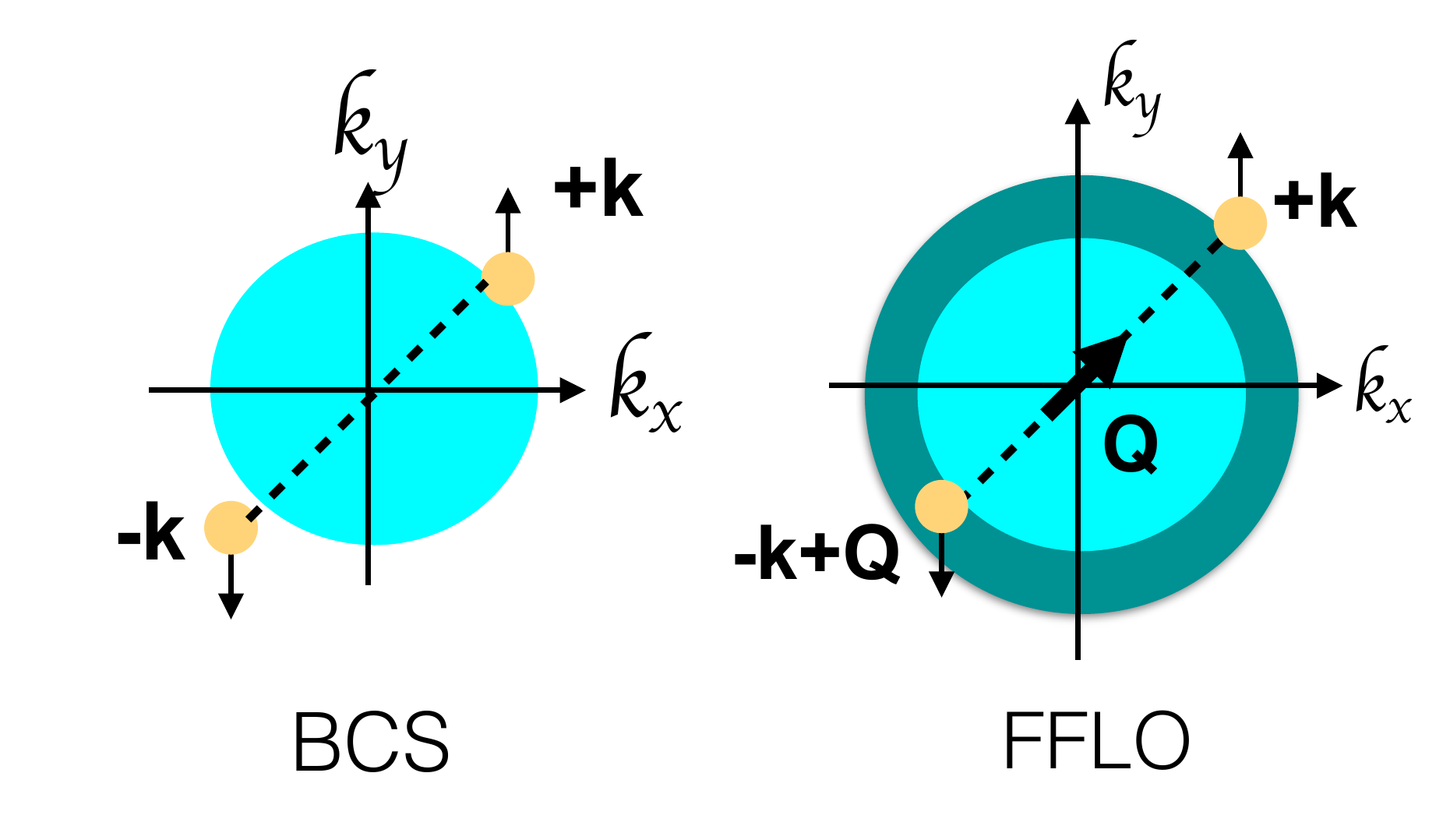}
\caption{Color online: Schematic diagram showing the Fermi surface mismatch in the FFLO 
phase in comparison with the Fermi surface in the BCS state.}
\end{figure}

The symmetry breaking arising due to the finite momentum ${\bf Q}$ gives rise to spatial modulations in the 
superconducting order parameter. There had been several proposals regarding the form of this spatially modulated 
order parameter with the first one suggested by Fulde and Ferrell being \cite{ff1964}, 

\begin{eqnarray}
\Delta({\bf r}) & = & \vert\Delta\vert e^{i{\bf Q}.{\bf r}}
\end{eqnarray}
Here, the superconducting amplitude is homogeneous but the phase undergoes periodic modulations in the real space. 
In a similar spirit Larkin and Ovchinnikov suggested that a linear combination of two plane wave solutions 
corresponding to degenerate superconducting states as \cite{lo1964}, 
\begin{eqnarray}
\Delta({\bf r}) & = & \vert \Delta\vert(e^{i{\bf Q}.{\bf r}}+e^{-i{\bf Q}.{\bf r}}) = 2\vert\Delta\vert\cos({\bf Q}.{\bf r})
\end{eqnarray}
can give rise to a spatially modulated superconducting state, such that the amplitude rather than the phase 
of the superconducting order parameter is modulated. In general, depending upon the symmetry of the system 
more than two plane waves can undergo linear combination to give rise to a spatially modulated 
superconducting order parameter \cite{lo1964,mora2004,mora2005,shimahara1998,bowers2002,mora2_2005,wang2006,combescot2005,shimahara2007}. It has been found that for a two-dimensional thin film the linear combination
of two and three plane waves give rise to square and hexagonal states, respectively \cite{shimahara1998,shimahara2007}. The spatial modulations give rise to nodes in the superconducting order parameter, which serve as hosts 
to the unpaired fermions arising out of the Pauli paramagnetic pair breaking. The FFLO phase can thus 
be envisaged as a coexistent phase wherein finite momentum superconductivity and finite magnetization 
(arising out of the unpaired fermions) compliment each other in real space. 

For a long time, the FFLO superconducting state though theoretically proposed early on, failed to garner the 
interest it deserved. One of the principal reasons for the same was its inaccessibility to the existing 
experimental probes. The FFLO state in two-dimensional materials is highly susceptible to impurities 
\cite{aslamazov1969,takada1970}, consequently 
the sample needs to be very clean in order to observe the FFLO state. Even for a clean sample the 
applied magnetic field should be applied precisely parallel to the plane because  
any small perpendicular component of the magnetic field would give rise to 
Abrikosov vortices and destroy the FFLO state \cite{shimahara1997}. Secondly, the transition 
temperature of the FFLO state 
is strongly suppressed and the state gets rapidly destroyed by thermal fluctuations even in three 
dimensions, which makes its experimental detection difficult \cite{shimahara2_1998,ohashi2002}. 
It was however suggested that crystal anisotropy of real materials can 
stabilize the FFLO phase against thermal fluctuations \cite{shimahara2_1998}. 

Though the FFLO phase continues to be the most sought after superconducting phase in systems with 
Fermi surface mismatch, there are other candidate phases which can be stabilized in these 
systems, viz. Breached Pair (BP) \cite{sarma1963,wilczek2003,gubankova2003} and deformed Fermi surface superfluid 
phases \cite{muther2002,sedrakian2005}. Unlike FFLO, both the BP and deformed Fermi surface phases 
comprise of zero momentum pairing between the fermions. The 
breached pair phase is characterized by the coexistence of finite magnetization and ``gapless'' 
zero momentum superconductivity. The issue of BP phase was first raised by Sarma in his seminal 
work where he discussed the possibility of self consistent mean field solution with gapless mode 
in $s$-wave superconductors, in presence of an applied magnetic field \cite{sarma1963}. He however found that such a
phase even if it exists would be energetically unfavorable as compared to the BCS state. Later Liu 
and Wilczek demonstrated that while a contact interaction between the species in a population imbalanced 
system does not allow 
the BP phase to be a stable ground state, the same can indeed be achieved under non trivial circumstances 
such as, imbalance in fermion effective masses with contact interaction between them, momentum dependent 
interaction, same species repulsion etc \cite{wilczek2003,gubankova2003,forbes2005,liu2004}. 
Another possible mechanism to achieve stable ground state in an imbalanced Fermi system is 
through deformed Fermi surfaces which spontaneously breaks the rotational symmetry \cite{muther2002,sedrakian2005}. 

\section{Experimental signatures}

The renewed interest on imbalanced superconducting systems is attributed to 
two key experimental breakthroughs viz. (i) the advent of ultracold atomic gases and (ii) improvisation 
of experiments on solid state materials. Before going into the details of the experimental observations 
we highlight the candidate systems amenable to the realization of imbalanced superconductivity 
and superfluidity. In general, the term superconductivity is used in the context of solid state 
systems containing charged electrons, while for the charge neutral ultracold atomic gases, superfluidity 
is the correct choice of nomenclature to describe the state with long range macroscopic phase coherence. 
In the remaining sections of this chapter we will follow this protocol to refer to the solid 
state and neutral cold atomic gases.   

We mentioned in the earlier sections that in order to realize FFLO phase the superconducting material should 
have large Maki parameter ($\alpha >$1). This requirement is stringent and is achievable only under 
specific conditions.
\begin{itemize}
\item{In a three dimensional superconductor if H$_{c2}^{orb}$ is significantly larger than 
H$_{c2}^{P}$ then it would lead to a large Maki parameter. In heavy fermion superconductors 
such as CeCoIn$_{5}$ the large mass of the electrons give rise to low electron velocity 
and thus a high H$_{c2}^{orb}$. The resulting high Maki parameter $\alpha \sim 5$ makes it 
a suitable candidate to realize FFLO superconductivity.}
\item{In layered two-dimensional materials the orbital effects are strongly suppressed when the 
magnetic field is applied, which makes a large Maki parameter plausible in such systems. 
A candidate material is quasi two-dimensional organic superconductor $\kappa$-(BEDT-TTF)$_{2}$Cu(NCS)$_{2}$, 
with Maki parameter of $\alpha \sim 8$.}
\item{If the system is a neutral Fermi gas, i. e. comprising of neutral atoms then the applied magnetic 
field can couple only to the spin degrees of freedom. This makes the ultracold atomic gases a promising 
candidate to realize FFLO phase. In these systems 
the superfluid pairing takes place between the fermions present in two hyperfine energy 
states of the neutral atoms or it can take place between the fermions in the energy states 
of the atoms belonging to two different atomic species or to two different isotopes of the same 
atomic species.}
\end{itemize}

We now discuss the experimental observations made on these candidate systems, which suggests 
that these systems host exotic phases such as, FFLO and BP, in presence of applied magnetic field 
or population imbalance of fermionic species.   

\subsection{Heavy fermion superconductor:- CeCoIn$_{5}$}

One of the earliest experiments which suggested the possibility of FFLO phase in CeCoIn$_{5}$ 
(T$_{c} \sim $2.3K) were the specific heat measurements reported by Bianchi {\it et al.} 
\cite{bianchi2003}. They demonstrated that below a critical temperature T$_{0} \sim$ 0.3T$_{c}$ and in presence 
of high magnetic field (H $>$ 10T) applied either in-plane or out of plane of the material, the thermal 
transition of superconducting to normal state changes from second to first order. They identified a 
specific heat anomaly in this regime of first order 
transition and attributed it to an underlying spatially modulated superconducting state. 
Similar observations were earlier made by Tayama {\it et al.} \cite{tayama2002} in their dc magnetization measurements 
on CeCoIn$_{5}$, which showed a first order phase transition, indicating strong Pauli paramagnetic 
suppression of the superconducting pairing in the regime of high magnetic field and low temperatures 
(T$_{0}\sim$ 0.3T$_{c}$). The possibility of FFLO phase in CeCoIn$_{5}$ was further established 
by thermal conductivity \cite{capan2004} as well as by resistivity and penetration depth 
measurements \cite{martin2005}. In presence of an in-plane applied magnetic field the thermal 
conductivity data showed a discontinuous jump indicating a change from second to first order phase 
transition below a critical temperature of T$_{0} \sim $1K. Similar signatures were detected 
by Martin {\it et al.} in their resistivity measurements \cite{martin2005}.

Based on the nuclear magnetic resonance (NMR) measurements carried out on single crystal of CeCoIn$_{5}$,  
Koutroulakis {\it et al.} \cite{koutroulakis2008} mapped out the (in-plane) magnetic field-temperature 
phase diagram of this system. Based on the results of NMR shift they identified three different regimes in 
the thermal phase diagram. In the regime of low magnetic field (H$\lesssim$ 9.2T) a long range ordered 
(LRO) homogeneous superconducting phase was identified. In the high magnetic field regime (H$\ge$ 10.2T), 
in-commensurate spin density wave (IC-SDW) order was found to coexist with FFLO at low temperatures. The 
intermediate magnetic field regime (9.2T $\le$ H $<$ 10.2T) was inferred to comprise of FFLO phase surviving 
with antiferromagnetic (AFM) correlations. This phase was found to undergo first order thermal transition. 
Similar inferences were
made earlier by Kumagai {\it et al.} \cite{kumagai2006} via NMR Knight shift measurements on CeCoIn$_{5}$. They 
observed that the FFLO superconductivity is realizable for both in-plane and out of plane magnetic 
fields and the corresponding phase diagrams are qualitatively different from each other. 

By carrying out high magnetic field neutron scattering experiments on CeCoIn$_{5}$ Gerber {\it et al.}
\cite{gerber2013} identified a ``Q-phase'' in the high magnetic field low temperature regime, corresponding 
to a finite 
momentum pairing. They further confirmed that the finite momentum pairing coexists with an SDW order 
and undergoes first order thermal phase transition at H$ \sim 9.8$T. In a relatively recent work Kim {\it et al.}
\cite{kim2016} suggested that though there is indeed a Q-phase in the high field low temperature regime of 
the CeCoIn$_{5}$ phase diagram in presence of an in-plane applied magnetic field, it does not correspond to a FFLO 
phase. It was inferred based on their thermal conductivity measurements that the Q-phase comprises 
of intertwinned SDW, $d$-wave superconducting and inhomogeneous $p$-wave pair density wave (PDW) 
orders. Evidently, the true nature of the Q-phase seems to be unsettled. A very recent 
thermal conductivity measurement by the same group suggests that the characteristic of the Q-phase 
is intimately connected to the direction of the applied magnetic field \cite{lin2020}. For a magnetic field 
applied along the ab-plane of the crystal the Q-phase is likely to host a SDW order. On the other hand 
for a field applied perpendicular to the ab-plane a FFLO order is realizable in the high field 
low temperature Q-phase of CeCoIn$_{5}$ \cite{lin2020}.   

\subsection{Layered organic superconductor:- \\ $\kappa$-(BEDT-TTF)$_{2}$Cu(NCS)$_{2}$}
 
In two-dimensional layered organic superconductor $\kappa$-BEDT (T$_{c} \sim$ 9.5K), the FFLO 
phase was inferred from specific heat measurements 
in presence of applied in-plane magnetic field \cite{lortz2007}. Unlike CeCoIn$_{5}$ it was observed that in $\kappa$-BEDT the thermal transitions from both the zero momentum and FFLO superconducting phases are of 
first order. A Zeeman field driven phase transition was detected in $\kappa$-BEDT via NMR measurements
by Wright {\it et al.} \cite{wright2011}. They suggested that this transition and the associated discontinuity in the 
derivative of magnetization, at H$\sim$213$\pm$3kOe amounts to a possible indication of spatially 
inhomogeneous superconducting phase. Similar inference was made by Bergk {\it et al.} \cite{bergk2011} 
and Agosta {\it et al.} \cite{agosta2012} based on magnetic torque and penetration depth measurements, 
respectively. The high magnetic field (H $\ge$ 21T) regime was suggested to host FFLO superconducting 
phase.  The most clear evidence of FFLO phase in 
$\kappa$-BEDT was given by Mayaffre {\it et al.} through NMR measurements \cite{mayaffre2014}. They observed 
that the NMR spin-lattice relaxation rate gets significantly enhanced in the high magnetic field 
regime, and attributed this observation to the Andreev bound states arising out of the unpaired 
quasiparticles residing in the nodes of the spatially modulated FFLO state.   

In addition, the presence of FFLO superconducting phase has been recently proposed in 
iron superconductors such as KFe$_{2}$As$_{2}$, LiFeAs etc. \cite{zocco2013,cho2011,khim2011}, 
which are suggested to be two-band superconductors. 

\subsection{Cold atomic gases and Fermi superfluids}

Over the past decade ultracold Fermi gases have proved to be the testbed of research on imbalanced 
superfluid systems, owing to the flexibility of tuning the interaction parameters (via Feshbach resonance), 
optical lattice and trap geometries, that these systems provide. Unlike the solid state superconducting 
systems where a magnetic 
field is applied, population imbalance in ultracold atomic gases is created by loading unequal populations 
of fermionic species in the optical lattice. Alternatively, equal populations of different atomic 
species with unequal effective masses 
can be loaded in the optical lattice to create mass imbalanced Fermi gases. Clear evidence of 
superfluid pair correlations have been found in case of both balanced \cite{gaebler2010,chin2004} as well 
as imbalanced \cite{ketterle_nature2008} Fermi gases. Most of these experiments are carried out at or close 
to unitarity. 

The concept of unitarity is intimately related to the $s$-wave scattering 
length which can be controlled via the Feshbach resonance. In three-dimensional (3D) continuum Fermi gas the 
scattering length $a_{3D} \rightarrow \infty$ as $g \rightarrow g_{c}$. Here, $g_{c}$ is finite in 3D 
and defines the interaction strength at which the first two-body bound state formation takes place in 
vacuum. The dimensionless coupling $1/k_{F}a_{3D}=0$ (where, $k_{F}$ is the Fermi momentum) at $g_{c}$ 
and also corresponds to the regime of maximum T$_{c}$ of the 3D Fermi gas. For two-dimensional 
(2D) Fermi gases unitarity corresponds to the coupling strength at which 
$ln(k_{F}a_{2D}) \rightarrow 0$ ($a_{2D}$ is $s$-wave scattering 
length in 2D), which once again corresponds to the interaction for maximum T$_{c}$. For a lattice the 
notion of $k_{F}$ is not very meaningful particularly if one is away from the regime of small filling. 
In the context of lattice fermion models, the crossover regime or unitarity corresponds to the interaction 
strength at which T$_{c}$ is maximum, where neither the fermionic nor bosonic description of the 
system holds true completely \cite{gaebler2010,sewer2002,paiva2010,ries2015}. We will revisit the issue of 
unitarity in the 
later sections of this chapter, when we discuss the BCS-BEC crossover picture.     
We now discuss some of the important experiments on imbalanced Fermi gas in ultracold  
cold atomic gas setups, and their principal observations.

Wolfgang Ketterle and his group mapped out the thermal phase diagram of strongly interacting imbalanced spin 
mixture created between two hyperfine states of ${}^{6}$Li atoms \cite{ketterle_science2007}. 
Based on radio frequency (rf) spectroscopy 
measurements they identified a gap in the single particle excitation spectra which indicated that 
the minority species is paired up. It was shown that in presence of 
large imbalance, at high temperature the spectra shows only the atomic peak which was considered to be 
the evidence of unpaired fermions in the system. As the temperature is lowered, in addition to the 
atomic a pairing peak also appeared in the spectra indicating a coexistent phase with superfluid 
pairs and unpaired fermions. 
Further decreasing the temperature left behind only the pairing peak. The work showed high temperature 
coexistence of paired and unpaired fermions but did not find any indication of superfluidity at low temperatures 
in the regime of large imbalance. Note that at large imbalance there would not be enough minority 
spins to pair up with the majority species. Consequently, signatures of pair correlations (if any) 
would be strongly suppressed.   

Shin {\it et al.} \cite{shin2006} reported the experimental observation of phase separation between 
superfluid and normal regions in a strongly interacting Fermi gas of ${}^{6}$Li with population 
imbalance. They used a special phase contrast imaging technique and 3D image reconstruction to 
determine the density difference between 
the two components of the gas in an optical trap. A shell structure was observed,  in which the superfluid 
region containing equal densities of the components is surrounded by a normal region of unequal densities. 
Furthermore, using phase contrast imaging they characterized the normal to superfluid phase transition. 

A quantum phase transition was detected in the unitary imbalanced Fermi gas of ${}^{6}$Li at a critical polarization 
(magnetization) of $\sim 36 \%$ by Shin {\it et al.} \cite{ketterle_nature2008}. Further, they detected the change 
in the order of thermal  phase transition as a function of polarization. In the regime of low temperatures 
and increasing polarization 
a first order phase transition was detected via a jump in the polarization. On the other hand the transition 
at high temperatures and weak polarization was found to be of second order. These two orders of phase transition 
are connected at a tricritical point. While they carried out the experiment in presence of trapping potential, 
using tomographic reconstruction of local Fermi temperatures and spin polarization they determined the phase 
diagram of the homogeneous system free from any inhomogeneity arising due to the trapping potential. The resulting 
phase diagram comprised of a superfluid phase in the regime of weak polarization characterized by  
spatial coexistence of superfluid order and non zero polarization at finite temperatures, akin to the 
BP phase. This regime undergoes second order thermal 
phase transition to a normal phase. The large polarization regime is a phase separated (unstable) regime 
which undergoes a first order thermal phase transition. At still larger polarizations the system is in the normal 
phase. Note that signatures of FFLO phase was not detected in this experiment, however Shin {\it et al.} did not 
rule out the possibility of such a phase at large polarizations \cite{ketterle_nature2008}. 

In a different experimental set up,  containing arrays of one-dimensional tubes constructed using a two- 
dimensional optical lattice, Liao {\it et al.} \cite{liao2010} investigated the phases of an imbalanced Fermi gas of 
${}^{6}$Li. Due to nesting of Fermi surfaces the FFLO phase is expected to have a larger regime of stability 
in one dimension as compared to the higher dimensions. Liao {\it et al.} \cite{liao2010} mapped out the phase 
diagram of this system in polarization-temperature plane and demonstrated that, at low imbalance a 
partially polarized phase is realized at the centre of the trap, which radially spreads out with increasing 
imbalance. This partially polarized region extends to the edge of the cloud at a critical imbalance. 
Further increase in imbalance leads the edges of the cloud to become completely polarized. Liao {\it et al.} 
found their experimental phase diagram to be in agreement with theoretical predictions and suggested that even 
though signatures of FFLO phases was not visible in their experiment it was certainly a possibility \cite{liao2010}. 

Experimental investigations of mass imbalanced Fermi-Fermi mixtures are relatively fewer. 
A mass imbalanced Fermi-Fermi mixture is achievable for example in a ${}^{6}$Li-${}^{40}$K mixture. While 
superfluidity in such a system is yet to be attained in experiments, the Fermi degenerate 
regime \cite{tagleiber2008,naik2011} as well as the Feshbach resonance between ${}^{6}$Li 
and ${}^{40}$K atoms \cite{naik2011,wille2008,costa2010} and formation 
of ${}^{6}$Li-${}^{40}$K heteromolecules \cite{voigt2009} are already a reality. Furthermore, experimental 
realization of mixtures of other fermion species (such as,  ${}^{161}$Dy, ${}^{163}$Dy, ${}^{167}$Er) 
are expected in future \cite{lu2012,frisch2013}. It has been reported that for a double-degenerate 
${}^{6}$Li-${}^{40}$K mixture the Fermi temperatures are T$_{Li}$=390 nK and T$_{K}$=135 nK, 
for Li and K species, respectively \cite{spiegelhalder2010}. In comparison, 
for a balanced Fermi gas of ${}^{6}$Li, the Fermi temperature is known to be T$_{F}$=1.0$\mu$K 
\cite{pieri2011} with the corresponding T$_{c}$ scale being T$_{c} \sim$ 0.15T$_{F}$ \cite{nascimbne2010}.
In other words, imbalance in mass strongly suppresses the superfluid T$_{c}$ of the Fermi-Fermi mixtures, 
making their experimental detection non trivial. However, short range superfluid pair correlations are 
likely to survive at temperatures significantly higher than the superfluid T$_{c}$ and accessible to 
experimental probes. 
 
\section{Theoretical formalism}

The experimental observations across different classes of solid state materials and ultracold atomic gases 
lead to a flurry of activities in the theoretical front. The theoretical studies can broadly 
be classified in two categories as, (i) the lattice fermion models and (ii) the continuum models. 
In this chapter we will mainly focus on (i). For a detail discussion on continuum 
models with imbalance the readers are referred to \cite{radzhiovsky_pra,gurarie2007,casellbouni_rmp}. 
Before getting into the details of the results obtained from different lattice fermion based calculations, 
let us first set up the theoretical framework which can be used to investigate imbalanced Fermi superfluids and 
superconductors. For this we resort to the simplest tool of mean field theory. We will use the 
framework of BCS theory and Bogoliubov-Valatin transformation 
to set up the mean field formalism. The generic Hamiltonian of the two component Fermi gas 
with an on-site interaction between the fermion species is used to model the system. The BCS 
mean field approximation is used such that the order parameters are striped of all 
fluctuations. We take into account the imbalance between the population of the fermion species 
and allow for the possibilities of exotic superfluid phases such as, FFLO and BP, in addition to 
the homogeneous BCS phase. Bogoliubov-Valatin transformations \cite{de_gennes_book} are used to 
diagonalize the Hamiltonian and the resulting eigenvalues and eigenfunctions are used to compute different observables 
relevant from the point of view of experiments. Once the basic formalism is in place we 
will discuss the modifications required to address different families of imbalanced
superconductors and superfluids. 

\subsection{Mean field theory}

We consider the Hamiltonian of a two component Fermi gas comprising of up ($\uparrow$) 
and down ($\downarrow$) fermion species, with an attractive interaction between them. 
For the sake of simplicity we do not take into account any external potential acting 
on the fermions. The Hamiltonian corresponding to such a system reads as, 

\begin{eqnarray}
\hat H & = & \int d{\bf r} \sum_{\sigma} \hat \psi_{\sigma}^{\dagger}({\bf r}) (\hat K_{\sigma}({\bf r})-\mu_{\sigma})
\hat \psi_{\sigma}({\bf r}) \nonumber \\ && + \int \int d{\bf r}d{\bf r}^{\prime} V_{\uparrow \downarrow}({\bf r},{\bf r}^{\prime}) 
\hat \psi_{\uparrow}^{\dagger}({\bf r})\hat \psi_{\downarrow}^{\dagger}({\bf r}^{\prime})\hat \psi_{\downarrow}({\bf r}^{\prime})\hat \psi_{\uparrow}({\bf r})  
\end{eqnarray}   
where, $\sigma = \uparrow, \downarrow$ correspond to the two fermion species. $V_{\uparrow \downarrow}({\bf r},{\bf r}^{\prime})$ is the interaction between the species, with ${\bf r}$ and ${\bf r}^{\prime}$ being the position coordinates, corresponding to a two dimensional system. The kinetic energy contribution to the Hamiltonian is given as, 
$\hat K_{\sigma}({\bf r}) = -\frac{\hbar^{2}}{2m_{\sigma}}\frac{\partial^{2}}{\partial {\bf r}^{2}}$ and 
$\mu_{\sigma}$ is the species dependent chemical potential. Note that the effective mass ($m_{\sigma}$) 
in the kinetic energy contribution is dependent on the species. For a system with only imbalance in fermionic 
populations $m_{\uparrow}=m_{\downarrow}=m$ since the mass of the fermions of both the species are the same. 
On the other hand, for a system comprising of two different isotopes of the same atom or of two different 
atomic species, $m_{\uparrow} \neq m_{\downarrow}$. Such systems can be realized in ultracold atomic gases 
and are called ``mass imbalanced'' Fermi-Fermi mixtures. For the present discussion we consider equal effective 
masses for both the species and set $m_{\uparrow} = m_{\downarrow} = m$.

We simplify the model by approximating the interaction to be a contact interaction as, 
\begin{eqnarray}
V_{\uparrow \downarrow}({\bf r},{\bf r}^{\prime}) & = & V_{0}\delta({\bf r}-{\bf r}^{\prime})
\end{eqnarray}
This approximation is particularly good for ultracold atomic gases in which the interaction is short ranged 
\cite{bloch2008}. With this approximation the Hamiltonian takes the form, 

\begin{eqnarray}
\hat H & = & \int d{\bf r} \sum_{\sigma} \hat \psi_{\sigma}^{\dagger}({\bf r}) (\hat K_{\sigma}({\bf r})-\mu_{\sigma})
\hat \psi_{\sigma}({\bf r}) \nonumber \\ && + V_{0}\int d{\bf r}\hat \psi_{\uparrow}^{\dagger}({\bf r})\hat \psi_{\downarrow}^{\dagger}({\bf r})\hat \psi_{\downarrow}({\bf r})\hat \psi_{\uparrow}({\bf r})
\end{eqnarray}

We consider an underlying lattice potential and for that expand the continuum Fermi operators in terms of Wannier 
functions \cite{kohn1959} as, 
\begin{eqnarray}
\hat \psi_{\sigma}^{\dagger}({\bf r}) & = & \sum_{{\bf n}, i} \hat c_{\sigma, {\bf n}, i}^{\dagger}w_{\bf n}^{*}({\bf r}-{\bf r}_{i})
\end{eqnarray}
where, $\hat c_{\sigma,{\bf n},i}^{\dagger}$ creates a $\sigma$ particle in the lattice site $i$ in the energy band 
${\bf n}$, and $w_{\bf n}^{*}({\bf r}-{\bf r}_{i})$ are the Wannier functions. The Wannier functions can be 
approximated by harmonic oscillator states in the tight-binding limit. At low temperatures and weak interactions only 
the lowest energy band ${\bf n}=0$ would be occupied by the fermions, we can thus omit the band index henceforth. 
Using the above expansion the continuum Hamiltonian is re-written as, 

\begin{eqnarray}
\hat H & = & -\sum_{\sigma}\sum_{\langle ij\rangle}t_{\sigma, ij}\hat c_{\sigma,i}^{\dagger}\hat c_{\sigma,j} + \sum_{\sigma,i}(\epsilon_{\sigma}-\mu_{\sigma})\hat c_{\sigma,i}^{\dagger}\hat c_{\sigma,i}\nonumber \\ && + U\sum_{i}\hat c_{\uparrow,i}^{\dagger}\hat c_{\downarrow,i}^{\dagger}\hat c_{\downarrow,i}\hat c_{\uparrow,i} 
\end{eqnarray}

Here, $\langle ij\rangle$ corresponds to the summation over nearest neighbors of a lattice and $t_{\sigma,ij}$ is the hopping amplitude between the nearest neighbors; $\epsilon_{\sigma}$ is the onsite energy and $U$ is the on-site interaction, and are expressed as \cite{jaksh1998}, 

\begin{eqnarray}
t_{\sigma,ij} & = & -\int d{\bf r}w^{*}({\bf r}-{\bf r}_{i})\hat K_{\sigma}({\bf r})w({\bf r}-{\bf r}_{j}) \\ 
\epsilon_{\sigma} & = & \int d{\bf r}w^{*}({\bf r}-{\bf r}_{i})\hat K_{\sigma}({\bf r})w({\bf r}-{\bf r}_{i}) \\ 
U & = & V_{0}\int d{\bf r}\vert w({\bf r}-{\bf r}_{i})\vert^{4}
\end{eqnarray}

Equation (8) corresponds to the standard single band Hubbard model for a two component Fermi gas. In order to 
take into account the imbalance in the population of fermionic species, we set $\mu_{\uparrow} \neq \mu_{\downarrow}$.
We consider the system to be mass balanced and the hopping amplitude to be a 
constant, such that $t_{\uparrow,ij} = t_{\downarrow,ij}=t_{ij}=t$. We also consider $\epsilon_{\uparrow}=\epsilon_{\downarrow}=\epsilon = 0$, i. e. the on-site energy is taken to be a constant and set to zero. The resulting Hamiltonian 
thus takes the form, 
\begin{eqnarray}
\hat H & = & -t\sum_{\sigma}\sum_{\langle ij\rangle}\hat c_{\sigma,i}^{\dagger}\hat c_{\sigma,j}-\sum_{\sigma,i}\mu_{\sigma}\hat c_{\sigma,i}^{\dagger}\hat c_{\sigma,i}+U\sum_{i}\hat c_{\uparrow,i}^{\dagger}\hat c_{\downarrow,i}^{\dagger} \hat c_{\downarrow,i}\hat c_{\uparrow,i} \nonumber \\
\end{eqnarray} 
Here, $U<0$ corresponding to the attractive interaction between the fermions. 

\subsubsection{BCS theory of uniform superconductors}

Equation (12) comprises of a four-fermion interaction term which is not readily solvable. In order to make the 
Hamiltonian tractable we use the BCS mean field approximation. Within this approximation we introduce a pairing field
($\hat \Delta_{i}$) to describe the Cooper pairs. We do not go into the details of the BCS theory here and guide the 
interested readers to \cite{tinkham_book,de_gennes_book,mahan_book,fetter_book}. The pairing field thus introduced serves 
as the superconducting ``order parameter'' and is related to the expectation value of $\hat \psi_{\downarrow}\hat \psi_{\uparrow}$ as, 

\begin{eqnarray}
\langle\hat \Delta_{i} \rangle \equiv \Delta_{i} & = & V_{0}\langle \hat \psi_{\downarrow}({\bf r}) \hat \psi_{\uparrow}({\bf r})\rangle = U\langle \hat c_{i\downarrow}\hat c_{i\uparrow}\rangle
\end{eqnarray} 

The standard BCS theory describes the transition from a normal Fermi gas to a zero momentum uniform superconducting 
state, and the pairing field is considered to be a constant, $\Delta_{i}=\Delta$. In order to capture the 
inhomogeneous superconducting state such as FFLO or BP, one needs to retain the spatial dependence 
of the pairing field. 

Note that the pairing field defined so far is on-site and does not take into account any directional dependence. 
In the language of superconducting pairing state symmetry, such an on-site pairing correspond to a $s$-wave 
superconducting state. For the sake of simplicity we will discuss the theoretical formalism in detail for a 
$s$-wave superconducting state and will outline the modifications of this formalism for a non $s$-wave pairing 
state, at the end of this section.

The BCS approximation in addition to the pairing field introduces Hartree field corresponding to the 
fermionic number densities, which is defined as, $n_{\sigma}=\langle \hat c_{\sigma}^{\dagger}\hat c_{\sigma}\rangle$.
Including the Hartree field in the formalism renormalizes the chemical potential as, 

\begin{eqnarray}
(-\mu_{\sigma}+Un_{-\sigma})\hat c_{\sigma}^{\dagger}\hat c_{\sigma} & \equiv & -\tilde \mu_{\sigma}\hat c_{\sigma}^{\dagger}\hat c_{\sigma}
\end{eqnarray}

Here, $\tilde \mu_{\sigma}$ is the renormalized chemical potential and the densities are considered to be uniform. 
In terms of the average pairing and the Hartree fields the mean field Hamiltonian describing a  
$s$-wave superconducting pairing of the fermions in a two component Fermi gas, reads as, 

\begin{eqnarray}
\hat H & = & -t\sum_{\sigma}\sum_{\langle ij\rangle}\hat c_{\sigma,i}^{\dagger}\hat c_{\sigma,j} -\sum_{\sigma}\sum_{i}\tilde \mu_{\sigma}\hat c_{\sigma,i}^{\dagger}\hat c_{\sigma,i} \nonumber \\ && + \sum_{i}(\hat c_{\uparrow,i}^{\dagger}\hat c_{\downarrow,i}^{\dagger}\Delta_{i} + \Delta_{i}^{*}\hat c_{\downarrow,i}\hat c_{\uparrow,i} - \frac{\vert \Delta_{i}\vert^{2}}{U})
\end{eqnarray}

In the Hamiltonian given by Equation (15) the pairing field is replaced by its approximated average value 
(see Equation 13), but we do not assume any specific form of this average. 

Note that the approximation of uniform density should be used carefully. The approximation breaks down in case 
of superconductors with spatial inhomogeneities, such as those arising in a FFLO of BP phase or in disordered 
superconductors. In such cases the correct tool of choice is Bogoliubov-de-Gennes (BdG) formalism which takes into 
account the spatial inhomogeneities of the pairing field amplitude ($\vert \Delta_{i}\vert$) as well as of 
the number densities \cite{tinkham_book,de_gennes_book}.  

\subsubsection{Mean field theory of FFLO superconductivity}

The FFLO phase comprises of spatially modulated superconducting state 
coexistent with non zero magnetization. The corresponding Cooper pair centre of mass momentum 
${\bf Q}$ is determined by minimizing the energy of the system. While the FF state corresponds to 
a superconducting state with modulation in the pairing field phase 
($\Delta({\bf r}) = \Delta e^{i{\bf Q}.{\bf r}}$), the LO ($\Delta({\bf r}) = \Delta \cos({\bf Q}.{\bf r})$) 
state on the other hand contains modulation in the pairing field amplitude. 

There is now a consensus that at the ground state of two component population imbalanced Fermi gas the LO 
phase is energetically favorable as compared to the FF phase, both 
in case of lattice as well as continuum systems \cite{mora2005,yoshida2007,batrouni2008,baarsma_jmodopt}. 
An exception however is the population imbalanced system with spin-orbit 
coupling, where the FF phase is found to have lower energy as compared to that of the LO phase. 
Strictly speaking, the regime of 
stability of the FF and LO phases should depend upon the system under consideration, however for many systems 
they are found to overlap. 

In this section we consider the simpler FF ansatz to discuss the mean field theory of the spatially modulated 
superconducting phase. The theory can however be easily generalized to describe the LO superconducting phase. 
Using the ansatz for the FF state ($\Delta_{i} = \Delta_{i} e^{i{\bf Q}.{\bf r}} = \Delta_{i} e^{i(Q_{x}x+Q_{y}y)}$) 
the superconducting Hamiltonian for the population imbalanced two component Fermi gas in two-dimensions is given as, 

\begin{eqnarray}
\hat H & = & -t\sum_{\sigma}\sum_{\langle ij\rangle} \hat c_{\sigma,i}^{\dagger}\hat c_{\sigma,j}-\sum_{\sigma}\sum_{i}\tilde \mu_{\sigma}\hat c_{\sigma,i}^{\dagger}\hat c_{\sigma,i} \nonumber \\ && + \sum_{i}(\Delta_{i} e^{i(Q_{x}.x+Q_{y}.y)}\hat c_{\uparrow,i}^{\dagger}\hat c_{\downarrow,i}^{\dagger}+\Delta_{i} e^{-i(Q_{x}.x+Q_{y}.y)}\hat c_{\downarrow,i}\hat c_{\uparrow,i} - \frac{\Delta_{i}^{2}}{U}) \nonumber \\ 
\end{eqnarray}  

Here, $Q_{x}$ and $Q_{y}$ correspond to the components of the Cooper pair centre of mass momentum 
${\bf Q}$,  in two dimensions.
In order to determine the superconducting gap and number equations this Hamiltonian needs to be diagonalized. 
Since the Hamiltonian is not diagonal in the electronic basis we carry out a basis transformation using 
Bogoliubov-Valatin transformations defined as \cite{de_gennes_book},   

\begin{eqnarray}
\hat c_{\uparrow,i} & = & \sum_{n}(u_{n,\uparrow}^{i}\gamma_{n,\uparrow}-v_{n,\uparrow}^{*i}
\gamma_{n,\downarrow}^{\dagger}) \\ 
\hat c_{\downarrow,i} & = & \sum_{n}(u_{n,\downarrow}^{i}\gamma_{n,\downarrow}+v_{n,\downarrow}^{*i}
\gamma_{n,\uparrow}^{\dagger})
\end{eqnarray}

Here, $n$ is a complete set of states, $u_{n}$ and $v_{n}$ are the eigenfunctions of the Hamiltonian $\hat H$, 
which satisfies the condition $u_{n}^{i2}+v_{n}^{i2}=1$ \cite{de_gennes_book}. The new fermionic operators 
$\gamma_{n}$ are quasiparticle operators which satisfies the anti commutation relation 
$\gamma_{n}^{\dagger}\gamma_{n}+\gamma_{n}\gamma_{n}^{\dagger} = 1$ and are chosen in a way such that,

\begin{eqnarray}
\hat H & = & E_{0} + \sum_{n,\sigma}\epsilon_{n}\gamma_{n,\sigma}^{\dagger}\gamma_{n,\sigma}
\end{eqnarray}  
Thus, we have, 
\begin{eqnarray}
[\hat H,\gamma_{n,\sigma}] & = & -\epsilon_{n}\gamma_{n,\sigma} \cr 
[\hat H,\gamma_{n,\sigma}^{\dagger}] & = & \epsilon_{n}\gamma_{n,\sigma}^{\dagger}
\end{eqnarray} 
\noindent where, $\epsilon_{n}$ corresponds to the eigenvalues of the Hamiltonian $\hat H$. Using the 
above expressions one can determine the coefficients $u_{n}^{i}$ and $v_{n}^{i}$ by solving the matrix eigenvalue 
equation. For this we define the matrices A and $\Psi$, such that, 
\begin{eqnarray}
A & = &
\begin{bmatrix}
-t-\mu_{\uparrow} & \hat \Delta_{i} \\ 
\hat \Delta_{i}^{*} & t+\mu_{\downarrow}
\end{bmatrix}
\end{eqnarray} 
and 
\begin{eqnarray}
\Psi & = & 
\begin{bmatrix}
\{u_{n}\} \\ 
\{v_{n}\}
\end{bmatrix}
\end{eqnarray}
The resulting eigenvalue equation is then given as, 
\begin{eqnarray}
A\Psi & = & \epsilon_{n}\Psi
\end{eqnarray}
 
In order to solve the eigenvalue equations, one first needs to determine the superconducting pairing field 
($\Delta_{i}$) as well as the chemical potentials $\mu_{\uparrow}$ and $\mu_{\downarrow}$. For this the set 
of coupled equations corresponding to $\Delta_{i}$ and fermionic number densities $n_{i\sigma}$ is to be solved 
self consistently. In terms of the BdG eigenfunctions and eigenvalues the self consistent mean field equations are 
given as, 

\begin{eqnarray}
\Delta_{i} & = & U\sum_{n}\langle \hat c_{i,\downarrow}\hat c_{i,\uparrow}\rangle \nonumber \\ & = & U\sum_{n}\{v_{n,\downarrow}^{*i}u_{n,\uparrow}^{i}f(\epsilon_{n})+u_{n,\downarrow}^{*i}v_{n,\uparrow}^{i}f(\epsilon_{-n})\} 
\end{eqnarray}
\begin{eqnarray}
n_{i\uparrow} & = & \sum_{n}\langle \hat c_{i,\uparrow}^{\dagger}\hat c_{i,\uparrow}\rangle \nonumber \\ & = & 
 \sum_{n}\{\vert u_{n,\uparrow}^{i}\vert^{2}f(\epsilon_{n})+\vert v_{n,\uparrow}^{i} \vert^{2}f(\epsilon_{-n})\} 
\end{eqnarray}
\begin{eqnarray} 
n_{i\downarrow} & = & \sum_{n}\langle \hat c_{i,\downarrow}^{\dagger}\hat c_{i,\downarrow}\rangle \nonumber \\ &=&
\sum_{n}\{\vert u_{n,\downarrow}^{i}\vert^{2}[1-f(\epsilon_{n})] + \vert v_{n,\downarrow}^{i}\vert^{2}
[1-f(\epsilon_{-n})]\}
\end{eqnarray}

where, $f(\epsilon_{n})$ and $f(\epsilon_{-n})$ are the Fermi functions corresponding to the positive 
and negative eigenvalues, respectively. 
Starting from an initial guess these equations are iterated to obtain self consistent values of the pairing field and 
the number densities. 

For a lattice of dimension $L \times L$ the dimension of the matrix is $2N \times 2N$, where, $N=L^{2}$. 
The net chemical potential of a population imbalanced Fermi gas is defined as, 
$\mu=\frac{\mu_{\uparrow}+\mu_{\downarrow}}{2}$. The difference between the chemical 
potentials act as a Zeeman field defined as, $h=\frac{\mu_{\uparrow}-\mu_{\downarrow}}{2}$. In terms 
of $\mu$ and $h$ we can thus write the spin resolved chemical potentials as, $\mu_{\uparrow}=\mu+h$ and 
$\mu_{\downarrow}=\mu-h$. While working in grand canonical ensemble we work at a fixed chemical potential 
$\mu$ and tune the Zeeman field $h$ such that, the system undergoes phase transition between an uniform 
superconducting phase and inhomogeneous FFLO or BP phases, at a fixed interaction $U$. 

By examining the real space self consistent mean field solutions of the pairing field and fermionic number 
densities one can analyze the underlying state. A non zero $\Delta_{i}$ corresponds to a superconducting 
state, while a state with $\Delta_{i}=0$ is non superconducting. In the same spirit, 
$m_{i}=(n_{i\uparrow}-n_{i\downarrow})\neq 0$ corresponds to a state with finite magnetization. Before analyzing 
the self consistent solution it is however essential to determine whether the particular solution is 
stable, in other words whether it corresponds to the minimum energy state for the given choice of parameters. 

We thus next calculate the free energy of the system and for that we go back to Equation 19, which gives the 
Hamiltonian in the quasiparticle basis. Here $E_{0}$ correspond 
to the ground state energy and in order to determine $E_{0}$ we note that, 

\begin{eqnarray}
\langle \Psi_{G}\vert \hat H \vert \Psi_{G} \rangle & = & E_{0}
\end{eqnarray}
where, $\vert \Psi_{G}\rangle$ correspond to the ground state. The free energy of the system reads as,
 
\begin{eqnarray}
F & = & E_{0} + \sum_{\epsilon_{n}}(-\epsilon_{-n}-\frac{1}{\beta}[\ln(1+e^{-\beta \epsilon_{n}})+\ln(1+e^{-\beta \epsilon_{-n}})] \nonumber \\ && -\frac{\Delta_{i}^{2}}{U})
\end{eqnarray}

The global minima of this free energy functional gives the stable ground state of the system. Based on 
the pairing field amplitude and average magnetization ($\langle m_{i}\rangle$) the ground state of 
two component imbalanced Fermi gas can be broadly classified as, (i) uniform (BCS) superconducting 
phase ($\Delta_{i}({\bf Q}=0) \neq 0$, $\langle m_{i}\rangle = 0$), (ii) (partially) polarized Fermi 
liquid (PPFL) ($\Delta_{i}=0$, $\langle m_{i}\rangle \neq 0$), (iii) FFLO phase ($\Delta_{i}({\bf Q}\neq 0)\neq 0$, 
$\langle m_{i}\rangle \neq 0$), (iv) BP phase ($\Delta_{i}({\bf Q}=0)\neq 0$, $\langle m_{i}\rangle \neq 0$). 
Note that even though both BCS and BP phases are characterized by ${\bf Q}=0$ pairing state, the BP phase 
comprises of gapless excitations unlike the BCS superconducting state \cite{sarma1963,wilczek2003,gubankova2003,forbes2005,liu2004}. These gapless excitations are 
essential to accommodate the unpaired fermions which in turn gives rise to finite magnetization.    

\subsection{Beyond the mean field approximation}

So far we have taken into account only the inhomogeneity in the amplitude of the pairing field 
(BdG formalism). 
The complex superconducting pairing field defined as, $\Delta_{i}=\vert \Delta_{i}\vert e^{i\theta_{i}}$ 
however comprises of amplitude and phase, both of which in principle can fluctuate and become inhomogeneous. 
The fluctuations in the pairing field amplitude and phase becomes progressively important with increasing 
temperature where thermal 
fluctuations can serve as pair breaking agent. In order to understand the importance of fluctuations 
on the superconductivity of a system let us take a detour to briefly summarise the BCS-BEC crossover in 
Fermi gases \cite{gurarie2007,zwerger_book,parish_book,randeria_taylor}. 

\subsubsection{BCS-BEC crossover}

Figure 3 shows the schematic phase diagram of the BCS-BEC crossover of the two component Fermi gas 
comprising of equal populations of up and down fermion species, in the interaction-temperature (U-T) 
plane. The phase diagram comprises of four distinct regimes viz. superconductor, (bosonic) insulator, 
Fermi liquid, non Fermi liquid and two thermal scales as T$_{c}$ and T$_{pg}$. While in a solid state 
system tuning the interaction strength is non trivial in ultracold atomic gases, Feshbach resonance is 
used to control the $s$-wave scattering length and in turn the interspecies interaction 
\cite{bloch2008,zwerger_book,lewenstein_book}. 
The ground state of the BCS-BEC crossover can be captured by the BCS theory with suitably modified coefficients, 
as shown by Leggett \cite{leggett}.
The two thermal scales that we show in this phase diagram can naively 
be understood to be the one determined through the experiments on ultracold atomic gases (T$_{c}$) and 
the one determined through BCS-meanfield approximation (T$_{pg}$). Evidently, they do not agree with 
each other except for the weak coupling regime (U$\ll$t) i. e. deep inside the BCS regime. 
\begin{figure}
\centering
\includegraphics[width=0.68\linewidth]{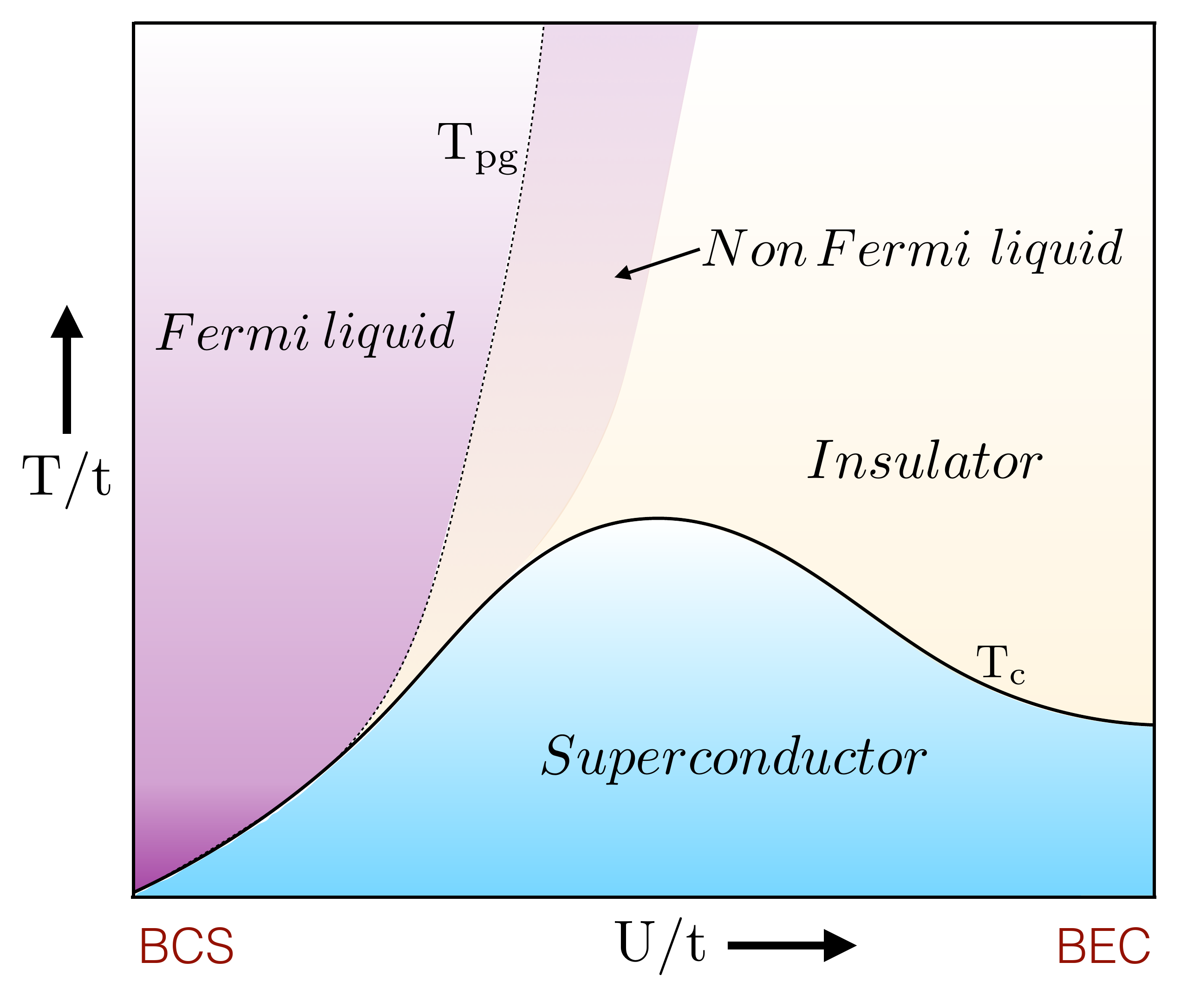}
\caption{Color online: Schematic phase diagram showing the BCS-BEC crossover picture in lattice fermion model.}
\end{figure}

According to the BCS mean-field theory, superconductivity of a system vanishes due to the 
suppression of the superconducting pairing field amplitude.  This obeys the thermal scale 
T$_{BCS} \sim e^{-1/U}$ and the corresponding superconducting 
pairing field amplitude varies as $\Delta \sim$ T$_{BCS}$. Thus, T$_{BCS}$ increases linearly with increasing 
$\Delta$ and in turn with $U$ which governs the magnitude of $\Delta$ \cite{tinkham_book,bcs,de_gennes_book}. 
As shown in Figure 3,  T$_{pg}$ bears 
out this linear relation with respect to $U$. The agreement of T$_{pg}$ with the experimentally 
determined T$_{c}$ in the weak coupling regime suggests that the BCS theory is well suited to capture 
the finite temperature behavior of the system in the weak coupling regime. The high temperature regime 
which can be captured by the BCS theory is marked as the Fermi liquid. 

Unlike the BCS prediction the 
experimental T$_{c}$ however behaves non monotonically and in the strong coupling regime T$_{c} \sim 1/U$, 
as shown in Figure 3. 
It is obvious that the BCS theory ceases to be applicable in the strong coupling regime where the pair 
of fermions tightly bound together and undergoes Bose-Einstein condensation, unlike the weak coupling 
regime where the Cooper pairs are characterized by large coherence length \cite{randeria_taylor,eagles1969,rohe2001,micnas_rmp1990}. An efficient way to understand the BEC regime of this crossover picture is by 
mapping the fermionic attractive Hubbard model on to an equivalent spin model. It is now established 
that the attractive Hubbard model yields an effective field theoretic description in terms of an 
$SO(3)$ non linear sigma model. In the strong coupling regime this model maps on to the Heisenberg 
model in an uniform field. Except for the regime of weak magnetic field (i. e. low filling of the attractive 
Hubbard model) the Heisenberg model reduces to the quantum XY model \cite{dupuis2004,dupuis2_2004}.
In the regime of strong coupling, the system undergoes transition from the superfluid state to a bosonic 
insulator with increasing temperature. The bosonic insulator is characterized by phase uncorrelated large 
pairing field amplitudes and a hard single particle spectral gap at the Fermi level.        

The regime of intermediate coupling is the most complex part of this phase diagram and none of the 
perturbative schemes are suitable to capture its behavior. In this regime marked as the ``non Fermi liquid'' 
in Figure 3,  neither a completely fermionic nor a bosonic description holds true because the phase comprises of 
coexisting unpaired fermions and preformed pairs (bosons). The intuitive basis of understanding the BCS-BEC crossover was 
provided by the earlier works of Leggett \cite{leggett} and Nozeires and Schmitt-Rink \cite{rink}. 
Later powerful semi analytic \cite{dupuis2004,deisz2002,tamaki2008,kyung2001,kopec2002}
and sophisticated numerical schemes such as determinant quantum Monte Carlo (DQMC) \cite{sewer2002,paiva2010,scalettar1998,moreo1991,moreo1992,randeria1992,trivedi1995,allen1999,singer1999,paiva2004}, dynamical mean field 
theory (DMFT) \cite{keller2001,capone2002,garg2005,bauer2009,bauer2_2009,koga2011} 
and static path approximation (SPA) \cite{tarat_epjb} were utilized with success to investigate the physics 
of the intermediate coupling regime of the BCS-BEC crossover picture. This non Fermi liquid regime of 
the BCS-BEC crossover is now known as the {\it pseudogap phase} and leaves distinct signatures in 
the quasiparticle spectrum of the system. The quasiparticle spectra of the pseudogap regime (T$_{c} < $T$ < $T$_{pg}$) 
is characterized by renormalized phase coherence peaks arising out of the existing short range pair 
correlations in this region. Further, at the Fermi level the quasiparticle spectra is no longer 
hard gapped (unlike T $<$ T$_{c}$) and contains finite spectral weight due to unpaired fermions arising out 
of thermal pair breaking. 

The two thermal scales are thus defined as the T$_{c}$, at which the superconducting state loses its long 
range macroscopic phase coherence and T$_{pg}$, at which short range pair correlations are lost. The 
pseudogap regime with short range pair correlations but no long range order, is characterized by spatial 
fragmentation of the state into isolated islands akin to uncorrelated Josephson junctions. 
The finite temperature behavior 
of the BCS-BEC crossover shows that in the regime of intermediate (and strong) coupling it is the thermally 
induced phase fluctuations which leads to the loss of long range phase coherence and of the global 
superconductivity,  unlike the weak coupling regime where superconductivity is destroyed via the suppression 
of pairing field amplitude. One thus needs to take into account the effect of phase fluctuations 
in order to correctly capture the behavior of superconductors or superfluids with interactions away from 
the weak coupling regime, at finite temperatures. 

\subsubsection{BCS-BEC crossover and FFLO systems}
 
Having established the importance of fluctuations in superconducting 
systems we now once again take a look at the candidate systems where FFLO or BP phases are reported. The key 
candidates as discussed earlier are, three dimensional heavy fermion superconductor CeCoIn$_{5}$, 
two-dimensional layered organic superconductor $\kappa$-BEDT, both with large Maki parameter of 
$\alpha>1$ and neutral ultracold atomic gases. 

Apart from the large Maki parameter there is another property which serves as a common link between 
these apparently very distinct systems. An important characteristic of any superconductor is 
the ratio between the superconducting gap at zero temperature and T$_{c}$. The BCS estimate of 
the same is $2\Delta_{0}/k_{B}T_{c} \sim 3.5$ \cite{tinkham_book,de_gennes_book}, which corresponds 
to the weak coupling regime of Figure 3. The BCS estimate sets the scale,  based on which the interaction 
regime of other superconducting systems are determined. Based on the point contact \cite{sumiyama2008} 
and scanning tunnelling spectroscopy \cite{fasano_physicab} used to determine the magnitude of the gap 
of any superconducting material,  it has been observed that for CeCoIn$_{5}$ ($\alpha \sim 5$) 
$2\Delta_{0}/k_{B}T_{c} \sim 7.73$ \cite{fasano_physicab}, 
while for $\kappa$-BEDT ($\alpha \sim 8$) it is $2\Delta_{0}/k_{B}T_{c}\sim 4.4$ \cite{wosnitza_crystals}. 
The iron superconductor is conjectured to be a two gap superconductor with the ratio for the smaller gap 
being $2\Delta_{0}/k_{B}T_{c} \sim 1.4$ and that of the larger gap being $2\Delta_{0}/k_{B}T_{c}\sim 4.6$ 
\cite{abdel2012}. 
These experimental observations indicate that all the candidate Pauli limited solid state 
superconductors are situated away from the weak coupling regime. In case of the ultracold 
atomic gases the interaction can be tuned externally and most of the experiments are carried out 
at unitarity, which roughly corresponds to the regime of maximum T$_{c}$ in the 
BCS-BEC crossover, shown in Figure 3. 

The experimental observations thus bring forth a common 
link between the various Pauli limited superconducting systems, viz. all these systems belong 
to the regime of intermediate coupling. While at one hand this sets up the stage for complex 
physical properties to play out, on the other hand it calls for a non perturbative approach to 
address these systems. It is thus imperative to take into account the effect of fluctuations 
while addressing the finite temperature behavior of the imbalanced Fermi superconductors and 
superfluids. Consequently,  a mean field approach to the problem will not suffice. 

We discussed above that apart from the semi analytic and T-matrix approaches 
\cite{dupuis2004,deisz2002,tamaki2008,kyung2001,kopec2002} there are powerful numerical techniques 
which takes into account the effect of fluctuations in quantum many body systems 
\cite{sewer2002,scalettar1998,moreo1991,moreo1992,randeria1992,trivedi1995,allen1999,singer1999,paiva2004,paiva2010,keller2001,capone2002,garg2005,bauer2009,bauer2_2009,koga2011}.  
We now discuss a path integral formalism based on which the different techniques and their 
regimes of validity can be understood. 

\subsubsection{Path integral formalism for imbalanced Fermi systems}

We begin by rewriting the two-dimensional attractive Hubbard model with population 
imbalance, introduced in Equation (12), which can be used to model the solid state as well as 
the ultracold atomic gas systems. We will discuss about the system specific modifications in the later 
sections and for the sake of simplicity discuss here the theoretical formalism in detail for a square 
lattice with on-site attractive interaction $\vert U \vert$ which gives rise to $s$-wave superconductivity. 
The corresponding Hamiltonian reads as, 
\begin{eqnarray}
\hat H & = & \sum_{\langle ij\rangle,\sigma}(t_{ij}-\mu\delta_{ij})\hat c_{i\sigma}^{\dagger}\hat c_{j\sigma}-h\sum_{i}\sigma_{iz}-\vert U\vert \sum_{i}\hat n_{i\uparrow}\hat n_{i\downarrow} \nonumber \\
\end{eqnarray}  
where, $t_{ij} = -t$ only for the nearest neighbor hopping and is zero otherwise; $\sigma_{iz}$ is the Pauli matrix. 
The hopping amplitude $t$ is set to unity and is chosen to be the 
reference energy scale for the problem. $\mu$ is 
the net chemical potential and the Zeeman field $h$ is applied along the $\hat z$ direction; $\vert U\vert > 0$. 

The Hubbard partition function is a functional integral over the Grassman fields $\psi_{i\sigma}(\tau)$ 
and $\bar \psi_{i\sigma}(\tau)$, 
\begin{eqnarray}
Z  =  \int {\cal D} \psi {\cal D} {\bar \psi}e^{-\int^{\beta}_{0} d\tau {\cal L}(\tau)} \cr
{\cal L}(\tau) = {\cal L}_{0}(\tau) + {\cal L}_{U}(\tau) \cr
{\cal L}_{0}(\tau)=\sum_{\langle ij\rangle, \sigma}\{\bar \psi_{i\sigma}((\partial_{\tau}-\mu_{\sigma})
\delta_{ij}+t_{ij})\psi_{j\sigma}\} \cr
{\cal L}_{U}(\tau) = -\vert U \vert\sum_{i\sigma \sigma^{\prime}}\bar \psi_{i\sigma}
\psi_{i\sigma}\bar\psi_{i\sigma^{\prime}}\psi_{i\sigma^{\prime}}
\end{eqnarray}
where, $\beta$ is the inverse temperature, $\mu_{\sigma}$ takes into account the field $h$. 
Our strategy is to decompose the interaction in terms of the 
bosonic auxiliary $s$-wave pairing field $\Delta_{i}(\tau) = \vert \Delta_{i}(\tau)\vert e^{i\theta_{i}(\tau)}$,  
using Hubbard-Stratonovich transformation \cite{hs,hs1}. Here $\vert \Delta_{i}(\tau)\vert$ and 
$\theta_{i}(\tau)$ corresponds 
to the amplitude and phase of the pairing field, respectively, and encodes spatial ($i$) and imaginary 
time ($\tau$) dependence. This leads to,   
\begin{eqnarray}
Z = \int {\cal D}\psi {\cal D} {\bar \psi} {\cal D}\Delta {\cal D}{\Delta}^{*}e^{-\int^{\beta}_{0} 
d\tau {\cal L}(\tau)} \cr
{\cal L}(\tau) = {\cal L}_{0}(\tau) + {\cal L}_{U}(\tau) + {\cal L}_{cl}(\tau) \cr
{\cal L}_{0}(\tau)=\sum_{\langle ij\rangle, \sigma}\{\bar \psi_{i\sigma}(\tau)((\partial_{\tau}-\mu_{\sigma})\delta_{ij}+
t_{ij})\psi_{j\sigma}(\tau)\} \cr
{\cal L}_{U}(\tau) = \sum_{i} [\Delta_{i}(\tau)\bar \psi_{i\uparrow}(\tau) \bar \psi_{i\downarrow}(\tau) + h. c.] \cr
{\cal L}_{cl}(\tau) = \sum_{i}\frac{\vert \Delta_{i}(\tau)\vert^{2}}{\vert U\vert}
\end{eqnarray}

The $\psi$ integral is now quadratic, but an additional integration over the field $\Delta_{i}(\tau)$ has been 
introduced. The weight factor for the $\Delta$ configurations can be determined by integrating out the $\psi$,
$\bar \psi$ and using these weighted configurations, one goes back and compute the fermionic properties. Formally, 
\begin{eqnarray}
Z = \int {\cal D}\Delta{\cal D}{\Delta}^{*} e^{-S_{eff}\{\Delta,\Delta^{*}\}} \cr
S_{eff} = \ln Det [{\cal G}^{-1}\{\Delta,\Delta^{*}\}] + \int^{\beta}_{0}d\tau{\cal L}_{cl}(\tau)
\end{eqnarray}
where, ${\cal G}$ is the electronic Green's function in the $\{\Delta\}$ background.
The weight factor for an arbitrary space-time configuration $\Delta_{i}(\tau)$ involves computation 
of the fermionic determinant in that background. If we write the auxiliary field $\Delta_{i}(\tau)$ in 
terms of its Matsubara modes as, $\Delta_{i}(\Omega_{n})$ then the different options are readily 
recognized, 
\begin{itemize}
\item{Quantum Monte Carlo retains the full ``$i,\Omega_n$'' dependence of $\Delta$ computing 
$\ln Det[{\cal G}^{-1}\{\Delta\}]$ iteratively for importance sampling. The approach is valid 
at all T, but does not readily yield real frequency spectra.}
\item{Mean field theory (MFT) is time independent, neglects the phase fluctuations completely but 
can handle spatial inhomogeneity in amplitude of the pairing field. Thus, 
$\Delta_{i}(i\Omega_n) \rightarrow \mid\Delta_{i}\mid$. When the mean field order parameter 
vanishes at high temperature the theory trivializes.}
\item{Dynamical mean field theory (DMFT) retains the full dynamics
but keeps $\Delta$ at  effectively one site, {\it i.e},
$\Delta_{i}(\Omega_n) \rightarrow \Delta(\Omega_n)$.
This is exact when dimensionality $D \rightarrow \infty$.}
\item{Static path approximation (SPA) \cite{wang1969,evenson1970,dubi2007} 
approach retains the full spatial dependence
in $\Delta$ but keeps only the $\Omega_n=0$ mode,
{\it i.e}, $\Delta_{i}(\Omega_n) \rightarrow \Delta_{i}$.
It thus includes classical fluctuations of arbitrary magnitude but no quantum ($\Omega_n \neq 0$)
fluctuations.  One may consider different
temperature regimes. (1) $T = 0$: Since classical fluctuations die off at $T = 0$, SPA reduces to
standard Bogoliubov-de-Gennes (BdG) MFT.
(2) At $T\neq 0$ we consider not just the saddle-point configuration but {\it all configurations} 
following the weight $e^{-S_{eff}}$ above.
These involve the classical amplitude and phase fluctuations of the order parameter, and the BdG equations
are solved in {\it all these configurations} to compute the thermally averaged properties.
This approach suppresses the order much quicker than in MFT. (3) High T : Since the $\Omega_{n}=0$
mode dominates the exact partition function, the SPA approach becomes exact as $T\rightarrow \infty$. 
It is thus akin to the MFT {\it only} at $T=0$ but captures the thermal behavior 
of the system accurately. Within this approach the system can be envisaged as free fermions moving in a 
random background of $\Delta_{i}$.}
\end{itemize} 

\subsubsection{Static path approximation (SPA)}

Evidently, all the numerical approaches have different set of merits and caveats, and consequently 
different regimes of applicability. We now discuss the numerical implementation of the SPA technique in 
some detail \cite{sanjeev_pinaki,mpk2016,mpk2018,mpk_epjd,mpk_jpcm2020}. The merits of this technique 
rests on its applicability, in principle to systems of arbitrary 
dimensions and lattice sizes, with moderate computation cost and its access to real frequency 
properties of the system. The technique takes into account spatial fluctuations and agreement
with the results of numerically exact DQMC suggests that it can capture the thermal scales and high 
temperature phases accurately \cite{tarat_epjb}. Using the Hubbard Stratonovich decomposition to 
decompose the interaction term in Equation (29) and then dropping the imaginary time dependence 
of the bosonic auxiliary field ($\Delta_{i}(\tau)$) the effective Hamiltonian reads as, 
\begin{eqnarray}
H_{eff} & = & -t\sum_{\langle ij\rangle, \sigma}\hat c_{i\sigma}^{\dagger}\hat c_{j\sigma}
+ \sum_{i}\Delta_{i} \hat c_{i\uparrow}^{\dagger}\hat c_{i\downarrow}^{\dagger} + h. c. -\mu\sum_{i, \sigma} 
\hat n_{i\sigma} \nonumber \\ && - h\sum_{i}\sigma_{i}^{z} + 
\sum_{i}\frac{\mid \Delta_{i}\mid^{2}}{\mid U \mid}
\end{eqnarray}
where, the last term $H_{cl}=\frac{\vert \Delta_{i}\vert^{2}}{\vert U\vert}$ is the stiffness 
cost associated with the now ``classical'' auxiliary field. In the standard mean field approach the
superconducting pairing field $\Delta_{i}$ is assumed to be a real number, but here we retain the degrees of freedom
associated with the pairing field phase and amplitude. 

The background field $\Delta_{i}$ obeys the Boltzmann distribution,
\begin{eqnarray}
P\{\Delta_{i}\} \propto Tr_{cc^{\dagger}}e^{-\beta H_{eff}}
\end{eqnarray}

\noindent which is connected to the free energy of the system. For large and random $\Delta_{i}$ the 
trace is taken numerically. We generate the random background of $\{\Delta_{i}\}$ by using Monte Carlo, 
diagonalizing $H_{eff}$ for each attempted update of $\Delta_{i}$. The relevant fermionic correlators are computed
on the optimized configurations at different temperatures. 

Evidently, the technique is numerically expensive 
and involves large computation cost in diagonalizing the matrix to compute the free energy for each 
attempted update of the auxiliary field. For an $N \times N$ matrix the computation cost scales as  
$\sim {\cal O}(N^{3})$ per update and $\sim {\cal O}(N^{4})$ per lattice sweep. This imposes constraint 
on the lattice size that can be accessed within reasonable computation time. In order to access larger 
system sizes a cluster based update scheme could be utilized wherein instead of diagonalizing for the entire 
lattice at each attempted update, a smaller cluster of size $N_{c}=L_{c}\times L_{c}$ surrounding the 
update site is diagonalized \cite{mpk2016,mpk2018,mpk_jpcm2020}. This reduces the cost per update 
to $\sim {\cal O}(N_{c}^{3})$ and to 
$\sim N\times {\cal O}(N_{c}^{3})$ per lattice sweep, which scales linearly with the system size. 
This allows one to access system sizes of upto $\sim 40\times 40$, which is beyond the reach of DQMC. 
Access to larger system sizes aid in to explore lower interaction regimes where the coherence length 
of the Cooper pairs are large. Moreover, in order to faithfully capture spatially inhomogeneous 
orders, access to lager lattice sizes is essential. 

We have now set up a theoretical framework to investigate the behavior of spatially inhomogeneous 
superconducting and superfluid phases. Different numerical tools to access these systems 
both at the level of ground state as well as at finite temperature through mean field and non perturbative 
techniques are discussed. In order to understand the theoretical formalism we considered that the 
interaction between the fermion species is onsite and the resulting pairing state symmetry is 
$s$-wave. This is a valid approximation as far as ultracold atomic gas set ups are concerned 
and most of the experiments carried out on imbalanced Fermi systems in ultracold atomic gases 
are for $s$-wave superfluids. Furthermore, here a single channel decomposition of the interaction term 
was carried out. In principle, along with the pairing channel the interaction can be decomposed 
in density and spin channels as well. This however would add to the computation cost and more 
importantly would not lead to any qualitative difference in the behavior of the present 
system. 

The behavior of the imbalanced Fermi superfluid (or superconductor) is analyzed based on 
different thermodynamic and quasiparticle indicators, some of which are,
\begin{itemize}
\item{Magnetization:- $m_{i}=n_{i\uparrow}-n_{i\downarrow}$.}
\item{Pairing field structure factor:- $S_{\Delta}({\bf q}) = \frac{1}{N^{2}}\sum_{i,j}
\langle \Delta_{i}\Delta_{j}^{*}\rangle e^{i{\bf q}.({\bf r}_{i}-{\bf r}_{j})}$}
\item{Spin resolved single particle density of states (DOS):- $N_{\uparrow}(\omega)=
\frac{1}{N}\langle \sum_{i}\vert u_{n\uparrow}^{i}\vert^{2}\delta(\omega-\epsilon_{n})\rangle$, 
$N_{\downarrow}(\omega) = \frac{1}{N}\langle \sum_{i}\vert v_{n\downarrow}^{i}\vert^{2}
\delta(\omega-\epsilon_{n})\rangle$}
\item{Spectral functions and lineshapes:- $A_{\sigma}({\bf k},\omega)=(-1/\pi)Im G_{\sigma}({\bf k},\omega)$.}
\item{Low energy spectral weight distribution:- $A({\bf k},0)$.}
\end{itemize} 
Here, $G({\bf k},\omega) = lim_{\delta \rightarrow 0}G({\bf k},i\omega_{n})\vert_{i\omega_{n}\rightarrow 
\omega+i\delta}$, with $G({\bf k},\omega)$ being the imaginary frequency transform of $\langle c_{\bf k}(\tau)
c_{\bf k}^{\dagger}(0)\rangle$.

Note that the static path approximation approach, as mentioned above does not take into account the 
quantum fluctuations at the ground state. While in continuum this could be a poor approximation, on 
a lattice it is reasonable. The low energy fluctuations in a FFLO like state arises from, (i) the phase 
symmetry of the $U(1)$ order parameter, (ii) the translational symmetry breaking and (iii) rotational 
symmetry breaking. As a result, in continuum a long range order can not survive even at T=0 in two 
dimensions. The mean field theory, which predicts such order is therefore invalid. 

On a lattice, the spatial symmetry is already broken and thus the translational and rotational modes are 
gapped out. The only low energy excitations thus left are the XY type, in presence of which the two-dimensional 
system can have long range order at T=0 and undergoes Berezinskii-Kosterlitz-Thousless (BKT) transition at the 
finite temperatures. The issue of fluctuations thus reduces to how well the $U(1)$ phase fluctuations 
and the corresponding T$_{c}$ are captured by this approach.
 
\subsubsection{Non $s$-wave pairing}
 
The superconducting pairing state symmetry in solid state materials depend on 
the material parameters and the coupling constants involved. As discussed earlier, the all the
known Pauli limited superconductors belong to the regime of intermediate coupling. This in 
turn naturally categorises them to be superconductors with {\it short coherence length}, whose 
pairing state symmetry can no longer be $s$-wave. Indeed the pairing state symmetry of 
CeCoIn$_{5}$ and $\kappa$-BEDT are established to be $d$-wave \cite{koutroulakis2008,mayaffre2014}. Unlike  
$s$-wave, the $d$-wave pairing state requires an intersite interaction between the fermions. 

The Hubbard Hamiltonian can be suitably modified to take into account the nearest 
neighbor interaction which gives rise to $d$-wave pairing state symmetry as, 
\begin{eqnarray}
\hat H & = & \sum_{\langle ij\rangle, \sigma}t_{ij}\hat c^{\dagger}_{i,\sigma}\hat c_{j,\sigma}-U\sum_{\langle ij\rangle}\hat n_{i} 
\hat n_{j} -\mu\sum_{i,\sigma}\hat n_{i,\sigma} - h\sum_{i}\sigma_{i}^{z} \nonumber \\ 
\end{eqnarray}
where, the attractive interaction $\vert U \vert > 0$ is now between the nearest neighbors, 
on a square lattice. After carrying out the Hubbard-Stratonovich transformation to decompose the 
four-fermion interaction, the corresponding partition function reads as, 
\begin{eqnarray}
Z = \int {\cal D}\psi {\cal D} {\bar \psi} {\cal D}\Delta {\cal D}{\Delta}^{*}e^{-\int^{\beta}_{0} d\tau {\cal L}(\tau)
} \cr
{\cal L}(\tau) = {\cal L}_{0}(\tau) + {\cal L}_{U}(\tau) + {\cal L}_{cl}(\tau) \cr
{\cal L}_{0}(\tau)=\sum_{\langle ij\rangle, \sigma}\{\bar \psi_{i\sigma}((\partial_{\tau}-\mu_{\sigma})
\delta_{ij}+t_{ij})\psi_{j\sigma}\} \cr
{\cal L}_{U}(\tau) = -\sum_{i \neq j} \Delta_{ij}(\bar \psi_{i\uparrow} \bar \psi_{j\downarrow}+
\bar \psi_{j\uparrow} \bar\psi_{i\downarrow}) + h. c. \cr
{\cal L}_{cl}(\tau) = 4\sum_{i \neq j}\frac{\mid \Delta_{ij}\mid^{2}}{\mid U\mid}
\end{eqnarray}
The bosonic auxiliary field $\Delta_{ij}$ encodes the direction dependence of the $d$-wave 
superconducting order parameter.

Apart from the superconducting pairing state symmetry there is another aspect which needs to be taken 
into account while constructing a model Hamiltonian for the solid state Pauli limited superconductors, 
viz the lattice structure. The heavy fermion CeCoIn$_{5}$ comprises of a three dimensional cubic lattice 
structure \cite{mayaffre2014}, while on the other hand the quasi two-dimensional $\kappa$-BEDT is a 
material with triangular lattice structure \cite{wosnitza_crystals}. 
KFe$_{2}$As$_{2}$ contains an added level of complexity which needs to be taken into 
account while setting up a theoretical framework, this material is conjectured to be a multiband 
superconductor with two superconducting gaps (and thus the pairing fields) of different magnitudes 
\cite{abdel2012}. 

We conclude this section by stating that a realistic theoretical model of solid state Pauli limited 
superconductors would be significantly complicated owing to the various interaction parameters and complex 
band structure that needs to be taken into account. However, a relatively simplistic model which takes into 
account the dominant coupling scales and energy bands is amenable. Such a theoretical framework should 
be able to capture the experimental observations on these materials with qualitative and in some cases 
with quantitative accuracy. It should however be kept in mind that solving even a minimal quantum many body 
Hamiltonian is a demanding and computationally extensive task. Consequently, there still remains several 
open areas to be explored in the playground of Pauli limited superconductors as well as of Fermi superfluids
in ultracold atomic gas setups.   

\section{Mean field and beyond mean field studies}

This section discusses some of the studies carried out on lattice fermion models to understand population 
imbalanced Fermi gases, Pauli limited superconductors and mass imbalanced Fermi-Fermi mixtures. Most of 
the T=0 investigations are carried out within the framework of mean field theory and there is a large 
volume of literature discussing the same. We do not attempt to be comprehensive and would discuss only 
some of the important results and the inferences drawn based on them. 

Regarding the beyond mean field attempts, broadly the observations made by three different numerical 
approaches are covered in this section, viz. (i) dynamical mean field theory (DMFT) 
\cite{kim2011,heikkinen2013,heikkinen2014}, (ii) determinant quantum Monte Carlo (DQMC) 
\cite{gukelberger2016,wolak2012} and (iii) static path approximation (SPA) \cite{mpk2016,mpk_epjd}.    

The thermal transitions pertaining to two-dimensional systems, discussed in this section 
are Berezinskii-Kosterlitz-Thouless (BKT) transitions corresponding to algebraic decay 
of quasi long range order in two dimensions \cite{bkt1,bkt2,bkt3}. 

\subsection{Population imbalanced $s$-wave superfluids} 

Within the framework of lattice mean field theory the two-component Fermi mixture with population 
imbalance in an optical lattice was studied by Koponen {\it et al.} \cite{koponen_njp2009}. 
The authors found that both the FF and BP
phases correspond to stable ground state of the system, with the FF state corresponding to 
the global minimum in a fixed chemical potential calculation, and the BP phase corresponding 
to the minimum energy state for a fixed fermionic number density calculation. The same authors further 
investigated the finite temperature behavior of this system, wherein they accessed the finite 
temperature physics by using the mean field theory. The resulting temperature-magnetization 
phase diagram showed that at fixed number density the FF phase constitutes a large part of the 
phase diagram along with the BP and phase separated regimes \cite{koponen2007}.   

Using an elaborate variational mean field scheme based on BdG method with Hartree corrections, 
Loh and Trivedi \cite{loh2010} investigated the LO phase in a cubic lattice. Their mean field scheme comprised 
of six variational parameters corresponding to the complex valued pairing field, chemical potential 
and three Zeeman fields. The study found that for a cubic lattice the LO phase is a stable ground state 
over a significant parameter regime. The authors further discussed the quasiparticle behavior of the 
LO state and showed that the state is characterized by additional van-Hove singularities in the 
single particle density of states (DOS). Based on the momentum 
resolved fermionic occupation number ($n_{\sigma}({\bf k}) = 
\langle \hat c_{{\bf k},\sigma}^{\dagger}\hat c_{{\bf k},\sigma}\rangle$) they inferred that the LO phase breaks 
the four fold symmetry but preserves the inversion symmetry, unlike the FF phase.  

\begin{figure}
\centering
\includegraphics[width=0.6\linewidth]{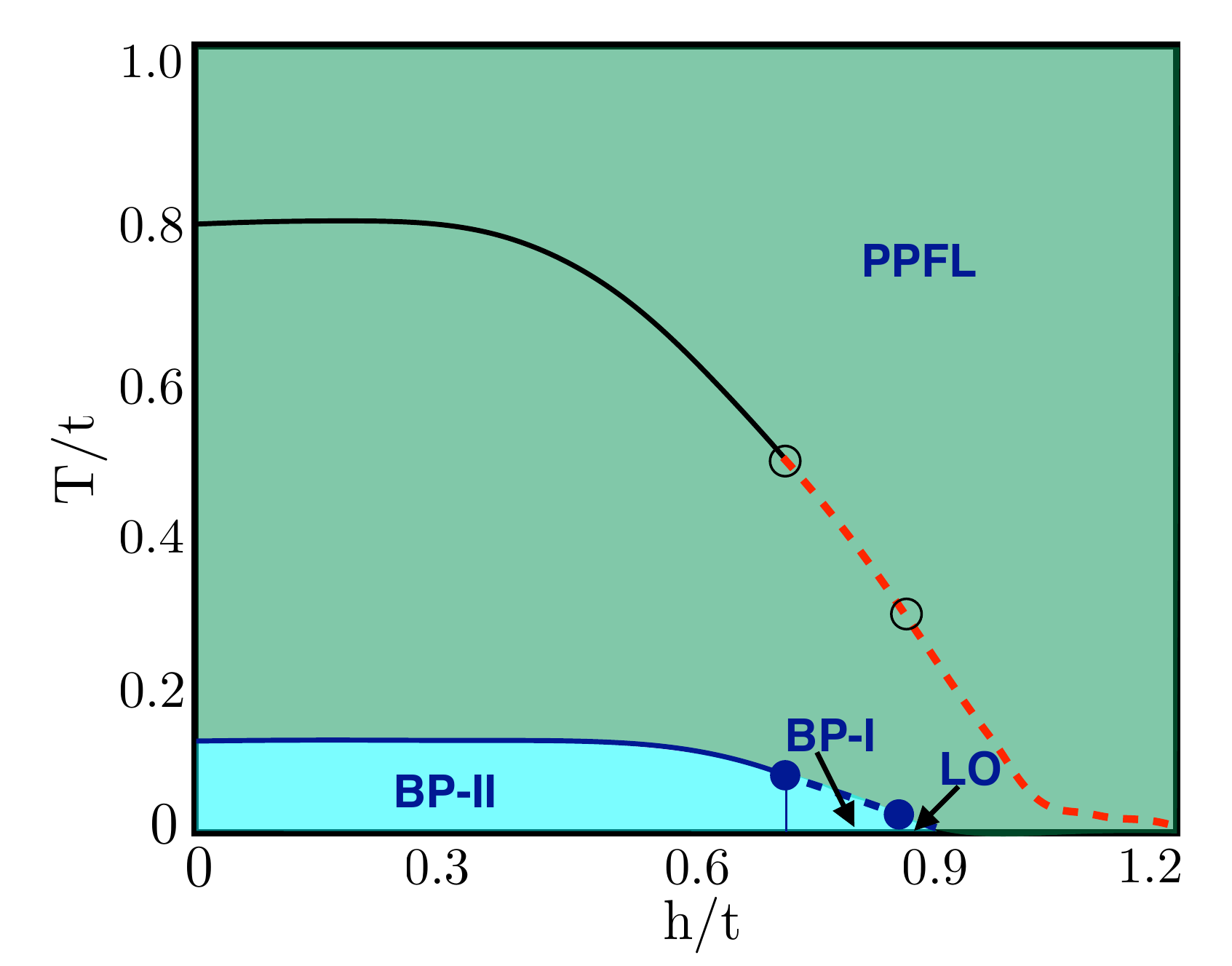}
\caption{Color online: Comparison of T$_{c}$ scales obtained from the mean-field calculation 
(upper curve) and static path approximation (SPA) Monte Carlo technique (lower curve). 
The solid and dashed curves in the mean field result corresponds to second and first order
phase transitions, respectively. In the 
SPA results BP-II represents a breached pair state that undergoes a second-order transition to the 
partially polarized Fermi liquid (PPFL), while BP-I undergoes a first-order thermal transition 
to the PPFL. Beyond BP-I the system exhibits FFLO order up to a critical field. Figure adapted 
with permission from \cite{mpk2016}, Copyright (2016) by the American Physical Society.}
\end{figure}

An extensive numerical study of the ground state was carried out by Chiesa and Zhang as well as by Rosenberg 
{\it et al} on a two-dimensional square lattice, using Hartree-Fock-Bogoliubov-de-Gennes 
(HFBDG) theory \cite{chiesa2013,rosenberg2015}. 
It was observed that an uniaxial LO phase covered a large part of the phase diagram 
in the magnetization-number density ($m-n$) plane, at an intermediate interaction strength. 
In the regime of small magnetization and large 
number density, diagonal stripe orders were found for intermediate interaction, corresponding to 
a ``supersolid'' phase, in which charge and pairing orders coexisted. 
Apart from these,  mean field based comparative study with LO and FF ansatz showed that for both 
two and three dimensional systems the LO state correspond to lower energy phase as compared to 
the FF \cite{baarsma_jmodopt}. The observation was in agreement with the preceding attempts along 
this direction \cite{loh2010,sheehy2007,xu2014}. 

Mean field investigations are carried out in one dimensional systems as well, for example by 
Sun and Bolech, who studied the system of coupled one dimensional tubes \cite{sun2013}. It was found that the 
stability of the FFLO state was dependent on the magnitude of pair tunnelling between the tubes. 
In addition to the FFLO phase, an exotic $\eta$-phase was realized in the mean field study of this 
one dimensional system. In the $\eta$-phase the FFLO wave vector was found to occur at the edge 
of the first Brillouin zone.

The behavior of the LO state across dimensional crossover between one to three dimensions was 
investigated using DMFT by T{\"o}rm{\"a} and co-authors \cite{heikkinen2013,heikkinen2014}. 
It was shown that at the ground state the LO phase is stable over large regime of parameters, 
extending well into the 3D-like regime, unlike the predictions of mean field \cite{parish2007_mft} 
and effective field theory \cite{zhao_liu_ft2008}. 
When extended to finite temperatures by using QMC solver in single site DMFT impurity problem, the 
same system showed that the transition temperature of the LO phases (T$_{FFLO}$) is 
suppressed to upto one third of that of uniform superfluid at zero imbalance \cite{heikkinen2013}. 
Effect of non local fluctuations were taken into 
account in this framework by using cluster-DMFT and it was demonstrated that CDMFT correctly 
captured the behavior of the system in terms of decreasing robustness of long range phase 
coherence with reducing dimensionality of the system \cite{heikkinen2014}. 

In their DQMC study of imbalanced $s$-wave superfluid Wolak {\it et al.} \cite{wolak2012} 
observed signatures of finite momentum pairing, suggesting instability towards the FFLO phase. 
At high temperature indications of a magnetized paired phase was observed. The work however failed to explore 
the low temperature regime (T $<$ 0.1t, with $t$ being the hopping scale), owing to the fermionic sign 
problem of DQMC. Note that the access to low temperature regime is important in case of FFLO phase, for which 
the transition temperature is expected to be strongly suppressed, due to the population imbalance. This 
work therefore failed to give an estimate of the T$_{c}$ scale associated with the FFLO phase. 
The authors predicted the zero imbalance T$_{c}$ to be $\sim 0.1$t but suggested that FFLO 
correlations survive up to a scale T $>>$ T$_{c}$. 

\begin{figure*}
\centering
\includegraphics[width=1.05\linewidth]{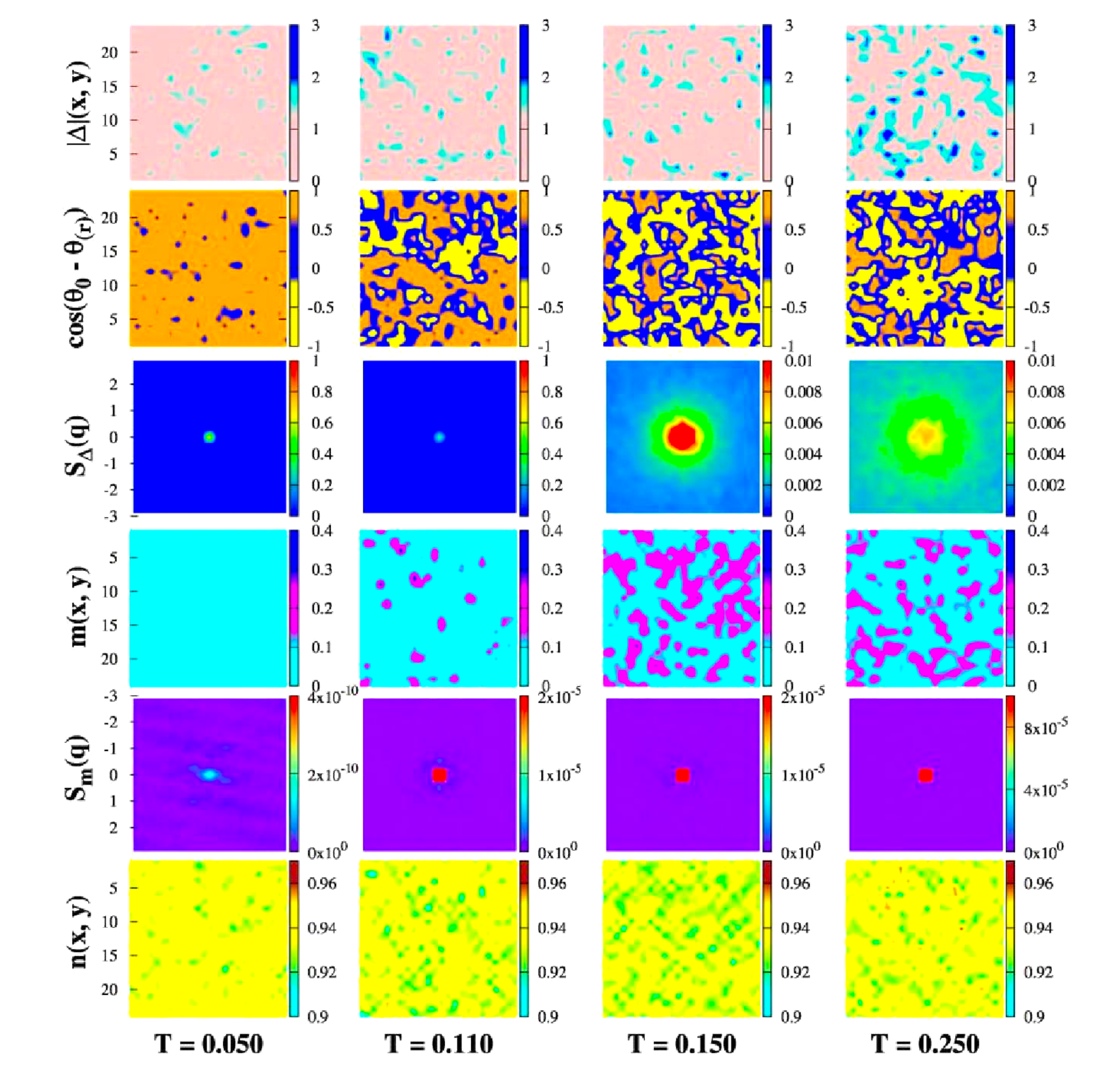}
\caption{Color online: Thermal evolution of the various indicators in the BP phase. Starting from the top row, 
we show maps of superconducting amplitude ($\vert \Delta_{i}\vert$), phase correlation [($\cos{\theta_{0}-
\theta_{i}}$), $\theta_{0}$ is a reference site], pairing structure factor ($S_{\Delta}({\bf q})$), 
magnetization $m_{i}$, magnetic structure factor ($S_{m}({\bf q})$), and fermionic number density ($n_{i}$). 
The temperature, along the rows, is marked at the bottom of the figure. Reprinted figure with permission 
from \cite{mpk2016}, Copyright (2016) by the American Physical Society.}
\end{figure*}

Gukelberger {\it et al.} \cite{gukelberger2016} used diagrammatic QMC to map out the thermal phase 
diagram of imbalanced $s$-wave superfluid in the magnetization-temperature plane at intermediate lattice filling, 
based on their observation of the divergence of the pair susceptibility. 
Interestingly, they found that the weak magnetization low temperature regime of the phase diagram 
comprises of a zero momentum pairing state with finite magnetization. As discussed in the earlier 
sections, such a coexistence of superfluidity and magnetization corresponds to the BP phase. The 
intermediate magnetization regime was found to be a FFLO phase, followed by the high magnetization 
regime corresponding to a Fermi liquid. The lowest temperature that was probed in this study 
was T$\sim 0.02$t which in principle can be higher than the T$_{FFLO}$. Thus, it remains unclear 
whether the BP phase shown in the phase diagram by Gukelberker {\it et al.} \cite{gukelberger2016} 
corresponds to a true ground state or a finite temperature phase. It must however be noted that earlier mean field 
studies of similar system, carried out at fixed lattice filling suggested the BP phase to be a 
possible stable ground state in the regime of weak magnetization \cite{baarsma_jmodopt}. 

A comprehensive analysis of this system in grand canonical ensemble (fixed chemical potential) was 
carried out using SPA, close to the half filling and at an intermediate interaction \cite{mpk2016,mpk_epjd}. 
Apart from mapping out the ground state and finite 
temperature phase diagrams and capturing the thermal scales accurately, the work discussed real space and 
quasiparticle signatures of the LO phase as well as their thermal evolution. The ground state of the 
system was found to comprise of uniform superfluid (USF), LO and partially polarized Fermi 
liquid (PPFL) phases, as the function of increasing magnetic field.  
It was demonstrated, as in Figure 4, that inclusion of thermal fluctuations significantly suppresses 
the T$_{c}$ corresponding to both uniform and LO superfluid phases. Comparison with the results of mean 
field calculations showed that the mean field theory overestimates the T$_{c}$ by a factor of $\sim 4$. 
In the LO phase the T$_{c}$ was about $20\%$ of its mean field estimate and in the absolute term only $2\%$ of 
the hopping scale. The T$_{c}$ corresponding to the two component balanced Fermi gas in two dimensions 
is estimated to be T$_{c}\sim$ 0.15t, based on DQMC \cite{paiva2010} as well as SPA \cite{tarat_epjb,mpk2016} 
calculations. The maximum T$_{c}$ of the LO phase is thus $\sim$0.03t, which is far lower than the lowest 
temperature accessed by Wolak {\it et al.} \cite{wolak2012} and very close to that accessed by Gukelberger 
{\it et al.} \cite{gukelberger2016}.  

\begin{figure*}
\centering
\includegraphics[width=0.8\linewidth]{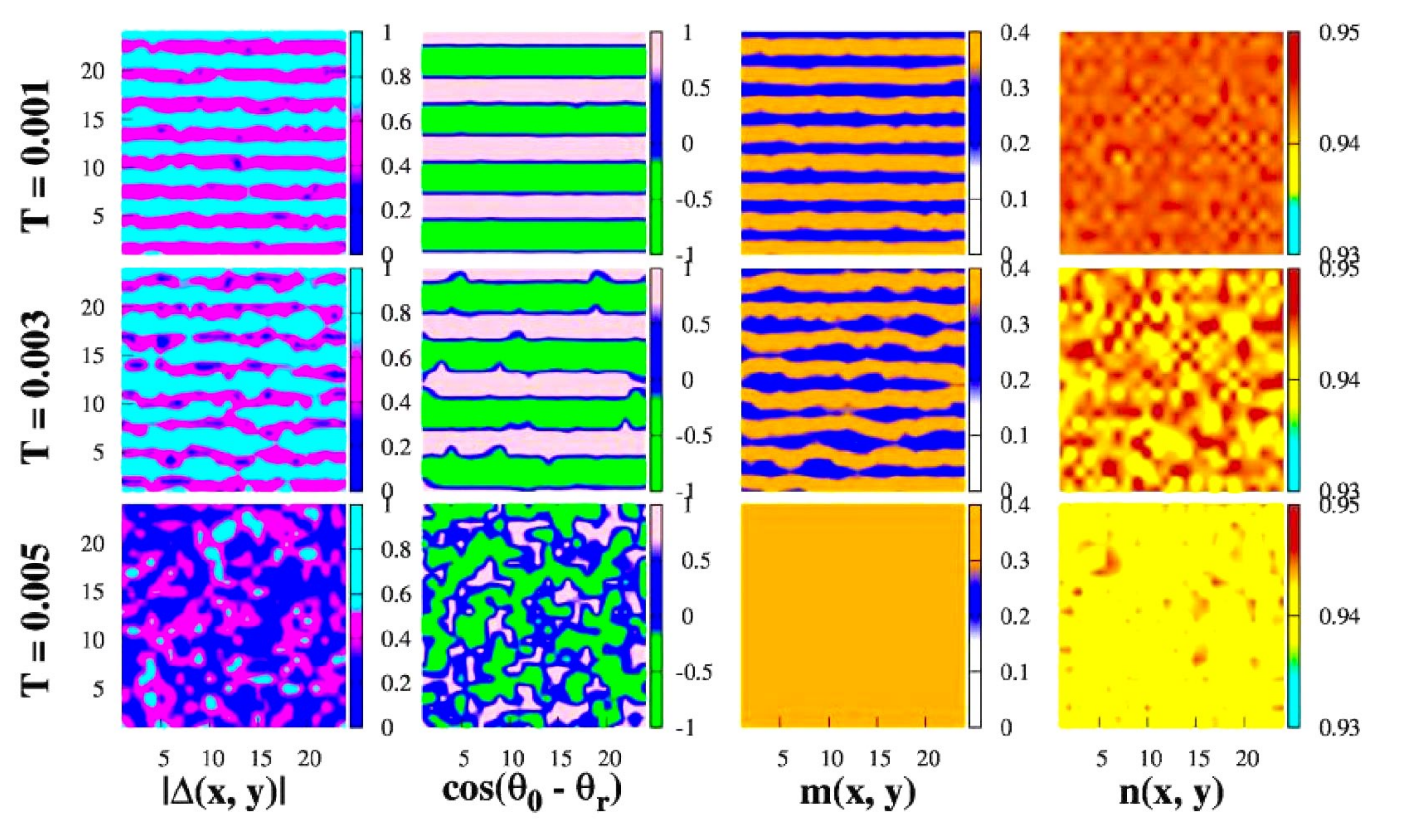}
\caption{Color online: Spatial maps characterizing the thermal evolution of the LO state, 
through (a) pairing field amplitude, (b) phase correlation, (c) magnetization and (d) fermionic number density.
Reprinted figure with permission from \cite{mpk2016}, Copyright (2016) by the American Physical Society.}
\end{figure*}

The work showed that in the regime of weak magnetic field the system loses uniform superfluid order 
to PPFL via a second order thermal transition, while for the strong magnetic field LO phase, the thermal 
transition is of first order. Moreover, a first order thermal transition regime was identified near the 
quantum critical point between the uniform superfluid and 
FFLO phases. In this regime, the uniform (${\bf Q}$=0) superfluid phase undergoes first order thermal 
transition to the PPFL phase. Over the entire regime of magnetic field (magnetization) signatures of superfluid 
pairing were observed at the high temperatures as short range pairing correlations, which survives upto 
temperatures T $\sim$ 1.5T$_{c}$ and gives rise to the pseudogap phase. Akin to the 
T$_{c}$ the pseudogap scale is strongly suppressed by the magnetic field.   

Based on the real space maps the low magnetic field high temperature phase was identified to be a BP 
phase with spatial coexistence of zero momentum superfluid correlations and non zero 
magnetization, as shown in Figure 5 \cite{mpk2016}. In Figure 5, along with the pairing field 
amplitude ($\vert \Delta_{i}\vert$), phase coherence ($\cos(\theta_{0}-\theta_{i})$) and magnetization 
($m_{i}$), the momentum resolved pairing field ($S_{\Delta}({\bf q})$) and magnetic ($S_{m}({\bf q})$) 
structure factors, as well as spatially resolved fermionic number density ($n_{i}$) are shown.
The lowest temperature state corresponds to a phase cohered uniform superfluid with vanishingly 
small magnetization. Increase in temperature leads to suppression of long range phase coherence 
and fragmentation of the superfluid state. Concomitantly, magnetization begins to develop 
spatially in the system, in the form of isolated islands, which are roughly complementary to 
the suppression of the pairing field. The real space observation is further attested by the momentum 
resolved structure factor corresponding to the pairing field and magnetization.  

\begin{figure*}
\centering
\includegraphics[width=1.0\linewidth]{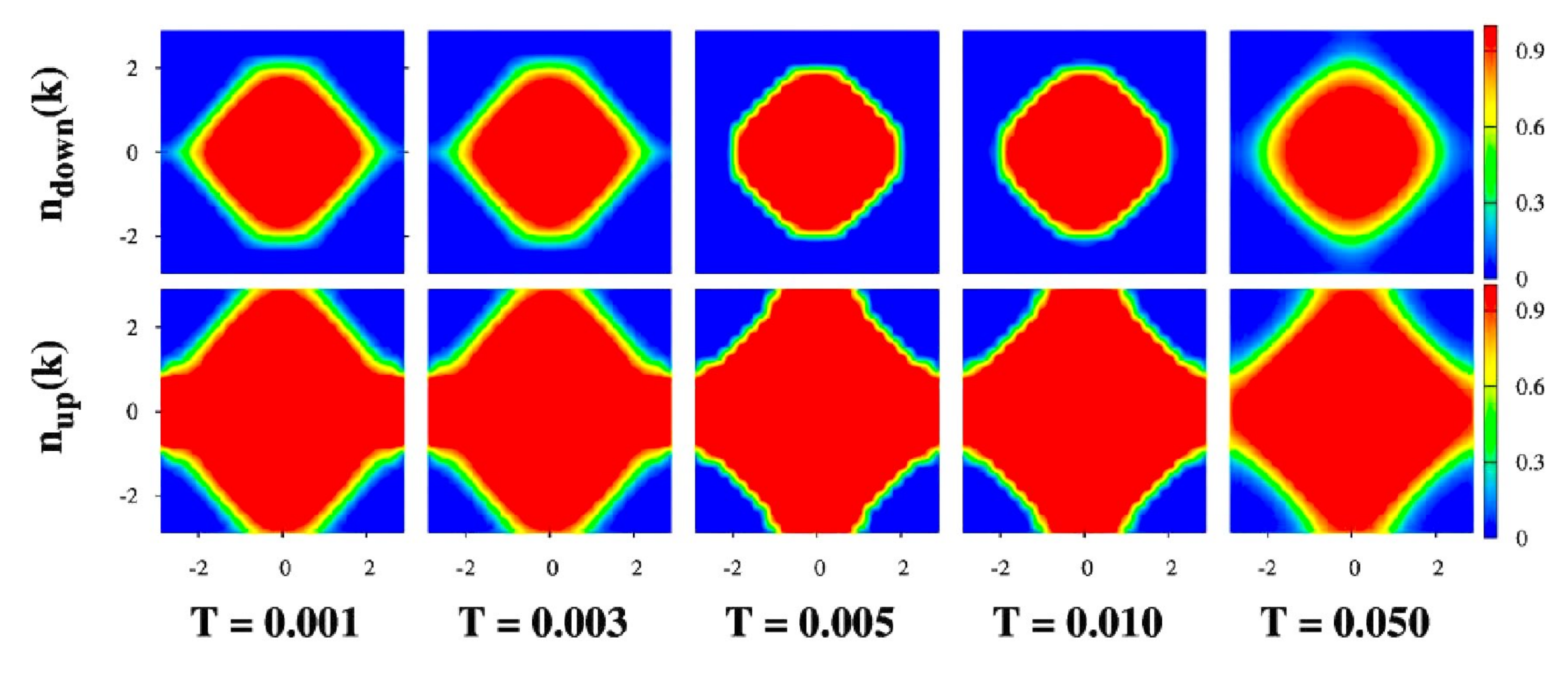}
\caption{Color online: Thermal evolution of the momentum and spin resolved fermionic occupation number $n_{\sigma}({\bf k})$ 
in the LO phase. Reprinted figure with permission from \cite{mpk2016}, Copyright (2016) by the American 
Physical Society.}
\end{figure*}

At the ground state the LO phase was characterized by uniaxial stripes with 
the exact pairing momentum being dictated by the magnitude of the Zeeman field for a fixed 
interaction and chemical potential. The thermal disordering of the LO phase is demonstrated via 
the real space maps in Figure 6 \cite{mpk2016}. The maps show that the thermal scales are strongly suppressed 
in the LO phase and the long range order is lost even at T$\sim$0.005t, even though short range 
correlations survive upto high temperatures. The superfluid order and magnetization are spatially 
complementary to each other and the minimum of one corresponds to the maximum of the other.

Distribution of spin and momentum resolved fermionic occupation number which maps out the Fermi 
surface shows that in the LO phase the Fermi surface is two-fold 
symmetric, rather than the four-fold symmetry of the uniform superconducting state. Signatures of 
LO phase was observed upto T $\sim$ 2T$_{c}$, at still higher temperatures the signatures of pairing 
correlations are lost and the distribution was found to be akin to the PPFL phase, as shown in 
Figure 7.  

\begin{figure*}
\centering
\includegraphics[width=0.93\linewidth]{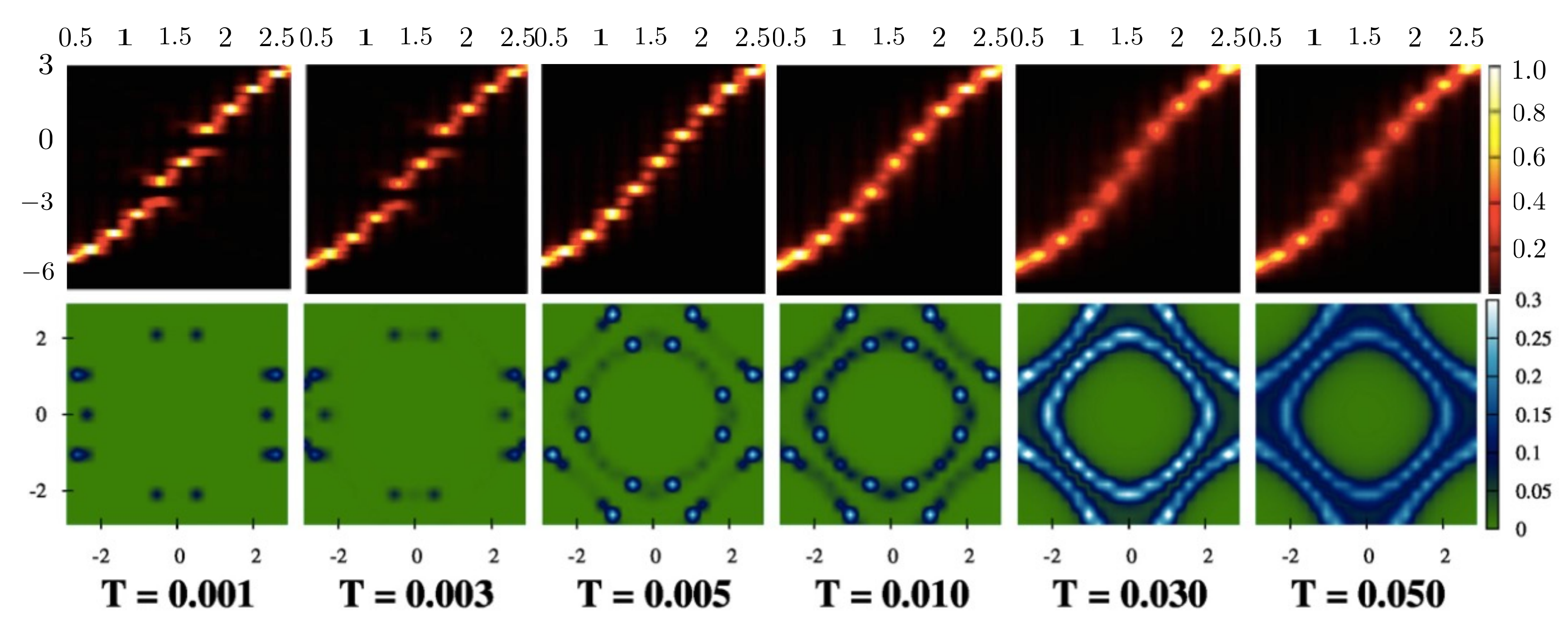}
\caption{Color online: Thermal evolution of spectral function ($A_{\uparrow}({\bf k},\omega)$) (top row) and low 
energy spectral weight distribution ($A({\bf k}, 0)$) (bottom row), in the LO phase. The spectral function is 
plotted along the $(0, 0)-(\pi, \pi)$ trajectory across the Brillouin zone. The $x$-axis corresponds to the momentum 
$k_{x}=k_{y}$ and the $y$-axis corresponds to the energy eigenvalues. The low energy spectral weight distribution 
in shown in the $k_{x}-k_{y}$-plane.}
\end{figure*}

The quasiparticle properties such as single particle DOS, spectral functions and distribution 
of low energy spectral weight carried intriguing imprints of the underlying modulated superfluid 
phase \cite{mpk2016,mpk_epjd}. The single particle DOS contained additional van-Hove singularities,  
in agreement with the mean field predictions \cite{loh2010}. This was the first and so far the only 
work on population imbalanced $s$-wave superfluids which mapped out the spectral function ($A({\bf k},\omega)$) 
of the LO phase, within a lattice fermion model. 
Owing to the multiple scatterings that the fermion undergoes,  the dispersion spectra of the LO phase 
was found to contain six branches, in contrast to the two branches of the BCS (uniform) superconductors. 
Each fermion species gives rise to three dispersive branches, separated by soft gaps, located at ($\omega=0$)
and away ($\omega = \pm h$), from the Fermi level. In agreement with such a non trivial 
dispersion spectra the corresponding low energy spectral weight distribution ($A({\bf k}, 0)$) 
was found to be highly intriguing. It was observed that in spite of a $s$-wave pairing state symmetry 
a {\it nodal} superfluid gap structure emerges out of the LO correlations. Figure 8 shows the thermal 
evolution of the spectral function (for one of the fermionic species) and low energy spectral weight 
distribution of the LO phase. Thermal fluctuations lead to the smearing out of the nodal architecture 
of the low energy spectral weight distribution and at high temperatures the four-fold symmetry of the 
$s$-wave superfluid gap is restored. 

\begin{figure}
\centering
\includegraphics[width=1.01\linewidth]{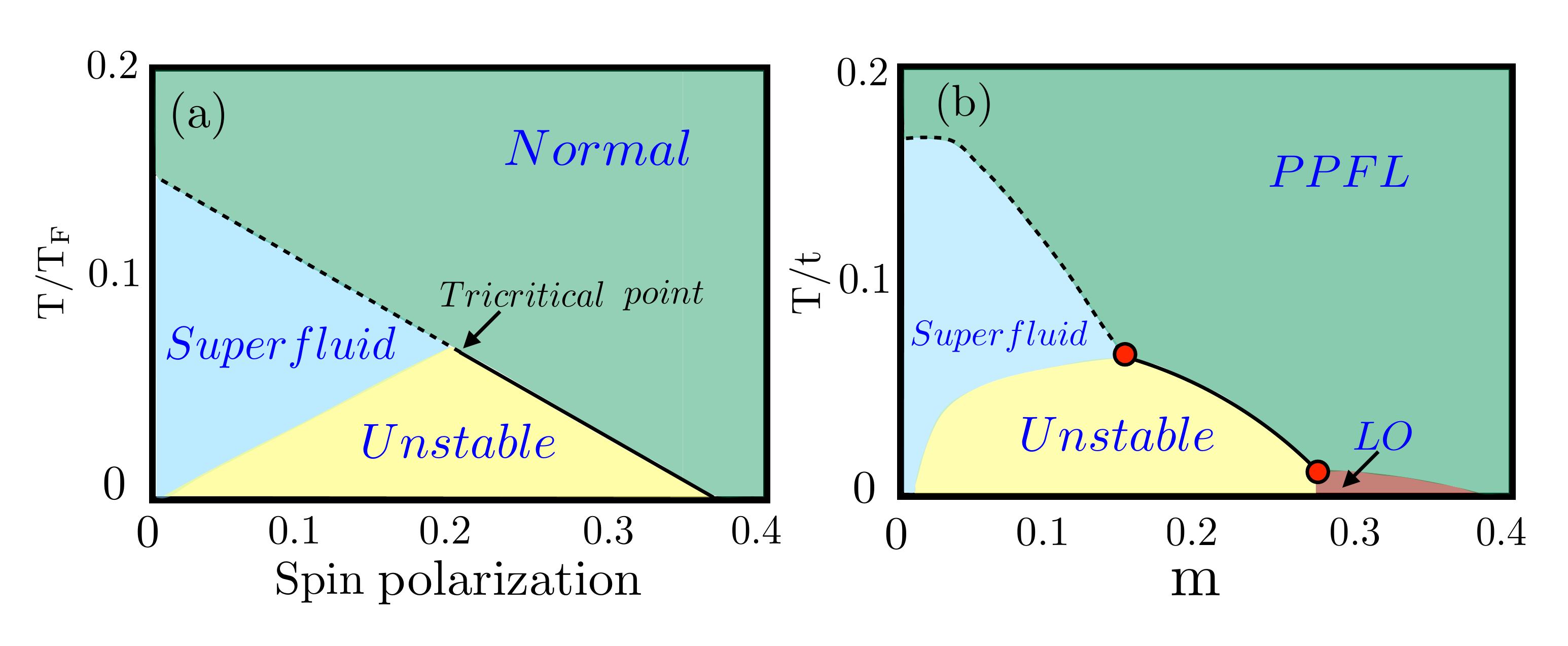}
\caption{Color online: The imbalance-temperature phase diagram inferred from measurements on the ${}^{6}Li$  
cold atomic gas at unitarity (left) \cite{ketterle_nature2008}, compared with the results obtained 
at an intermediate coupling (peak T$_{c}$) of 2D Hubbard model, using SPA (right). 
The normalization of the x axis is same in both panels, while the y axis have different reference scales.
Figure adapted with permission from \cite{mpk2016}, Copyright (2016) by the American Physical Society.}
\end{figure}

This work drew comparison between the numerically computed \cite{mpk2016} thermal phase diagram with that obtained 
experimentally in ultracold atomic gas set up \cite{ketterle_nature2008}. As shown in Figure 9, the agreement 
between the two, both in terms of the order of phase transitions as well as thermal scales, is fairly good. 
Numerical calculations suggest that the large magnetization regime hosts the LO phase. Though the signature of 
spatially modulated LO phase was not observed in the experiment, its possibility has not been ruled out either.
 In a related work the same authors determined an anomalous 
pseudogap phase with non trivial spectral signatures, near the finite temperature BP-LO boundary 
at low magnetizations and a re-entrant pseudogap phase at large magnetizations \cite{mpk_epjd}.   

\subsection{Pauli limited $d$-wave superconductors}

Studies carried out on the FFLO phase of non $s$-wave superconductors are relatively fewer, 
inspite of all the solid state Pauli limited superconductors being $d$-wave. Vorontsov {\it et al.} 
used self consistent theory to study the effect of in-plane Zeeman field on a two-dimensional 
superconducting system with $d_{x^{2}-y^{2}}$ pairing state symmetry \cite{vorontsov2005,vorontsov2006}. 
They solved Eilenberger equations in presence of magnetic field for the superconducting order parameter 
with spatial dependence, to map out the phase diagram in the temperature-magnetic field (T-H) plane. Their 
study showed the LO phase to be more stable as compared to the FF, over a large part of the parameter space. 
The single particle DOS as determined from their calculations showed van-Hove singularities akin 
to Andreev bound states, which are characteristic to FFLO state. 

\begin{figure}
\centering
\includegraphics[width=0.7\linewidth]{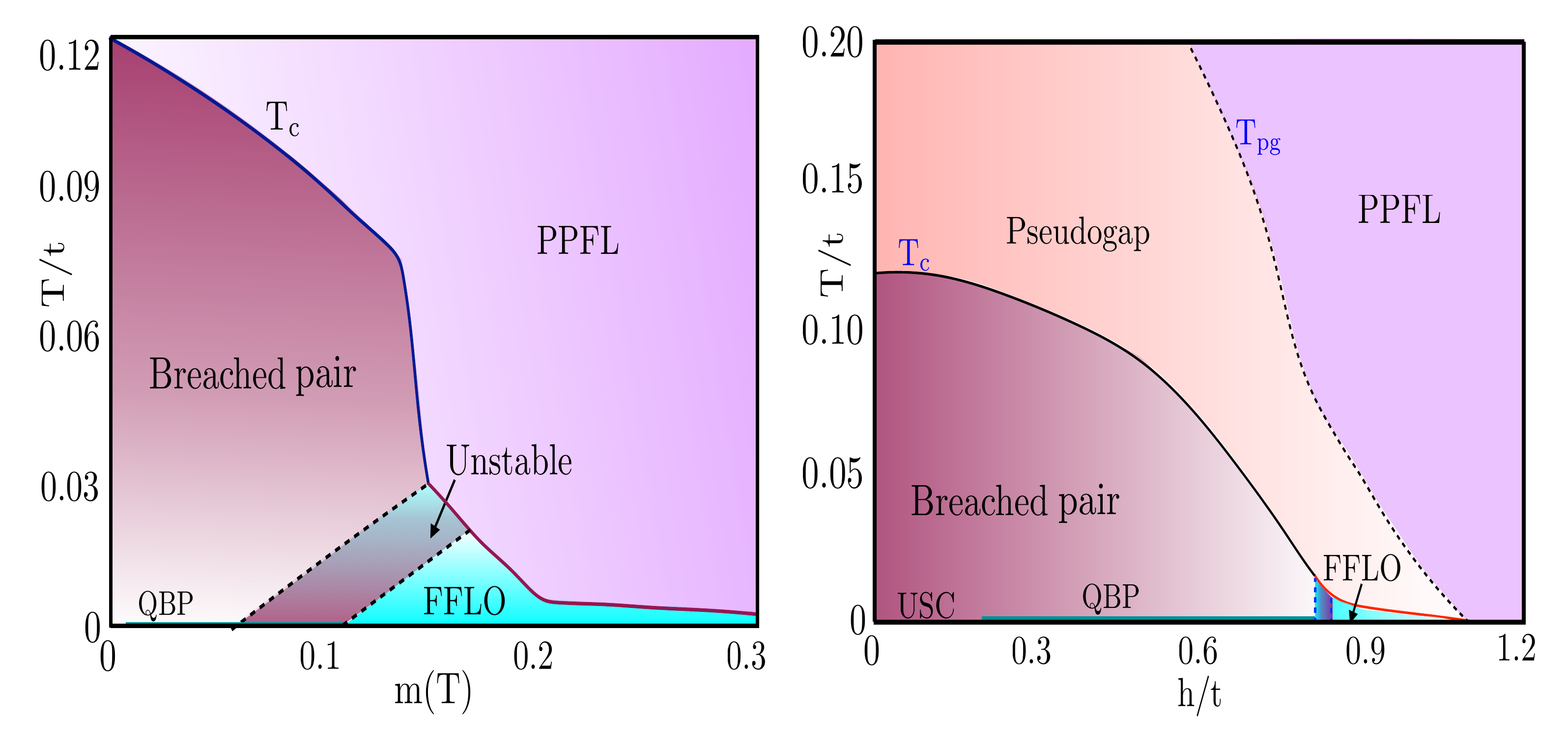}
\caption{Color online: (Left) Magnetization-temperature (m-T) and (Right) Zeeman field-temperature (h-T) phase 
diagram of population imbalanced $d$-wave superconductor. The solid curves correspond to T$_{c}$, with second and first 
order phase transitions being represented by black and red curves, respectively. The h-T phase diagram shows 
the thermal scale T$_{pg}$ in terms of dotted curve and the associated pseudogap regime. 
Reprinted figure with permission from $\cite{mpk_jpcm2020}$, Copyright (2020) IOP Publishing Ltd. Printed in the UK.}
\end{figure}

Within the framework of Ginzburg-Landau theory  Beaird {\it et al.} studied Pauli limited superconductors 
with $d$-wave pairing state symmetry \cite{beaird2010}. They included thermally induced classical 
magnetic fluctuations 
in their theoretical framework to explain the experimentally observed phase diagram of CeCoIn$_{5}$. 
Their work showed that the inclusion of magnetic fluctuations affect the thermodynamic phase 
diagram significantly such that, a regime of first order thermal transition from the normal state 
emerges near the quantum critical point where the system undergoes transition between the uniform 
$d$-wave superconducting state and a spatially modulated LO phase. Beaird {\it et al.} \cite{beaird2010} 
showed that the regime of first order thermal transition straddles both the uniform and LO superconducting 
phases in agreement with the experimental observations of the heavy fermion Pauli limited superconductor 
CeCoIn$_{5}$ \cite{kumagai2006}.

The effect of non magnetic impurities on Pauli limited $d$-wave superconductor was studied by 
Vorontsov {\it et al.} using quasiclassical Eilenberger equations in which the effect of impurities 
was included via self consistent T-matrix approximation \cite{vorontsov2008}. It was inferred that 
contrary to the naive expectations, the FFLO phase is robust against impurities in the limit of 
strong disorder. 

\begin{figure*}
\centering
\includegraphics[width=0.8\linewidth]{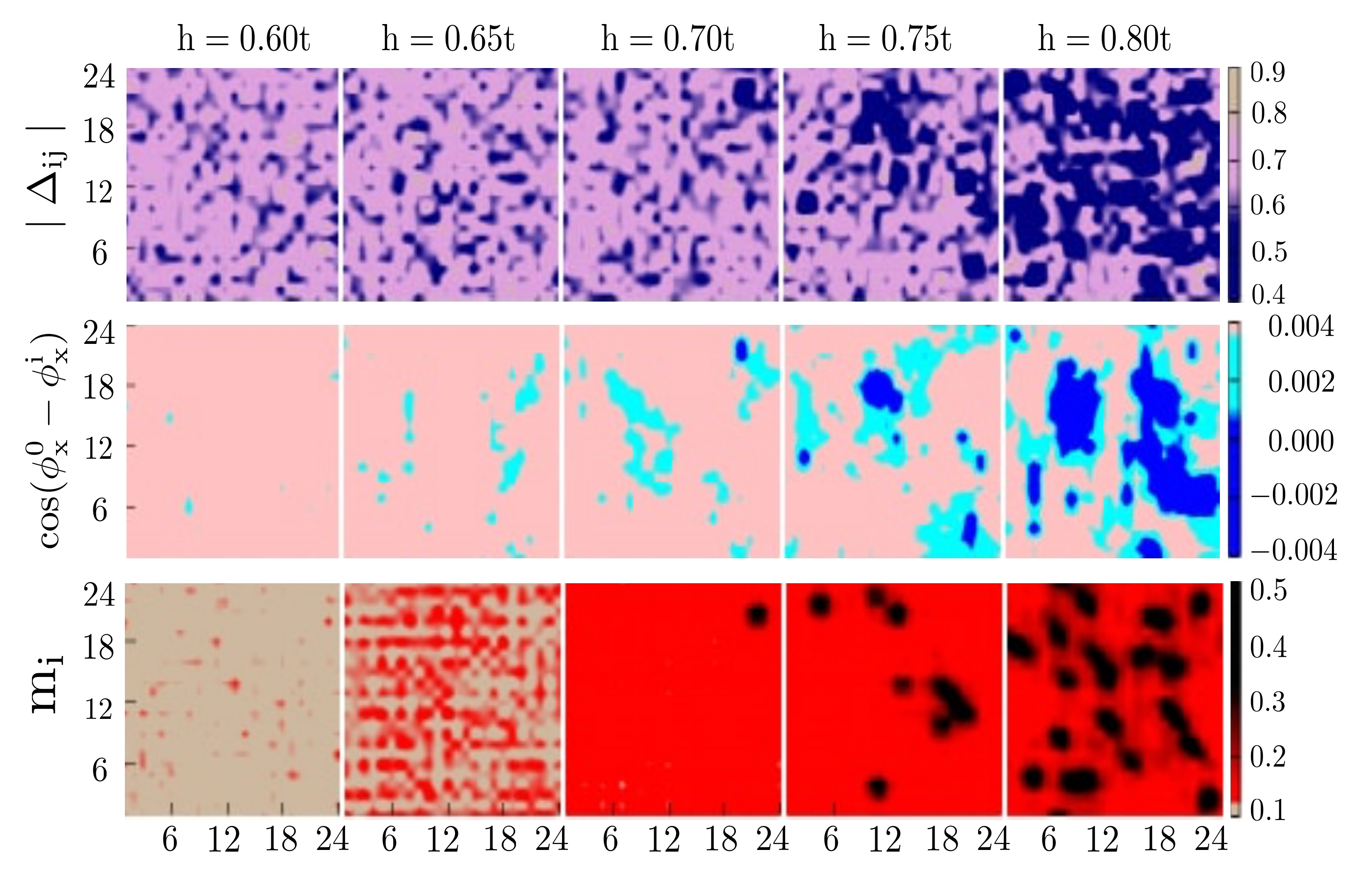}
\caption{Color online: Real space maps at T=0 showing the evolution of the QBP phase over a 
range of Zeeman field in terms of (i) the amplitude of superconducting pairing field, (ii) phase correlation of 
the superconducting pairing field and (iii) the magnetization. The color signifies the magnitude of the corresponding 
indicator. Reprinted figure with permission from $\cite{mpk_jpcm2020}$, Copyright (2020) IOP Publishing Ltd.
 Printed in the UK.}
\end{figure*}

Using the BCS weak coupling model Yang and Sondhi made certain intriguing predictions about $d$-wave 
superconductor subjected to an in-plane Zeeman field \cite{yang1998}. They showed that owing to the nodal architecture 
of the $d$-wave pairing state the Zeeman field would destroy superconductivity in parts of the Fermi 
surface and give rise to pockets of normal electrons. The authors suggested that this phenomenon should 
be observable in the quasiparticle signatures of the system.

Using the BdG mean field theory on a two-dimensional lattice model Zhou and Ting mapped out the thermal 
phase diagram of Pauli limited $d$-wave superconductor as a function of in-plane Zeeman field \cite{zhou2009}. 
Apart from the uniform superconducting phase in the low field regime they observed one and two-dimensionally 
modulated FFLO phases in the regime of high field and low temperatures. Since the approach does not take 
into account the effects of thermal fluctuations the stability of the modulated phases remain unknown. 
The single particle DOS at the ground state as obtained by the BdG mean field theory showed signatures 
of LO state in terms of Andreev bound states. 

The beyond mean field studies of $d$-wave Pauli limited superconductors are much lesser than 
that of $s$-wave. In a series of papers Ikeda and co-authors studied the problem of FFLO phase 
in $d$-wave superconductor in presence of antiferromagnetic (AFM) fluctuations, relevant for materials 
such as CeCoIn$_{5}$ \cite{ikeda2014,ikeda2015,ikeda2013,ikeda2011,ikeda2017}. Within the 
framework of repulsive Hubbard model which gives rise to $d$-wave superconductivity as well as AFM order, 
the effect of paramagnetic pair breaking was taken into account through an in-plane Zeeman field. Using fluctuation 
exchange approximation (FLEX) the authors showed that the high magnetic field low temperature regime 
comprises of coexisting AFM and $d$-wave superconducting orders, as well as a FFLO phase. The authors claimed 
their results to be in agreement with the experimental observations on CeCoIn$_{5}$. 
A Ginzburg Landau theory based study was carried out by Adachi and Ikeda on two-band Pauli limited 
superconductors, which is expected to provide useful insights to the behavior of Pauli limited iron superconductors
such as KFe$_{2}$As$_{2}$ \cite{adachi2015}. 

\begin{figure*}
\centering
\includegraphics[width=0.8\linewidth]{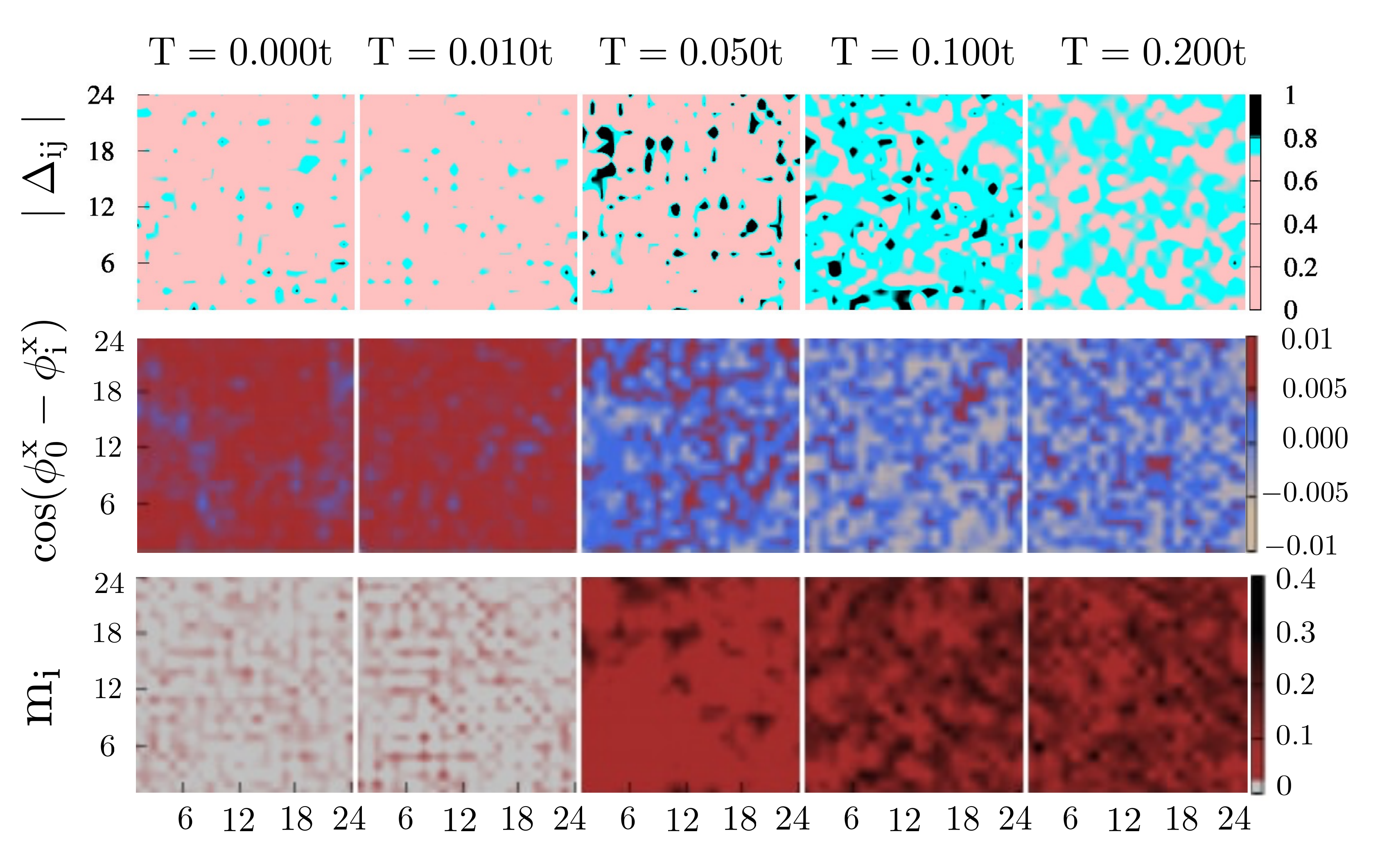}
\caption{Color online: Real space maps as the system evolves in temperature, at a selected Zeeman field in the 
QBP phase.  Reprinted figure with permission from $\cite{mpk_jpcm2020}$, Copyright (2020) IOP Publishing Ltd.
 Printed in the UK.}
\end{figure*}

The static path approximation (SPA) technique was used at a fixed chemical potential and intermediate 
interactions to address the behavior of two-dimensional population imbalanced $d$-wave 
superconductor, in detail \cite{mpk_jpcm2020}. The corresponding phase diagram as shown in Figure 10 
was mapped out in the magnetization-temperature (m-T) as well as in the field-temperature 
(h-T) plane. It was demonstrated 
that the low field (magnetization) regime comprises of a ``quantum breached pair'' (QBP) phase, 
characterized by spatial coexistence of ``gapless'' $d$-wave superconductivity and non zero 
magnetization. This was the first numerical realization of the coexistent phase predicted by 
Yang and Sondhi, in a lattice fermion model \cite{yang1998}. The QBP phase can be understood as the T=0 counterpart 
of the finite temperature BP phase realized in the population imbalanced $s$-wave superconductors/superfluids. 
It was observed that increase in field (magnetization) leads to a {\it Lifshitz transition} from the uniform 
$d$-wave superconductor to the QBP phase, quantified by the average magnetization of the system. 
Both the uniform $d$-wave superconductor and QBP phases 
undergo second order thermal phase transition to the PPFL phase. The high magnetic field (magnetization) 
regime hosts FFLO state at low temperatures, which undergoes first order thermal phase transition to PPFL. 
Across the quantum critical point corresponding to the transition between the QBP and FFLO phases the 
thermal phase transition was found to be of first order, in agreement with the earlier predictions \cite{beaird2010}. 
Inclusion of fluctuations showed significant suppression of the superconducting transition temperature, both for 
the uniform superconducting as well as FFLO regimes. 

Based on the real space signatures shown in Figure 11 and Figure 12 it,  was inferred that both field and temperature 
leads to spatial inhomogeneity in the superconducting state, which shows up in the quasiparticle signatures 
as the ``pseudogap'' behavior. The phase diagram thus comprises of an additional thermal 
scale T$_{pg}$ as shown in Figure 10, where short range superconducting pair correlations die out, long 
after the loss of long range superconducting phase coherence at T$_{c}$ \cite{mpk_jpcm2020}.  

\subsection{Mass imbalanced $s$-wave superfluids}

In ultracold atomic gases pairing between equal populations of two atomic species with unequal atomic 
masses (Fermi-Fermi mixture) can give rise to mass imbalanced Fermi superfluid. Experimentally a mass 
imbalanced Fermi-Fermi mixture can be achieved in ${}^{6}$Li-${}^{40}$K mixtures. The theoretical studies 
are primarily restricted to continuum models, using both mean field and beyond mean field techniques
\cite{takemori2012,ohashi2013,roscher2015,braun2014,ohashi2014,guo2009,gubbels2009}. 
The cumulative outcome of these works brought forth two important observations, (i) imbalance 
in mass promotes the instability of the system towards a supersolid phase accompanied by a Lifshitz 
transition, (ii) the mass imbalanced Fermi-Fermi mixture contains more than one pseudogap scales at 
high temperature. 

Works on lattice fermion models of mass imbalanced Fermi-Fermi mixtures are very limited. 
Wang {\it et al.} \cite{wang2009} studied this problem in one dimension using time-evolving block 
decimation technique and found that increasing mass imbalance ratio shrinks the FFLO regime of the 
phase diagram and at large imbalance the FFLO phase disappears. 

\begin{figure*}
\centering
\includegraphics[width=1.03\linewidth]{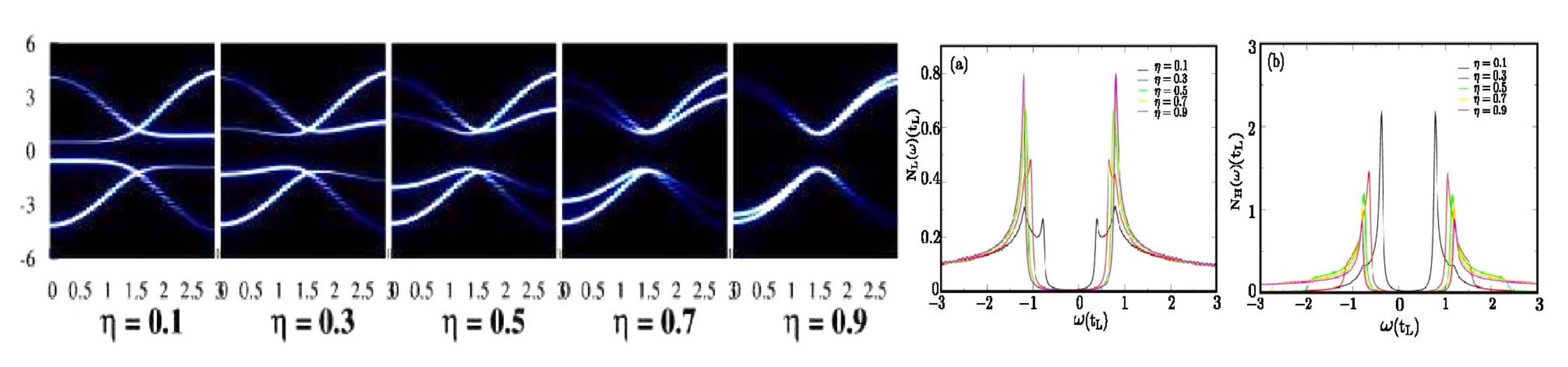}
\caption{Color online: Ground-state dispersion spectra at different mass imbalance ratio $\eta$.
The last two panels show the $\eta$ dependence of the density of states (DOS) for the (a) light- 
and (b) heavy-fermion species. Reprinted figure with permission from \cite{mpk2018}, Copyright (2018) 
by the American Physical Society.}
\end{figure*}

Using density matrix renormalization group (DMRG) and quantum Monte Carlo (QMC) techniques Dalmonte {\it et al.} 
\cite{dalmonte2012} showed that mass imbalance favors paired phases as compared to population imbalance,  and 
concluded that in mass imbalanced systems it might be possible to detect FFLO in experimentally realizable 
parameter regimes. A mean field study on two-dimensional square lattice by Pahl and Koinov revealed 
that the ground state phase diagram of this system comprises of normal, FF and BP regimes and that a majority of 
heavy species promotes the FF phase \cite{pahl2014}. A non perturbative lattice Monte Carlo study was carried 
out in two-dimensions, which revealed that a mean field approach to the problem grossly overestimates 
the ground state energy and therefore makes the stability of the phases questionable \cite{braun2015}.  

The only beyond mean field study of mass imbalanced Fermi-Fermi mixture in two-dimensions, at 
and above the ground state was carried out using the SPA \cite{mpk2018}. Within the 
framework of attractive Hubbard model the imbalance in mass was taken into account through imbalance 
in the hopping amplitude of the two species as, $t_{H}/t_{L}=m_{L}/m_{H}=\eta$. Here, $H$ and $L$ 
corresponds to the heavy and light fermion species, respectively and $\eta$ serves as the tuning parameter to 
control the ratio of imbalance. The work demonstrated that imbalance in mass significantly alters 
the dispersion spectra of the fermionic species and gives rise to subgap and supergap 
features in the species resolved single particle DOS, as shown in Figure 13. 

\begin{figure}
\centering
\includegraphics[width=0.7\linewidth]{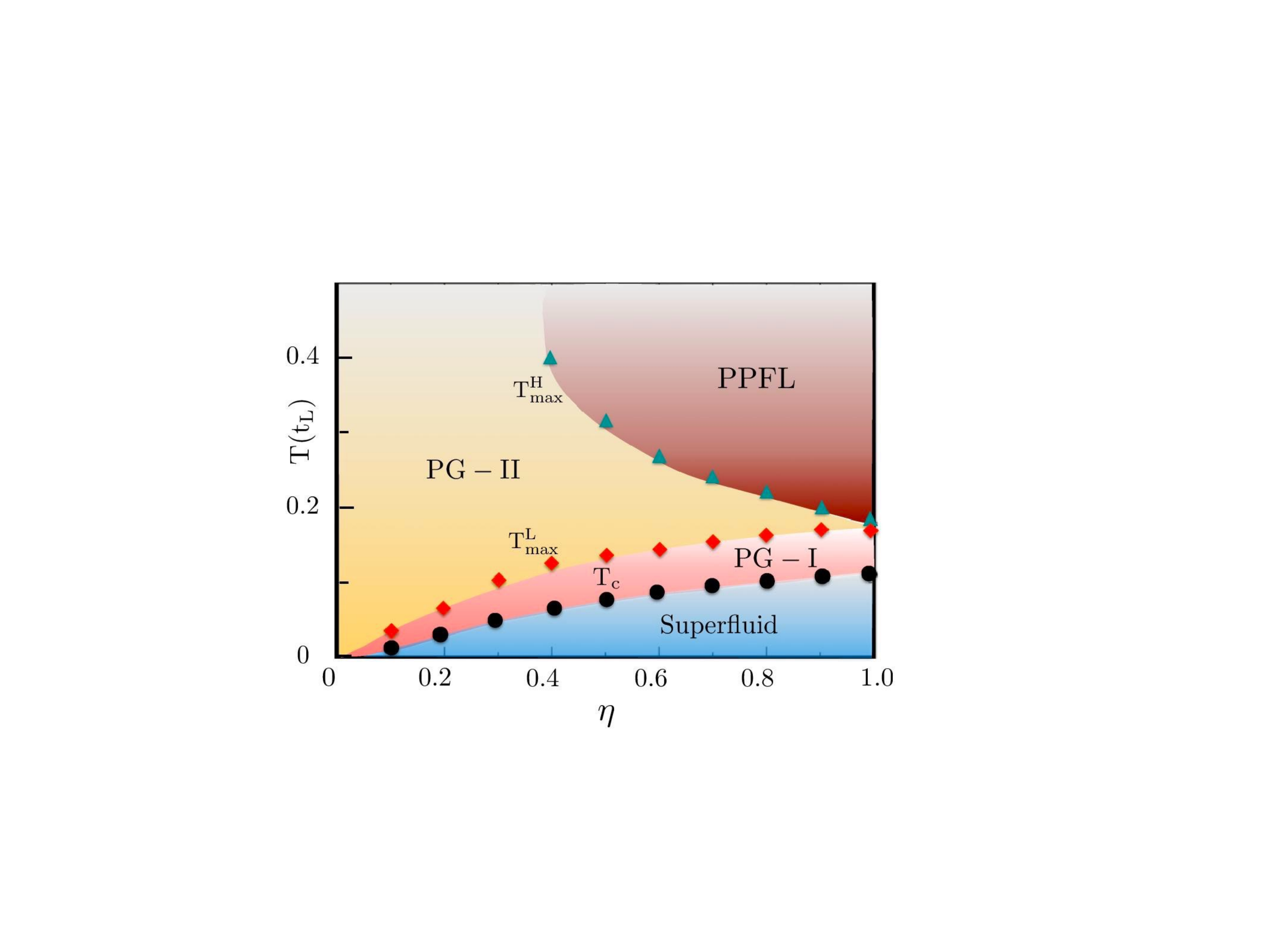}
\caption{Color online: Mass imbalance-temperature ($\eta$-T) phase diagram at fixed population 
imbalance. Along with the superfluid regime, the figure shows the pseudogap regimes
based on the species-resolved DOS. T$_{max}^{L}$ and T$_{max}^{H}$ correspond to the pseudogap scales. 
PG-I corresponds to the regime where both the fermionic 
species are pseudogapped and in the PG-II regime only the lighter species is pseudogapped. 
Reprinted figure with permission from \cite{mpk2018}, Copyright (2018) by the American Physical Society.}
\end{figure}

At a fixed imbalance of population, the phase diagram comprises of LO, BP and unstable (phase separated) 
regimes in the imbalance-temperature plane and a larger imbalance in mass reduces the regime of stability 
of these phases. The thermal phase diagram in the imbalance-temperature 
($\eta-T$)-plane presented in Figure 14 shows the intriguing feature of two pseudogap phases as PG-I and 
PG-II, in agreement 
with the observation of continuum models \cite{ohashi2013,ohashi2014}. Owing to the mismatch in mass and 
thus of the hopping amplitudes the two species are subjected different ``scaled'' temperatures. Consequently, 
while in the pseudogap regime PG-I both the fermion species are pseudogapped, in the pseudogap regime PG-II 
it is only the lighter 
species which is pseudogapped while the heavier species is a polarized Fermi liquid. At $\eta=1$ corresponding 
to zero imbalance in mass,  both the pseudogap scales collapses to one. The phase diagram shows that,  while 
akin to the population imbalanced system the T$_{c}$ is strongly suppressed by imbalance in mass, the short 
range pair correlations survive over a large temperature regime of T $\sim$ 2T$_{c}$ \cite{mpk2018}.

\begin{figure}
\centering 
\includegraphics[width=0.62\linewidth]{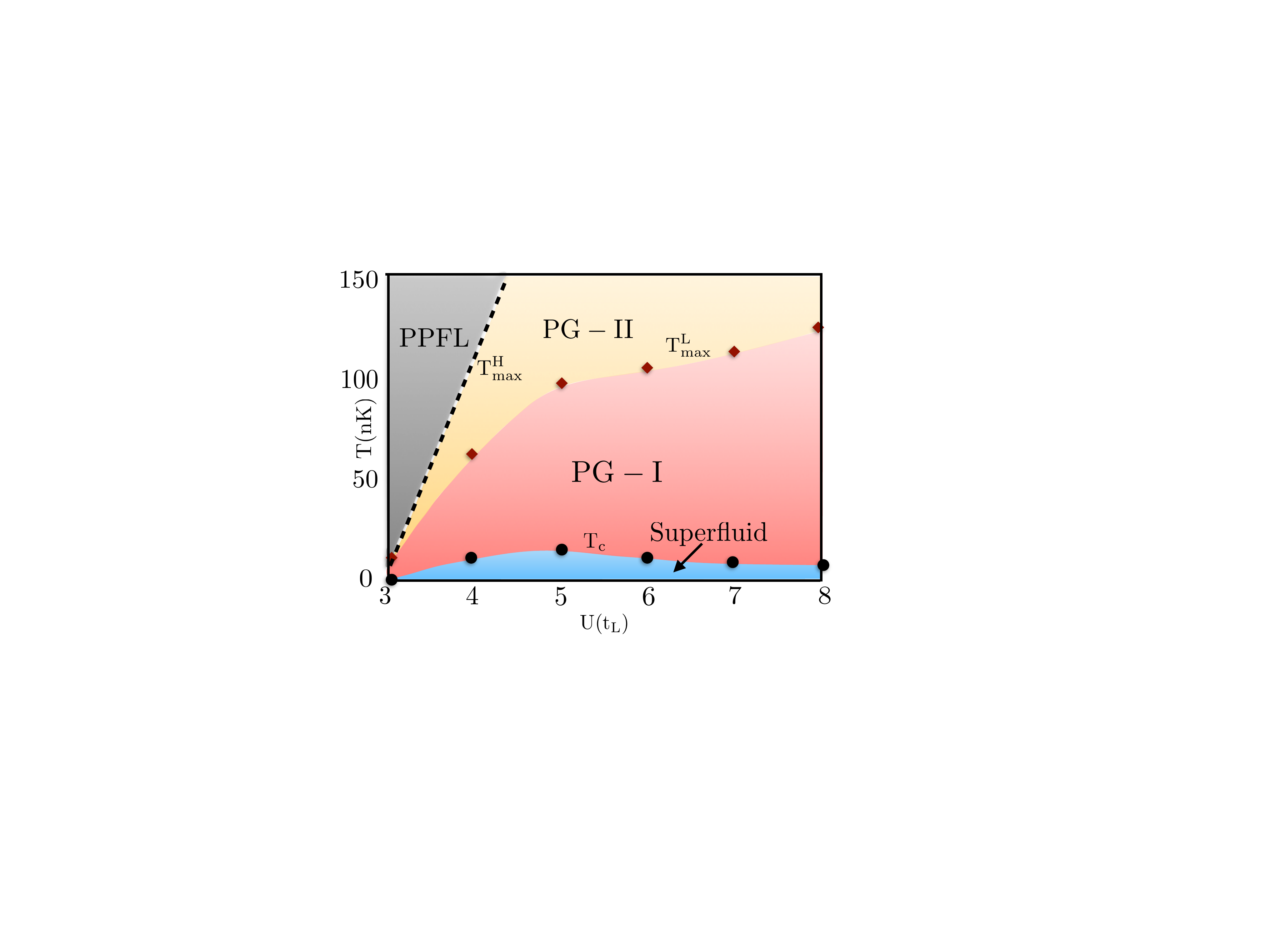}
\caption{Color online: Interaction-temperature (U-T) phase diagram at $\eta$=0.15 corresponding 
to the ${}^{6}Li-{}^{40}K$ mixture. The figure shows the T$_{c}$ scale along with the pseudogap 
scales for this mixture at a fixed population imbalance. 
Reprinted figure with permission from \cite{mpk2018}, Copyright (2018) by the American Physical Society.}
\end{figure}

The work further discussed the BCS-BEC crossover picture for mass imbalanced Fermi-Fermi mixture at a fixed 
population imbalance and established the presence of two pseudogap regimes across the crossover. The 
BCS-BEC crossover revealed a crucial observation. It showed that a critical interaction strength (U$_{c}$) 
is required to realize superfluid order in mass imbalanced Fermi-Fermi mixture in presence of a population imbalance. 
This observation is in striking contrast with those of two component balanced Fermi gases where superfluid 
phase is realized for any arbitrary strength of attractive interaction. Figure 15 presents the BCS-BEC crossover 
of the Fermi-Fermi mixture, which shows that the superfluid T$_{c}$ collapses to zero at U$\sim 3t_{L}$ for 
the parameter regime mentioned in the figure caption. 

A similar study was carried using SPA by Karmakar \cite{mpk_jpcm2020}, for Fermi-Fermi 
mixtures with non local $d$-wave interaction between the fermion species. In the context of solid state 
systems, materials with unequal masses of the fermions can be envisaged as different fermion species 
belonging to different electronic bands. The thermal phase diagram of this system was found to be similar 
to the one obtained for a mass imbalanced $s$-wave Fermi gas, with two finite temperature pseudogap regimes 
even for a population balanced system.
The ground state calculations showed that over a large regime of mass imbalance the QBP phase is 
stablized with coexisting gapless $d$-wave superfluidity and non zero magnetization, which undergoes 
Lifshitz transition to an uniform $d$-wave superfluid phase, with decreasing imbalance in mass. 

\section{Imbalance and spin-orbit coupling in Fermi systems}

Over the past few years there has been immense progress both in the experimental and theoretical 
front to understand the spin-orbit coupled ultracold atomic gases. Since the ultracold atomic gases are 
one of the prospective systems to realize exotic superfluid phases such as,  FFLO and BP, there has 
been obvious interest to investigate the effect of spin-orbit coupling on imbalanced Fermi gases. 
In this section we will summarize some of the major developments in this area and the observations 
therein. We will also touch upon some of the open problems in this area towards the end of this section. 
For further details on spin orbit coupled Fermi gases the readers are encouraged to consult the excellent 
reviews by Galitski and Spielman \cite{galitski2013} and Dalibard {\it et al.} \cite{dalibard2011}.
The subject is also discussed in a recent review by Kinnunen {\it et al.} \cite{torma_rpp_review2018}.  

Spin orbit coupling has remained a subject of interest since the early days of quantum mechanics and 
can be understood as follows. If an electron is allowed to move in an electric field and we 
move the electron's rest frame, this will lead to the generation of a magnetic field by the associated 
Lorentz transformation. This magnetic field will couple itself to 
the electron spin leading to spin-orbit coupling. The exact form of the spin-orbit coupling depends upon 
the direction of the electric field and the electron velocity; one of the commonly used form of the 
spin-orbit coupling is the Rashba spin-orbit coupling (RSOC), given as,
\begin{eqnarray}
H_{RSOC} & = & \alpha(k_{y}\hat \sigma_{x} - k_{x}\hat \sigma_{y})
\end{eqnarray} 
where, $\alpha$ is the strength of the spin-orbit coupling interaction, $\hat \sigma_{x}$ and $\hat \sigma_{y}$ 
are the Pauli matrices. In the absence of inversion symmetry the Rashba spin-orbit coupling takes the form 
of Dresselhaus spin-orbit coupling and is given as,
\begin{eqnarray}
H_{DSOC} & = & \alpha(k_{y}\hat \sigma_{x} + k_{x}\hat \sigma_{y})
\end{eqnarray}
In real space the Rashba spin-orbit coupling is given as, 
\begin{eqnarray}
H_{RSOC} & = & -\sum_{\langle ij\rangle, \sigma}t_{ij,\alpha}\hat c_{i,\sigma}^{\dagger}\{{\bf \sigma}\cdot 
{\bf \alpha}_{ij}\}_{\sigma,\sigma^{\prime}}\hat c_{j,\sigma^{\prime}}
\end{eqnarray}
where, ${\bf \sigma}$ is the vector of Pauli matrices and ${\bf \alpha}_{ij} = \alpha \hat z \times {\bf r}_{ij}$
is the spin-orbit vector on each bond of a two-dimensional square lattice. $t_{ij,\alpha}$ correspond 
to the strength of the spin-orbit coupling.

In ultracold atomic gas experiments spin-orbit coupling is generated using synthetic electric field \cite{dalibard2011}
and pseudospins of the atomic species. The basic principle that is followed in such experiments is to couple 
different internal energy states of atoms through a two-photon Raman transition. The atom first absorbs 
the photon from one laser beam then emits the photon to a second laser beam, by stimulated emission. 
The process conserves momentum by imparting some of the momentum to the atom. The Doppler shift between 
the laser beams and the atom gives rise to the momentum dependent spin-orbit coupling \cite{higbie2002}. 
Spin-orbit coupling in ultracold atomic gases was experimentally realised for bosons \cite{lin_nature2011} 
as well as for fermions \cite{wang_prl2012,cheuk_prl2012}. A detail discussion on the experimental realization 
of spin-orbit coupling in ultracold atomic gases can be found in 
\cite{dalibard2011,ruseckas2005,osterloh2005,campbell2011}. 

Theoretically, the effect of spin-orbit coupling on the FFLO superfluid state has largely been investigated 
using mean field theory. This has been carried out either by determining the ground state energy using 
the FF ansatz in presence of spin-orbit coupling \cite{zheng_pra2013}, or by solving the self consistent BdG 
equations for arbitrary 
choice of spatially inhomogeneous superfluid order parameter, so as to determine the global minima of the 
energy landscape \cite{xu_2014,iskin2012,iskin2013,seo2013}. Obviously, the second approach is computationally 
more demanding than the first one but allows for a larger family of possible ground state solutions. 

Based on BdG mean field theory in two-dimensions, Xu {\it et al.} showed that in a continuum system in 
presence of spin-orbit coupling the FF state is energetically favoured over 
the LO state, particularly at strong spin-orbit couplings \cite{xu_2014}. This observation was reconfirmed 
by several other works.  A continuum system in three-dimensions with spin-orbit coupling and in presence 
of in-plane Zeeman field was studied by Zheng {\it et al.} \cite{zheng_pra2013}. They showed that the FF state 
is stable over a large parameter regime.      
In a similar spirit, Hu and Liu pointed out that at sufficiently large spin-orbit coupling the superfluid 
state is always an FF state \cite{hu_njp2013}. They worked out the finite temperature phase diagram at 
a broad Feshbach resonance and predicted a superfluid transition temperature of T$_{c}\sim$ 0.2T$_{F}$, 
where, T$_{F}$ is the Fermi temperature. Similar estimates were made by these authors when they extended 
their model by including out of plane Zeeman field \cite{liu_hu2013}. 

Beyond mean field corrections were incorporated for the continuum model of three dimensional spin-orbit 
coupled Fermi gas by Liu \cite{liu2013}. Once again the superfluid transition temperature was found to be  
upto $\sim$ 0.2T$_{F}$. Similar conclusions were drawn by Dong {\it et al.} \cite{dong2013}
and Zhou {\it et al.} \cite{zhou2013} for spin-orbit coupled three-dimensional Fermi gases.

Within the framework of lattice fermion models, Iskin investigated the spin-orbit coupled Fermi gas in 
presence of an optical trap in two dimensions, using BdG mean field theory \cite{iskin2_2013}. 
He showed that in presence of spin-orbit coupling and in-plane Zeeman field the FF state is stabilized,  
in agreement with the observations of continuum model. By investigating the finite temperature behavior 
Iskin suggested that since the modulations of the superfluid order parameter are restricted near the edge 
of the trap, it would be difficult to detect them at high temperatures \cite{iskin2_2013}. 

The topological aspect of spin-orbit coupled Fermi gases was investigated by Tewari {\it et al.} \cite{tewari2011}, 
who showed that spin-orbit coupling along with Zeeman field can be used to drive a phase transition between a 
topologically trivial and a topologically non-trivial superfluid. Similar observations were reported 
for imbalanced Fermi gases in presence of both spin-orbit coupling and Zeeman field \cite{zhang2013,qu2013}. 
In their studies carried out on two dimensional continuum system Zhang {\it et al.} \cite{zhang2013} and 
Qu {\it et al.} \cite{qu2013} identified a topological FF state in addition to gapped FF and gapless 
nodal FF states. They further confirmed the existence of chiral edge states of the topological FF state, 
on a two-dimensional square lattice \cite{qu2013}. 
Observation of gapless topological FF states was also reported by Cao {\it et al.} \cite{cao2014}. 
They further confirmed the robustness of the Majorna modes by 
adding disorder potential to the system. Presence of topological 
FF phase has also been confirmed in one-dimensional lattices \cite{liu2012}. 

Mean field theory has been pushed to investigate the finite temperature behavior of spin-orbit coupled 
imbalanced Fermi gases, as well \cite{xu2015,cao2015}. It was inferred that together with in-plane and 
out of plane Zeeman field, spin orbit coupling gives rise to BKT transition in two dimensions, both for 
gapped as well as gapless FF phases \cite{xu2015,cao2015}.

In a recent work, Wang {\it et al.} investigated the spin-orbit coupled Fermi gas within the framework 
of continuum model in two and three dimensions \cite{wang2018}. They took into account the effect of pairing field 
fluctuations and studied the system using two schemes, viz. (i) T-matrix approximation and 
(ii) fluctuation exchange approximation (FLEX). Based on their observations from both these approaches 
Wang {\it et al.} inferred that a finite temperature realization of FFLO phase is not possible in this system. 
They remarked that the FFLO phase is unstable against pair fluctuations in continuum in both two and three 
dimensions \cite{wang2018}. Within the purview of lattice fermion models, an auxiliary field QMC study was 
carried out for balanced two-component Fermi gas with spin-orbit coupling at the ground state, by 
Rosenberg {\it et al} \cite{rosenberg2017}. It was shown that such a system comprises of exotic phases 
such as, supersolid phase containing both singlet and triplet pairings. Rosenberg {\it et al.} showed 
the presence of edge currents arising out of the spin-orbit coupling. So far no such investigation has 
been carried out in the context of spin-orbit coupled imbalanced Fermi gases.   

As is evident from the above discussions, the theoretical works on spin-orbit coupled imbalanced Fermi 
gases are largely the ones based on mean field theory. This leaves a void in our understanding as far 
as the effects of fluctuations are concerned. For example, the exact thermal scales of the spin-orbit 
coupled imbalanced Fermi gas in an optical lattice is so far unknown. The presence of lattice breaks 
the translational symmetry and can potentially stabilize phases which are unstable in continuum. It 
is worth investigating whether a modulated superfluid phase can be stabilized in a lattice with spin 
orbit coupling in presence of pair fluctuations. Moreover, spin-orbit coupling allows for the possibility of 
realizing triplet superfluid pairing. Most of the works we discussed here do not take this possibility 
into account. Even if a triplet pairing state does not lead to a long range order at the 
ground state, it is certainly possible that triplet pair correlations are present at high temperatures. 
In the same spirit, with the energy of the FF and LO phases being comparable,  even though the ground 
state hosts FF superfluid phase in presence of spin-orbit coupling, competing LO correlations 
can be present at finite temperatures. Signatures of the presence of such competing correlations would be borne out 
in the quasiparticle properties of the system. Investigations of such effects require one to include 
fluctuations beyond the mean field theory, and are certainly worth pursuing.  

\section{Conclusions and outlook}

The last few years have witnessed a great deal of activity in the theoretical as well as experimental 
front to unveil the mysteries of the exotic superconducting phases like FFLO and BP. Though 
the possibility of superconducting phases with finite centre of mass momentum of the Cooper pairs was 
predicted decades ago, it was the advent of ultracold atomic gases, and significant advancement of 
material science experiments, that has paved the way to access and understand such novel quantum many 
body phases. 

Amongst the solid state materials, so far three candidate materials have stood out as possible hosts of 
FFLO phase viz., heavy fermion superconductor CeCoIn$_{5}$, organic superconductor 
$\kappa$-(BEDT-TTF)$_{2}$Cu(NCS)$_{2}$ and iron superconductor KFe$_{2}$As$_{2}$. With the continuous 
development that is going on in material science, the hope is that other materials would join the class 
of Pauli limited superconductors in not so distant future. The ultracold atomic gas systems with the unprecedented 
control of the tuning parameters, that it provides, have taken the centre stage to engineer and analyze 
exotic quantum phases, such as FFLO. While a clear spatial signature of the modulated superfluid phase 
is yet to be realized in these systems, coexistence of superfluid pair correlations and unpaired fermions 
are indeed observed, suggesting a possible BP phase. 

The theoretical attempts are not far behind either, and the problem of imbalanced Fermi systems have  
been addressed by mean field as well as non mean field approaches. We now have a broad understanding 
of not just the ground state but also of how the effects of fluctuations play out and dictate the thermal 
scales associated with these systems. 

The world of quantum many body physics is however a fast evolving one and newer experimental breakthroughs 
are not infrequent. For example, experimental realization of spin-orbit coupling in fermionic and bosonic 
systems in ultracold atomic gas setups have been reported recently. In the theoretical front however, we 
have not moved beyond the realms of mean field theory so far, except occasionally. It is obvious that 
even for the population balanced scenario, these spin-orbit coupled bosonic and fermionic systems are theoretically 
not well understood at this moment and a concerted effort is required to achieve the same.    
In the same spirit, the effect of disorder adds another interesting dimension to these systems. While 
disorder is largely expected to be detrimental for the FFLO phase, it can possibly aid in to stabilize 
the BP phase, at the ground state of a $s$-wave superfluid. 

This chapter has covered some of the important theoretical and experimental advances and 
observations pertaining to imbalanced Fermi systems, and is no way comprehensive. The subject is not yet 
well understood and promises a plethora of avenues and areas which needs to be explored inquisitively.

\bibliographystyle{unsrt}
\bibliography{imbalance_chapter}

\begin{thebibliography}{100}

\bibitem{tinkham_book}
Michael Tinkham.
\newblock {\em Introduction to $\mathrm{S}$uperconductivity,
  $\mathrm{M}$c$\mathrm{G}$raw-$\mathrm{H}$ill $\mathrm{B}$ook $\mathrm{C}$o.,
  $\mathrm{N}$ew $\mathrm{Y}$ork, 1975.}

\bibitem{bcs}
J.~Bardeen, L.~N. Cooper, and J.~R. Schrieffer.
\newblock Theory of $\mathrm{S}$uperconductivity.
\newblock {\em Phys. Rev.}, 108:1175--1204, 1957.

\bibitem{muller_hightc}
J.~G. Bednorz and K.~A. M\"uller.
\newblock Possible high $\mathrm{T}$$_{c}$ superconductivity in the
  $\mathrm{B}$a-$\mathrm{L}$a$\mathrm{Cu}$-$\mathrm{0}$ system.
\newblock {\em Zeitschrift f\"ur Physik B Condensed Matter}, 64:189--193, 1986.

\bibitem{lee_rmp2006}
Patrick~A. Lee, Naoto Nagaosa, and Xiao-Gang Wen.
\newblock Doping a $\mathrm{M}$ott insulator: $\mathrm{P}$hysics of
  high-temperature superconductivity.
\newblock {\em Rev. Mod. Phys.}, 78:17--85, 2006.

\bibitem{pfleiderer_rmp2009}
Christian Pfleiderer.
\newblock Superconducting phases of $f$-electron compounds.
\newblock {\em Rev. Mod. Phys.}, 81:1551--1624, 2009.

\bibitem{stewart_rmp2011}
G.~R. Stewart.
\newblock Superconductivity in iron compounds.
\newblock {\em Rev. Mod. Phys.}, 83:1589--1652, 2011.

\bibitem{johnson_adv2010}
David~C. Johnston.
\newblock The puzzle of high temperature superconductivity in layered iron
  pnictides and chalcogenides.
\newblock {\em Advances in Physics}, 59(6):803--1061, 2010.

\bibitem{de_gennes_book}
P.~G. de~Gennes.
\newblock {\em Superconductivity Of $\mathrm{M}$etals $\mathrm{A}$nd
  $\mathrm{A}$lloys, $\mathrm{T}$aylor and $\mathrm{F}$rancis,
  $\mathrm{A}$bingdon, 1999}.

\bibitem{saint_james_book}
G.~Sarma D.~Saint~James and E.~J. Thomas.
\newblock {\em Type II $\mathrm{S}$uperconductivity, $\mathrm{P}$ergamon,
  $\mathrm{N}$ew $\mathrm{Y}$ork, 1969}.

\bibitem{abrikosov1957}
A.~A. Abrikosov.
\newblock On the $\mathrm{M}$agnetic $\mathrm{P}$roperties of
  $\mathrm{S}$uperconductors of the $\mathrm{S}$econd $\mathrm{G}$roup.
\newblock {\em Sov. Phys. JETP}, 5:1174, 1957.

\bibitem{chandrasekhar1962}
B.~S. Chandrasekhar.
\newblock A note on the maximum critical field of high$\ensuremath{-}$field
  superconductors.
\newblock {\em Appl. Phys. Lett.}, 1:7, 1962.

\bibitem{clogston1962}
A.~M. Clogston.
\newblock Upper $\mathrm{L}$imit for the $\mathrm{C}$ritical $\mathrm{F}$ield
  in $\mathrm{H}$ard $\mathrm{S}$uperconductors.
\newblock {\em Phys. Rev. Lett.}, 9:266--267, 1962.

\bibitem{bianchi2003}
A.~Bianchi, R.~Movshovich, C.~Capan, P.~G. Pagliuso, and J.~L. Sarrao.
\newblock Possible
  $\mathrm{F}$ulde-$\mathrm{F}$errell-$\mathrm{L}$arkin-$\mathrm{O}$vchinnikov
  $\mathrm{S}$uperconducting $\mathrm{S}$tate in
  $\mathrm{CeCo}\mathrm{In}_{5}$.
\newblock {\em Phys. Rev. Lett.}, 91:187004, 2003.

\bibitem{lortz2007}
R.~Lortz, Y.~Wang, A.~Demuer, P.~H.~M. B\"ottger, B.~Bergk, G.~Zwicknagl,
  Y.~Nakazawa, and J.~Wosnitza.
\newblock Calorimetric $\mathrm{E}$vidence for a
  $\mathrm{F}$ulde-$\mathrm{F}$errell-$\mathrm{L}$arkin- $\mathrm{O}$vchinnikov
  $\mathrm{S}$uperconducting $\mathrm{S}$tate in the $\mathrm{L}$ayered
  $\mathrm{O}$rganic $\mathrm{S}$uperconductor
  $\ensuremath{\kappa}\mathrm{\text{\ensuremath{-}}}(\mathrm{BEDT}\mathrm{\text{\ensuremath{-}}}\mathrm{TTF}{)}_{2}\mathrm{Cu}(\mathrm{NCS}{)}_{2}$.
\newblock {\em Phys. Rev. Lett.}, 99:187002, 2007.

\bibitem{zocco2013}
D.~A. Zocco, K.~Grube, F.~Eilers, T.~Wolf, and H.~v. L\"ohneysen.
\newblock Pauli-$\mathrm{L}$imited $\mathrm{M}$ultiband
  $\mathrm{S}$uperconductivity in $\mathrm{K}\mathrm{Fe}_{2}\mathrm{As}_{2}$.
\newblock {\em Phys. Rev. Lett.}, 111:057007, 2013.

\bibitem{ketterle_nature2008}
Yong-il Shin, Christian~H. Schunck, Andre Schirotzek, and Wolfgang Ketterle.
\newblock Phase diagram of a two-component $\mathrm{F}$ermi gas with resonant
  interactions.
\newblock {\em Nature}, 451:689, 2008.

\bibitem{ketterson_song_book}
J.~B. Ketterson and S.~N. Song.
\newblock {\em Superconductivity, $\mathrm{C}$ambrigde $\mathrm{U}$niversity
  $\mathrm{P}$ress, $\mathrm{C}$ambridge, 1999.}

\bibitem{ff1964}
Peter Fulde and Richard~A. Ferrell.
\newblock Superconductivity in a $\mathrm{S}$trong
  $\mathrm{S}$pin-$\mathrm{E}$xchange $\mathrm{F}$ield.
\newblock {\em Phys. Rev.}, 135:A550--A563, 1964.

\bibitem{lo1964}
A.~I. Larkin and Yu.~N. Ovchinnikov.
\newblock Nonuniform state of superconductors.
\newblock {\em Sov. Phys. JETP}, 20:762, 1965.

\bibitem{mora2004}
C~Mora and R~Combescot.
\newblock Nature of the
  $\mathrm{F}$ulde-$\mathrm{F}$errell-$\mathrm{L}$arkin-$\mathrm{O}$vchinnikov
  phases at low temperature in 2 dimensions.
\newblock {\em Europhysics Letters ({EPL})}, 66(6):833--839, 2004.

\bibitem{mora2005}
C.~Mora and R.~Combescot.
\newblock Transition to
  $\mathrm{F}$ulde-$\mathrm{F}$errell-$\mathrm{L}$arkin-$\mathrm{O}$vchinnikov
  phases in three dimensions: $\mathrm{A}$ quasiclassical investigation at low
  temperature with $\mathrm{F}$ourier expansion.
\newblock {\em Phys. Rev. B}, 71:214504, 2005.

\bibitem{shimahara1998}
Hiroshi Shimahara.
\newblock Structure of the
  $\mathrm{F}$ulde-$\mathrm{F}$errell-$\mathrm{L}$arkin-$\mathrm{O}$vchinnikov
  $\mathrm{S}$tate in $\mathrm{T}$wo-$\mathrm{D}$imensional
  $\mathrm{S}$uperconductors.
\newblock {\em Journal of the Physical Society of Japan}, 67(3):736--739, 1998.

\bibitem{bowers2002}
Jeffrey~A. Bowers and Krishna Rajagopal.
\newblock Crystallography of color superconductivity.
\newblock {\em Phys. Rev. D}, 66:065002, 2002.

\bibitem{mora2_2005}
Cornelis~J. van~der Beek, Marcin Konczykowski, Luc Fruchter, Ren\'e Brusetti,
  Thierry Klein, Jacques Marcus, and Christophe Marcenat.
\newblock Thermodynamics of the vortex liquid in heavy-ion-irradiated
  superconductors.
\newblock {\em Phys. Rev. B}, 72:214504, 2005.

\bibitem{wang2006}
Qian Wang, H.-Y. Chen, C.-R. Hu, and C.~S. Ting.
\newblock Local $\mathrm{T}$unneling $\mathrm{S}$pectroscopy as a
  $\mathrm{S}$ignature of the
  $\mathrm{F}$ulde-$\mathrm{F}$errell-$\mathrm{L}$arkin-$\mathrm{O}$vchinnikov
  $\mathrm{S}$tate in $s$-and $d$-$\mathrm{d}$wave $\mathrm{S}$uperconductors.
\newblock {\em Phys. Rev. Lett.}, 96:117006, 2006.

\bibitem{combescot2005}
R.~Combescot and G.~Tonini.
\newblock Cascade of two-dimensional
  $\mathrm{F}$ulde-$\mathrm{F}$errell-$\mathrm{L}$arkin-$\mathrm{O}$vchinnikov
  phases with anisotropy.
\newblock {\em Phys. Rev. B}, 72:094513, 2005.

\bibitem{shimahara2007}
Yuji Matsuda and Hiroshi Shimahara.
\newblock Fulde, $\mathrm{F}$errell, $\mathrm{L}$arkin, $\mathrm{O}$vchinnikov
  $\mathrm{S}$tate in $\mathrm{H}$eavy $\mathrm{F}$ermion
  $\mathrm{S}$uperconductors.
\newblock {\em Journal of the Physical Society of Japan}, 76(5):051005, 2007.

\bibitem{aslamazov1969}
L.~G. Aslamazov.
\newblock Influence of $\mathrm{I}$mpurities on the $\mathrm{E}$xistence of an
  $\mathrm{I}$nhomogeneous $\mathrm{S}$tate in a $\mathrm{F}$erromagnetic
  $\mathrm{S}$uperconductor.
\newblock {\em Sov. Phys. JETP}, 28:773, 1969.

\bibitem{takada1970}
Satoshi Takada.
\newblock Superconductivity in a $\mathrm{M}$olecular $\mathrm{F}$ield. ii:
  $\mathrm{S}$tability of $\mathrm{F}$ulde $\ensuremath{-}$$\mathrm{F}$errel
  $\mathrm{P}$hase.
\newblock {\em Prog. Theor. Phys.}, 43:27, 1970.

\bibitem{shimahara1997}
Hiroshi Shimahara and Dierk Rainer.
\newblock Crossover from $\mathrm{V}$ortex $\mathrm{S}$tates to the
  $\mathrm{F}$ulde-$\mathrm{F}$errell- $\mathrm{L}$arkin-$\mathrm{O}$vchinnikov
  $\mathrm{S}$tate in $\mathrm{T}$wo-$\mathrm{D}$imensional $s$-and
  $d$-$\mathrm{W}$ave $\mathrm{S}$uperconductors.
\newblock {\em Journal of the Physical Society of Japan}, 66(11):3591--3599,
  1997.

\bibitem{shimahara2_1998}
Hiroshi Shimahara.
\newblock Phase $\mathrm{F}$luctuations and
  $\mathrm{K}$osterlitz$\ensuremath{-}$$\mathrm{T}$houless
  $\mathrm{T}$ransition in $\mathrm{T}$wo-$\mathrm{D}$imensional
  $\mathrm{F}$ulde-$\mathrm{F}$errell- $\mathrm{L}$arkin-$\mathrm{O}$vchinnikov
  $\mathrm{S}$uperconductors.
\newblock {\em Journal of the Physical Society of Japan}, 67(6):1872--1875,
  1998.

\bibitem{ohashi2002}
Yoji Ohashi.
\newblock On the $\mathrm{F}$ulde-$\mathrm{F}$errell $\mathrm{S}$tate in
  $\mathrm{S}$patially $\mathrm{I}$sotropic $\mathrm{S}$uperconductors.
\newblock {\em Journal of the Physical Society of Japan}, 71(11):2625--2628,
  2002.

\bibitem{sarma1963}
G.~Sarma.
\newblock On the influence of a uniform exchange field acting on the spins of
  the conduction electrons in a superconductor.
\newblock {\em Journal of Physics and Chemistry of Solids}, 24(8):1029 -- 1032,
  1963.

\bibitem{wilczek2003}
W.~Vincent Liu and Frank Wilczek.
\newblock Interior $\mathrm{G}$ap $\mathrm{S}$uperfluidity.
\newblock {\em Phys. Rev. Lett.}, 90:047002, 2003.

\bibitem{gubankova2003}
Elena Gubankova, W.~Vincent Liu, and Frank Wilczek.
\newblock Breached $\mathrm{P}$airing $\mathrm{S}$uperfluidity:
  $\mathrm{P}$ossible $\mathrm{R}$ealization in $\mathrm{QCD}$.
\newblock {\em Phys. Rev. Lett.}, 91:032001, 2003.

\bibitem{muther2002}
H.~M\"uther and A.~Sedrakian.
\newblock Spontaneous $\mathrm{B}$reaking of $\mathrm{R}$otational
  $\mathrm{S}$ymmetry in $\mathrm{S}$uperconductors.
\newblock {\em Phys. Rev. Lett.}, 88:252503, 2002.

\bibitem{sedrakian2005}
Armen Sedrakian, Jordi Mur-Petit, Artur Polls, and Herbert M\"uther.
\newblock Pairing in a two-component ultracold $\mathrm{F}$ermi gas:
  $\mathrm{P}$hases with broken-space symmetries.
\newblock {\em Phys. Rev. A}, 72:013613, 2005.

\bibitem{forbes2005}
Michael~McNeil Forbes, Elena Gubankova, W.~Vincent Liu, and Frank Wilczek.
\newblock Stability $\mathrm{C}$riteria for $\mathrm{B}$reached-$\mathrm{P}$air
  $\mathrm{S}$uperfluidity.
\newblock {\em Phys. Rev. Lett.}, 94:017001, 2005.

\bibitem{liu2004}
W.~Vincent Liu, Frank Wilczek, and Peter Zoller.
\newblock Spin-dependent $\mathrm{H}$ubbard model and a quantum phase
  transition in cold atoms.
\newblock {\em Phys. Rev. A}, 70:033603, 2004.

\bibitem{tayama2002}
T.~Tayama, A.~Harita, T.~Sakakibara, Y.~Haga, H.~Shishido, R.~Settai, and
  Y.~Onuki.
\newblock Unconventional heavy-fermion superconductor
  $\mathrm{CeCo}\mathrm{In}_{5}$: dc magnetization study at temperatures down
  to 50 mk.
\newblock {\em Phys. Rev. B}, 65:180504, 2002.

\bibitem{capan2004}
C.~Capan, A.~Bianchi, R.~Movshovich, A.~D. Christianson, A.~Malinowski, M.~F.
  Hundley, A.~Lacerda, P.~G. Pagliuso, and J.~L. Sarrao.
\newblock Anisotropy of thermal conductivity and possible signature of the
  $\mathrm{F}$ulde-$\mathrm{F}$errell- $\mathrm{L}$arkin-$\mathrm{O}$vchinnikov
  state in $\mathrm{CeCo}\mathrm{In}_{5}$.
\newblock {\em Phys. Rev. B}, 70:134513, 2004.

\bibitem{martin2005}
C.~Martin, C.~C. Agosta, S.~W. Tozer, H.~A. Radovan, E.~C. Palm, T.~P. Murphy,
  and J.~L. Sarrao.
\newblock Evidence for the
  $\mathrm{F}$ulde-$\mathrm{F}$errell-$\mathrm{L}$arkin-$\mathrm{O}$vchinnikov
  state in $\mathrm{CeCo}\mathrm{In}_{5}$ from penetration depth measurements.
\newblock {\em Phys. Rev. B}, 71:020503, 2005.

\bibitem{koutroulakis2008}
G.~Koutroulakis, V.~F. Mitrovic, M.~Horvatic, C.~Berthier, G.~Lapertot, and
  J.~Flouquet.
\newblock Field $\mathrm{D}$ependence of the $\mathrm{G}$round $\mathrm{S}$tate
  in the $\mathrm{E}$xotic $\mathrm{S}$uperconductor
  $\mathrm{CeCo}\mathrm{In}_{5}$: $\mathrm{A}$ $\mathrm{N}$uclear
  $\mathrm{M}$agnetic $\mathrm{R}$esonance $\mathrm{I}$nvestigation.
\newblock {\em Phys. Rev. Lett.}, 101:047004, 2008.

\bibitem{kumagai2006}
K.~Kumagai, M.~Saitoh, T.~Oyaizu, Y.~Furukawa, S.~Takashima, M.~Nohara,
  H.~Takagi, and Y.~Matsuda.
\newblock Fulde-$\mathrm{F}$errell-$\mathrm{L}$arkin-$\mathrm{O}$vchinnikov
  $\mathrm{S}$tate in a $\mathrm{P}$erpendicular $\mathrm{F}$ield of
  $\mathrm{Q}$uasi-$\mathrm{T}$wo-$\mathrm{D}$imensional
  $\mathrm{CeCo}\mathrm{In}_{5}$.
\newblock {\em Phys. Rev. Lett.}, 97:227002, 2006.

\bibitem{gerber2013}
Simon Gerber, Marek Bartkowiak, Jorge~L. Gavilano, Eric Ressouche, Nikola
  Egetenmeyer, Christof Niedermayer, Andrea~D. Bianchi, Roman Movshovich,
  Eric~D. Bauer, Joe~D. Thompson, and Michel Kenzelmann.
\newblock Switching of magnetic domains reveals spatially inhomogeneous
  superconductivity.
\newblock {\em Nature Physics}, 10:126, 2013.

\bibitem{kim2016}
Duk~Y. Kim, Shi-Zeng Lin, Franziska Weickert, Michel Kenzelmann, Eric~D. Bauer,
  Filip Ronning, J.~D. Thompson, and Roman Movshovich.
\newblock Intertwined $\mathrm{O}$rders in $\mathrm{H}$eavy-$\mathrm{F}$ermion
  $\mathrm{S}$uperconductor $\mathrm{CeCo}\mathrm{In}_{5}$.
\newblock {\em Phys. Rev. X}, 6:041059, 2016.

\bibitem{lin2020}
Shi-Zeng Lin, Duk~Y. Kim, Eric~D. Bauer, Filip Ronning, J.~D. Thompson, and
  Roman Movshovich.
\newblock Interplay of the $\mathrm{S}$pin $\mathrm{D}$ensity $\mathrm{W}$ave
  and a $\mathrm{P}$ossible
  $\mathrm{F}$ulde-$\mathrm{F}$errell-$\mathrm{L}$arkin-$\mathrm{O}$vchinnikov
  $\mathrm{S}$tate in $\mathrm{CeCo}\mathrm{In}_{5}$ in $\mathrm{R}$otating
  $\mathrm{M}$agnetic $\mathrm{F}$ield.
\newblock {\em Phys. Rev. Lett.}, 124:217001, 2020.

\bibitem{wright2011}
J.~A. Wright, E.~Green, P.~Kuhns, A.~Reyes, J.~Brooks, J.~Schlueter, R.~Kato,
  H.~Yamamoto, M.~Kobayashi, and S.~E. Brown.
\newblock Zeeman-$\mathrm{D}$riven $\mathrm{P}$hase $\mathrm{T}$ransition
  within the $\mathrm{S}$uperconducting $\mathrm{S}$tate of
  $\ensuremath{\kappa}\mathrm{\text{\ensuremath{-}}}(\mathrm{BEDT}\mathrm{\text{\ensuremath{-}}}\mathrm{TTF}{)}_{2}\mathrm{Cu}(\mathrm{NCS}{)}_{2}$.
\newblock {\em Phys. Rev. Lett.}, 107:087002, 2011.

\bibitem{bergk2011}
B.~Bergk, A.~Demuer, I.~Sheikin, Y.~Wang, J.~Wosnitza, Y.~Nakazawa, and
  R.~Lortz.
\newblock Magnetic torque evidence for the
  $\mathrm{F}$ulde-$\mathrm{F}$errell-$\mathrm{L}$arkin- $\mathrm{O}$vchinnikov
  state in the layered organic superconductor
  $\ensuremath{\kappa}\ensuremath{-}(\mathrm{BEDT}\ensuremath{-}\mathrm{TTF}){}_{2}\mathrm{Cu}(\mathrm{NCS}){}_{2}$.
\newblock {\em Phys. Rev. B}, 83:064506, 2011.

\bibitem{agosta2012}
C.~C. Agosta, Jing Jin, W.~A. Coniglio, B.~E. Smith, K.~Cho, I.~Stroe,
  C.~Martin, S.~W. Tozer, T.~P. Murphy, E.~C. Palm, J.~A. Schlueter, and
  M.~Kurmoo.
\newblock Experimental and semiempirical method to determine the
  $\mathrm{P}$auli-limiting field in quasi-two-dimensional superconductors as
  applied to
  $\ensuremath{\kappa}\ensuremath{-}(\mathrm{BEDT}\ensuremath{-}\mathrm{TTF}){}_{2}\mathrm{Cu}(\mathrm{NCS}){}_{2}$:
  $\mathrm{S}$trong evidence of a $\mathrm{FFLO}$ state.
\newblock {\em Phys. Rev. B}, 85:214514, 2012.

\bibitem{mayaffre2014}
H.~Mayaffre, S.~Kramer, M.~HorvatiÄ, C.~Berthier, K.~Miyagawa, K.~Kanoda, and
  V.~F. Mitrovic.
\newblock Evidence of andreev bound states as a hallmark of the fflo phase in
  $\ensuremath{\kappa}\ensuremath{-}(\mathrm{BEDT}\ensuremath{-}\mathrm{TTF}){}_{2}\mathrm{Cu}(\mathrm{NCS}){}_{2}$.
\newblock {\em Nature Physics}, 10:928, 2014.

\bibitem{cho2011}
K.~Cho, H.~Kim, M.~A. Tanatar, Y.~J. Song, Y.~S. Kwon, W.~A. Coniglio, C.~C.
  Agosta, A.~Gurevich, and R.~Prozorov.
\newblock Anisotropic upper critical field and possible
  $\mathrm{F}$ulde-$\mathrm{F}$errell-$\mathrm{L}$arkin- $\mathrm{O}$vchinnikov
  state in the stoichiometric pnictide superconductor $\mathrm{LiFeAs}$.
\newblock {\em Phys. Rev. B}, 83:060502, 2011.

\bibitem{khim2011}
Seunghyun Khim, Bumsung Lee, Jae~Wook Kim, Eun~Sang Choi, G.~R. Stewart, and
  Kee~Hoon Kim.
\newblock Pauli-limiting effects in the upper critical fields of a clean
  $\mathrm{LiFeAs}$ single crystal.
\newblock {\em Phys. Rev. B}, 84:104502, 2011.

\bibitem{gaebler2010}
J.~P. Gaebler, J.~T. Stewart, T.~E. Drake, D.~S. Jin, A.~Perali, P.~Pieri, and
  G.~C. Strinati.
\newblock Observation of pseudogap behaviour in a strongly interacting fermi
  gas.
\newblock {\em Nat Phys}, 6:569, 2010.

\bibitem{chin2004}
C.~Chin, M.~Bartenstein, A.~Altmeyer, S.~Riedl, S.~Jochim, J.~Hecker Denschlag,
  and R.~Grimm.
\newblock Observation of the $\mathrm{P}$airing $\mathrm{G}$ap in a
  $\mathrm{S}$trongly $\mathrm{I}$nteracting $\mathrm{F}$ermi $\mathrm{G}$as.
\newblock {\em Science}, 305(5687):1128--1130, 2004.

\bibitem{sewer2002}
Alain Sewer, Xenophon Zotos, and Hans Beck.
\newblock Quantum $\mathrm{M}$onte $\mathrm{C}$arlo study of the
  three-dimensional attractive $\mathrm{H}$ubbard model.
\newblock {\em Phys. Rev. B}, 66:140504, 2002.

\bibitem{paiva2010}
Thereza Paiva, Richard Scalettar, Mohit Randeria, and Nandini Trivedi.
\newblock Fermions in 2$\mathrm{D}$ $\mathrm{O}$ptical $\mathrm{L}$attices:
  $\mathrm{T}$emperature and $\mathrm{E}$ntropy $\mathrm{S}$cales for
  $\mathrm{O}$bserving $\mathrm{A}$ntiferromagnetism and
  $\mathrm{S}$uperfluidity.
\newblock {\em Phys. Rev. Lett.}, 104:066406, 2010.

\bibitem{ries2015}
M.~G. Ries, A.~N. Wenz, G.~Z\"urn, L.~Bayha, I.~Boettcher, D.~Kedar, P.~A.
  Murthy, M.~Neidig, T.~Lompe, and S.~Jochim.
\newblock Observation of $\mathrm{P}$air $\mathrm{C}$ondensation in the
  $\mathrm{Q}$uasi-2$\mathrm{D}$ $\mathrm{BEC}\ensuremath{-}\mathrm{BCS}$
  $\mathrm{C}$rossover.
\newblock {\em Phys. Rev. Lett.}, 114:230401, 2015.

\bibitem{ketterle_science2007}
C.~H. Schunck, Y.~Shin, A.~Schirotzek, M.~W. Zwierlein, and W.~Ketterle.
\newblock Pairing $\mathrm{W}$ithout $\mathrm{S}$uperfluidity: $\mathrm{T}$he
  $\mathrm{G}$round $\mathrm{S}$tate of an $\mathrm{I}$mbalanced
  $\mathrm{F}$ermi $\mathrm{M}$ixture.
\newblock {\em Science}, 316:867--870, 2007.

\bibitem{shin2006}
Y.~Shin, M.~W. Zwierlein, C.~H. Schunck, A.~Schirotzek, and W.~Ketterle.
\newblock Observation of $\mathrm{P}$hase $\mathrm{S}$eparation in a
  $\mathrm{S}$trongly $\mathrm{I}$nteracting $\mathrm{I}$mbalanced
  $\mathrm{F}$ermi $\mathrm{G}$as.
\newblock {\em Phys. Rev. Lett.}, 97:030401, 2006.

\bibitem{liao2010}
Yean-an Liao, Ann Sophie~C. Rittner, Tobias Paprotta, Wenhui Li, Guthrie~B.
  Partridge, Randall~G. Hulet, Stefan~K. Baur, and Erich~J. Mueller.
\newblock Spin-imbalance in a one-dimensional $\mathrm{F}$ermi gas.
\newblock {\em Nature}, 467:567, 2010.

\bibitem{tagleiber2008}
M.~Taglieber, A.-C. Voigt, T.~Aoki, T.~W. H\"ansch, and K.~Dieckmann.
\newblock Quantum $\mathrm{D}$egenerate $\mathrm{T}$wo-$\mathrm{S}$pecies
  $\mathrm{F}$ermi-$\mathrm{F}$ermi $\mathrm{M}$ixture $\mathrm{C}$oexisting
  with a $\mathrm{B}$ose-$\mathrm{E}$instein $\mathrm{C}$ondensate.
\newblock {\em Phys. Rev. Lett.}, 100:010401, 2008.

\bibitem{naik2011}
D.~Naik, A.~Trenkwalder, C.~Kohstall, F.~M. Spiegelhalder, M.~Zaccanti,
  G.~Hendl, F.~Schreck, R.~Grimm, T.~M. Hanna, and P.~S. Julienne.
\newblock Feshbach resonances in the
  $^{6}\mathrm{Li}$$\ensuremath{-}$$^{40}\mathrm{K}$ $\mathrm{F}$ermi-
  $\mathrm{F}$ermi mixture: elastic versus inelastic interactions.
\newblock {\em The European Physical Journal D}, 65(1):55--65, 2011.

\bibitem{wille2008}
E.~Wille, F.~M. Spiegelhalder, G.~Kerner, D.~Naik, A.~Trenkwalder, G.~Hendl,
  F.~Schreck, R.~Grimm, T.~G. Tiecke, J.~T.~M. Walraven, S.~J. J. M.~F.
  Kokkelmans, E.~Tiesinga, and P.~S. Julienne.
\newblock Exploring an $\mathrm{U}$ltracold $\mathrm{F}$ermi-$\mathrm{F}$ermi
  $\mathrm{M}$ixture: $\mathrm{I}$nterspecies $\mathrm{F}$eshbach
  $\mathrm{R}$esonances and $\mathrm{S}$cattering $\mathrm{P}$roperties of
  $^{6}\mathrm{Li}$ and $^{40}\mathrm{K}$.
\newblock {\em Phys. Rev. Lett.}, 100:053201, 2008.

\bibitem{costa2010}
L.~Costa, J.~Brachmann, A.-C. Voigt, C.~Hahn, M.~Taglieber, T.~W. H\"ansch, and
  K.~Dieckmann.
\newblock $s$-$\mathrm{W}$ave $\mathrm{I}$nteraction in a
  $\mathrm{T}$wo-$\mathrm{S}$pecies $\mathrm{F}$ermi- $\mathrm{F}$ermi
  $\mathrm{M}$ixture at a $\mathrm{N}$arrow $\mathrm{F}$eshbach
  $\mathrm{R}$esonance.
\newblock {\em Phys. Rev. Lett.}, 105:123201, 2010.

\bibitem{voigt2009}
A.-C. Voigt, M.~Taglieber, L.~Costa, T.~Aoki, W.~Wieser, T.~W. H\"ansch, and
  K.~Dieckmann.
\newblock Ultracold $\mathrm{H}$eteronuclear $\mathrm{F}$ermi-$\mathrm{F}$ermi
  $\mathrm{M}$olecules.
\newblock {\em Phys. Rev. Lett.}, 102:020405, 2009.

\bibitem{lu2012}
Mingwu Lu, Nathaniel~Q. Burdick, and Benjamin~L. Lev.
\newblock Quantum $\mathrm{D}$egenerate $\mathrm{D}$ipolar $\mathrm{F}$ermi
  $\mathrm{G}$as.
\newblock {\em Phys. Rev. Lett.}, 108:215301, 2012.

\bibitem{frisch2013}
Albert Frisch, Kiyotaka Aikawa, Michael Mark, Francesca Ferlaino, Ekaterina
  Berseneva, and Svetlana Kotochigova.
\newblock Hyperfine structure of laser-cooling transitions in fermionic
  $\mathrm{E}$rbium-167.
\newblock {\em Phys. Rev. A}, 88:032508, 2013.

\bibitem{spiegelhalder2010}
F.~M. Spiegelhalder, A.~Trenkwalder, D.~Naik, G.~Kerner, E.~Wille, G.~Hendl,
  F.~Schreck, and R.~Grimm.
\newblock All-optical production of a degenerate mixture of $^{6}\mathrm{Li}$
  and $^{40}\mathrm{K}$ and creation of heteronuclear molecules.
\newblock {\em Phys. Rev. A}, 81:043637, 2010.

\bibitem{pieri2011}
P.~Pieri, A.~Perali, G.~C. Strinati, S.~Riedl, M.~J. Wright, A.~Altmeyer,
  C.~Kohstall, E.~R. S\'anchez~Guajardo, J.~Hecker~Denschlag, and R.~Grimm.
\newblock Pairing-gap, pseudogap, and no-gap phases in the radio-frequency
  spectra of a trapped unitary $^{6}\mathrm{Li}$ gas.
\newblock {\em Phys. Rev. A}, 84:011608, 2011.

\bibitem{nascimbne2010}
S.~Nascimbasne, N.~Navon, K.~J. Jiang, F.~Chevy, and C.~Salomon.
\newblock Exploring the thermodynamics of a universal $\mathrm{F}$ermi gas.
\newblock {\em Nature}, 463:1057--1060, 2010.

\bibitem{radzhiovsky_pra}
Leo Radzihovsky.
\newblock Fluctuations and phase transitions in
  $\mathrm{L}$arkin$\ensuremath{-}$$\mathrm{O}$vchinnikov
  liquid$\ensuremath{-}$crystal states of a
  population$\ensuremath{-}$imbalanced resonant $\mathrm{F}$ermi gas.
\newblock {\em Phys. Rev. A}, 84:023611, 2011.

\bibitem{gurarie2007}
V.~Gurarie and L.~Radzihovsky.
\newblock Resonantly paired fermionic superfluids.
\newblock {\em Annals of Physics}, 322(1):2 -- 119, 2007.
\newblock January Special Issue 2007.

\bibitem{casellbouni_rmp}
Roberto Casalbuoni and Giuseppe Nardulli.
\newblock Inhomogeneous superconductivity in condensed matter and
  $\mathrm{QCD}$.
\newblock {\em Rev. Mod. Phys.}, 76:263--320, 2004.

\bibitem{bloch2008}
Immanuel Bloch, Jean Dalibard, and Wilhelm Zwerger.
\newblock Many-body physics with ultracold gases.
\newblock {\em Rev. Mod. Phys.}, 80:885--964, 2008.

\bibitem{kohn1959}
W.~Kohn.
\newblock Analytic $\mathrm{P}$roperties of $\mathrm{B}$loch $\mathrm{W}$aves
  and $\mathrm{W}$annier $\mathrm{F}$unctions.
\newblock {\em Phys. Rev.}, 115:809--821, 1959.

\bibitem{jaksh1998}
D.~Jaksch, C.~Bruder, J.~I. Cirac, C.~W. Gardiner, and P.~Zoller.
\newblock Cold bosonic atoms in optical lattices.
\newblock {\em Phys. Rev. Lett.}, 81:3108--3111, 1998.

\bibitem{mahan_book}
Gerald~D. Mahan.
\newblock {\em Many particle physics, $\mathrm{S}$pringer-$\mathrm{V}$erlag
  $\mathrm{N}$ew $\mathrm{Y}$ork $\mathrm{I}$nc., $\mathrm{N}$ew york, 1990}.

\bibitem{fetter_book}
John Dirk~Walecka Alexander L.~Fetter.
\newblock {\em Quantum $\mathrm{T}$heory of $\mathrm{M}$any-$\mathrm{P}$article
  $\mathrm{S}$ystems, $\mathrm{M}$c$\mathrm{G}$raw-$\mathrm{H}$ill,
  $\mathrm{S}$an $\mathrm{F}$ransisco, 1971.}

\bibitem{yoshida2007}
Nobukatsu Yoshida and S.-K. Yip.
\newblock Larkin-$\mathrm{O}$vchinnikov state in resonant $\mathrm{F}$ermi gas.
\newblock {\em Phys. Rev. A}, 75:063601, 2007.

\bibitem{batrouni2008}
G.~G. Batrouni, M.~H. Huntley, V.~G. Rousseau, and R.~T. Scalettar.
\newblock Exact $\mathrm{N}$umerical $\mathrm{S}$tudy of $\mathrm{P}$air
  $\mathrm{F}$ormation with $\mathrm{I}$mbalanced $\mathrm{F}$ermion
  $\mathrm{P}$opulations.
\newblock {\em Phys. Rev. Lett.}, 100:116405, 2008.

\bibitem{baarsma_jmodopt}
J.~E. Baarsma and P.~T\"orm\"a.
\newblock Larkin-$\mathrm{O}$vchinnikov phases in two-dimensional square
  lattices.
\newblock {\em Journal of Modern Optics}, 63(18):1795--1804, 2016.

\bibitem{zwerger_book}
Orso~G Feiguin A~E, Heidrich-Meisner~F and Zwerger W.
\newblock {\em $\mathrm{BCS}$$\ensuremath{–}$$\mathrm{BEC}$
  $\mathrm{C}$rossover and $\mathrm{U}$nconventional $\mathrm{S}$uperfluid
  $\mathrm{O}$rder in $\mathrm{O}$ne $\mathrm{D}$imension ($\mathrm{B}$erlin:
  $\mathrm{S}$pringer), 2012.}

\bibitem{parish_book}
M.~M. Parish.
\newblock {\em The $\mathrm{BCS}$$\ensuremath{-}$$\mathrm{BEC}$ crossover
  Quantum Gas Experiments $\mathrm{C}$h 9 (London: Imperial College Press),
  2014.}

\bibitem{randeria_taylor}
Mohit Randeria and Edward Taylor.
\newblock Crossover from
  $\mathrm{B}$ardeen-$\mathrm{C}$ooper-$\mathrm{S}$chrieffer to
  $\mathrm{B}$ose- $\mathrm{E}$instein $\mathrm{C}$ondensation and the
  $\mathrm{U}$nitary $\mathrm{F}$ermi $\mathrm{G}$as.
\newblock {\em Annual Review of Condensed Matter Physics}, 5(1):209--232, 2014.

\bibitem{lewenstein_book}
A.~Sanpera M.~Lewenstein and V.~Ahufinger.
\newblock {\em Ultracold Atoms in Optical Lattices: Simulating Quantum
  Many-body Systems (Oxford: Oxford University Press), 2012.}

\bibitem{leggett}
A.~J. Leggett.
\newblock {\em Modern Trends in the Theory of Condensed Matter,
  Springer-Verlag, Berlin, 1979}.

\bibitem{eagles1969}
D.~M. Eagles.
\newblock Possible $\mathrm{P}$airing without $\mathrm{S}$uperconductivity at
  $\mathrm{L}$ow $\mathrm{C}$arrier $\mathrm{C}$oncentrations in
  $\mathrm{B}$ulk and $\mathrm{T}$hin-$\mathrm{F}$ilm
  $\mathrm{S}$uperconducting $\mathrm{S}$emiconductors.
\newblock {\em Phys. Rev.}, 186:456--463, 1969.

\bibitem{rohe2001}
Daniel Rohe and Walter Metzner.
\newblock Pair-fluctuation-induced pseudogap in the normal phase of the
  two-dimensional attractive hubbard model at weak coupling.
\newblock {\em Phys. Rev. B}, 63:224509, 2001.

\bibitem{micnas_rmp1990}
R.~Micnas, J.~Ranninger, and S.~Robaszkiewicz.
\newblock Superconductivity in narrow-band systems with local nonretarded
  attractive interactions.
\newblock {\em Rev. Mod. Phys.}, 62:113--171, 1990.

\bibitem{dupuis2004}
N.~Dupuis.
\newblock Berezinskii-$\mathrm{K}$osterlitz-$\mathrm{T}$houless transition and
  $\mathrm{BCS}$-$\mathrm{B}$ose crossover in the two-dimensional attractive
  $\mathrm{H}$ubbard model.
\newblock {\em Phys. Rev. B}, 70:134502, 2004.

\bibitem{dupuis2_2004}
K.~Borejsza and N.~Dupuis.
\newblock Antiferromagnetism and single-particle properties in the
  two-dimensional half-filled $\mathrm{H}$ubbard model: $\mathrm{A}$ nonlinear
  sigma model approach.
\newblock {\em Phys. Rev. B}, 69:085119, 2004.

\bibitem{rink}
P.~Nozi{\`e}res and S.~Schmitt-Rink.
\newblock Bose condensation in an attractive fermion gas: $\mathrm{F}$rom weak
  to strong coupling superconductivity.
\newblock {\em Journal of Low Temperature Physics}, 59(3):195--211, 1985.

\bibitem{deisz2002}
J.~J. Deisz, D.~W. Hess, and J.~W. Serene.
\newblock Phase diagram for the attractive hubbard model in two dimensions in a
  conserving approximation.
\newblock {\em Phys. Rev. B}, 66:014539, 2002.

\bibitem{tamaki2008}
H.~Tamaki, Y.~Ohashi, and K.~Miyake.
\newblock $\mathrm{BCS}$$\ensuremath{-}$$\mathrm{BEC}$ crossover and effects of
  density fluctuations in a two-component $\mathrm{F}$ermi gas described by the
  three-dimensional attractive $\mathrm{H}$ubbard model.
\newblock {\em Phys. Rev. A}, 77:063616, 2008.

\bibitem{kyung2001}
B.~Kyung, S.~Allen, and A.-M.~S. Tremblay.
\newblock Pairing fluctuations and pseudogaps in the attractive
  $\mathrm{H}$ubbard model.
\newblock {\em Phys. Rev. B}, 64:075116, 2001.

\bibitem{kopec2002}
T.~K. Kope\ifmmode~\acute{c}\else \'{c}\fi{}.
\newblock Superconducting phase coherence and pairing gap in the
  three-dimensional attractive $\mathrm{H}$ubbard model.
\newblock {\em Phys. Rev. B}, 65:054509, 2002.

\bibitem{scalettar1998}
R.~T. Scalettar, E.~Y. Loh, J.~E. Gubernatis, A.~Moreo, S.~R. White, D.~J.
  Scalapino, R.~L. Sugar, and E.~Dagotto.
\newblock Phase diagram of the two-dimensional negative-${U}$
  $\mathrm{H}$ubbard model.
\newblock {\em Phys. Rev. Lett.}, 62:1407--1410, 1989.

\bibitem{moreo1991}
A.~Moreo and D.~J. Scalapino.
\newblock Two-dimensional negative-${U}$ $\mathrm{H}$ubbard model.
\newblock {\em Phys. Rev. Lett.}, 66:946--948, 1991.

\bibitem{moreo1992}
Adriana Moreo, Douglas~J. Scalapino, and Steven~R. White.
\newblock Quasiparticle gap in a two-dimensional
  $\mathrm{K}$osterlitz-$\mathrm{T}$houless superconductor.
\newblock {\em Phys. Rev. B}, 45:7544--7546, 1992.

\bibitem{randeria1992}
Mohit Randeria, Nandini Trivedi, Adriana Moreo, and Richard~T. Scalettar.
\newblock Pairing and spin gap in the normal state of short coherence length
  superconductors.
\newblock {\em Phys. Rev. Lett.}, 69:2001--2004, 1992.

\bibitem{trivedi1995}
Nandini Trivedi and Mohit Randeria.
\newblock Deviations from $\mathrm{F}$ermi-$\mathrm{L}$iquid
  $\mathrm{B}$ehavior above ${T}_{c}$ in 2$\mathrm{D}$ $\mathrm{S}$hort
  $\mathrm{C}$oherence $\mathrm{L}$ength $\mathrm{S}$uperconductors.
\newblock {\em Phys. Rev. Lett.}, 75:312--315, 1995.

\bibitem{allen1999}
S.~Allen, H.~Touchette, S.~Moukouri, Y.~M. Vilk, and A.-M.~S. Tremblay.
\newblock Role of $\mathrm{S}$ymmetry and $\mathrm{D}$imension in
  $\mathrm{P}$seudogap $\mathrm{P}$henomena.
\newblock {\em Phys. Rev. Lett.}, 83:4128--4131, 1999.

\bibitem{singer1999}
{Singer, J. M.}, {Schneider, T.}, and {Meier, P. F.}
\newblock Spectral properties of the attractive $\mathrm{H}$ubbard model.
\newblock {\em Eur. Phys. J. B}, 7(1):37--51, 1999.

\bibitem{paiva2004}
Thereza Paiva, Raimundo~R. dos Santos, R.~T. Scalettar, and P.~J.~H. Denteneer.
\newblock Critical temperature for the two-dimensional attractive
  $\mathrm{H}$ubbard model.
\newblock {\em Phys. Rev. B}, 69:184501, 2004.

\bibitem{keller2001}
M.~Keller, W.~Metzner, and U.~Schollw\"ock.
\newblock Dynamical $\mathrm{M}$ean-$\mathrm{F}$ield $\mathrm{T}$heory for
  $\mathrm{P}$airing and $\mathrm{S}$pin $\mathrm{G}$ap in the
  $\mathrm{A}$ttractive $\mathrm{H}$ubbard $\mathrm{M}$odel.
\newblock {\em Phys. Rev. Lett.}, 86:4612--4615, 2001.

\bibitem{capone2002}
M.~Capone, C.~Castellani, and M.~Grilli.
\newblock First-$\mathrm{O}$rder $\mathrm{P}$airing $\mathrm{T}$ransition and
  $\mathrm{S}$ingle-$\mathrm{P}$article $\mathrm{S}$pectral $\mathrm{F}$unction
  in the $\mathrm{A}$ttractive $\mathrm{H}$ubbard $\mathrm{M}$odel.
\newblock {\em Phys. Rev. Lett.}, 88:126403, 2002.

\bibitem{garg2005}
Arti Garg, H.~R. Krishnamurthy, and Mohit Randeria.
\newblock $\mathrm{BCS}$$\ensuremath{-}$$\mathrm{BEC}$ crossover at
  $\mathrm{T}=0$: $\mathrm{A}$ dynamical mean-field theory approach.
\newblock {\em Phys. Rev. B}, 72:024517, 2005.

\bibitem{bauer2009}
J.~Bauer and A.~C. Hewson.
\newblock Quasiparticle excitations and dynamic susceptibilities in the
  {BCS}-{BEC} crossover.
\newblock {\em Euro Phys. Lett.}, 85(2):27001, 2009.

\bibitem{bauer2_2009}
J.~Bauer, A.~C. Hewson, and N.~Dupuis.
\newblock Dynamical mean-field theory and numerical renormalization group study
  of superconductivity in the attractive $\mathrm{H}$ubbard model.
\newblock {\em Phys. Rev. B}, 79:214518, 2009.

\bibitem{koga2011}
Akihisa Koga and Philipp Werner.
\newblock Low-temperature properties of the infinite-dimensional attractive
  $\mathrm{H}$ubbard model.
\newblock {\em Phys. Rev. A}, 84:023638, 2011.

\bibitem{tarat_epjb}
S.~Tarat and P.~Majumdar.
\newblock A real space auxiliary field approach to the
  $\mathrm{BCS}$$\ensuremath{-}$$\mathrm{BEC}$ crossover.
\newblock {\em Eur. Phys. Jour. B}, 88:68, 2015.

\bibitem{sumiyama2008}
A.~Sumiyama, R.~Onuki, Y.~Oda, H.~Shishido, R.~Settai, and Y.~Ånuki.
\newblock Point-contact study of the heavy-fermion superconductor
  $\mathrm{CeCo}$$\mathrm{In}_{5}$.
\newblock {\em Journal of Physics and Chemistry of Solids}, 69(12):3018 --
  3021, 2008.

\bibitem{fasano_physicab}
Y.~Fasano, P.~Szab{\"A}, J.~Ka{\"A}mar{\"A}k, Z.~Pribulov{\"A}, P.~Pedrazzini,
  P.~Samuely, and V.F. Correa.
\newblock Unconventional superconductivity in the strong-coupling limit for the
  heavy fermion system $\mathrm{CeCo}\mathrm{In}_{5}$.
\newblock {\em Physica B: Condensed Matter}, 536:798, 2018.

\bibitem{wosnitza_crystals}
Jochen Wosnitza.
\newblock Superconductivity in $\mathrm{L}$ayered $\mathrm{O}$rganic
  $\mathrm{M}$etals.
\newblock {\em Crystals}, 2(2):248--265, 2012.

\bibitem{abdel2012}
M.~Abdel-Hafiez, S.~Aswartham, S.~Wurmehl, V.~Grinenko, C.~Hess, S.-L.
  Drechsler, S.~Johnston, A.~U.~B. Wolter, B.~B\"uchner, H.~Rosner, and
  L.~Boeri.
\newblock Specific heat and upper critical fields in
  $\mathrm{K}$$\mathrm{Fe}_{2}$$\mathrm{As}_{2}$ single crystals.
\newblock {\em Phys. Rev. B}, 85:134533, 2012.

\bibitem{hs}
J.~Hubbard.
\newblock Calculation of $\mathrm{P}$artition $\mathrm{F}$unctions.
\newblock {\em Phys. Rev. Lett.}, 3:77--78, 1959.

\bibitem{hs1}
H.~J. Schulz.
\newblock Effective action for strongly correlated fermions from functional
  integrals.
\newblock {\em Phys. Rev. Lett.}, 65:2462--2465, 1990.

\bibitem{wang1969}
S.~Q. Wang, W.~E. Evenson, and J.~R. Schrieffer.
\newblock Theory of $\mathrm{I}$tinerant $\mathrm{F}$erromagnets
  $\mathrm{E}$xhibiting $\mathrm{L}$ocalized- $\mathrm{M}$oment
  $\mathrm{B}$ehavior above the $\mathrm{C}$urie $\mathrm{P}$oint.
\newblock {\em Phys. Rev. Lett.}, 23:92--95, 1969.

\bibitem{evenson1970}
W.~E. Evenson, J.~R. Schrieffer, and S.~Q. Wang.
\newblock New $\mathrm{A}$pproach to the $\mathrm{T}$heory of
  $\mathrm{I}$tinerant $\mathrm{E}$lectron $\mathrm{F}$erromagnets with
  $\mathrm{L}$ocal$\ensuremath{-}$$\mathrm{M}$oment $\mathrm{C}$haracteristics.
\newblock {\em Journal of Applied Physics}, 41(3):1199--1204, 1970.

\bibitem{dubi2007}
Yonatan Dubi, Yigal Meir, and Yshai Avishai.
\newblock Nature of the superconductor–insulator transition in disordered
  superconductors.
\newblock {\em Nature}, 449:876--880, 2007.

\bibitem{sanjeev_pinaki}
S.~Kumar and P.~Majumdar.
\newblock A travelling cluster approximation for lattice fermions strongly
  coupled to classical degrees of freedom.
\newblock {\em Eur. Phys. J. B}, 50:571--579, 2006.

\bibitem{mpk2016}
Madhuparna Karmakar and Pinaki Majumdar.
\newblock Population-imbalanced lattice fermions near the
  $\mathrm{BCS}$$\ensuremath{-}$$\mathrm{BEC}$ crossover: $\mathrm{T}$hermal
  physics of the breached pair and $\mathrm{F}$ulde-$\mathrm{F}$errell-
  $\mathrm{L}$arkin-$\mathrm{O}$vchinnikov phases.
\newblock {\em Phys. Rev. A}, 93:053609, 2016.

\bibitem{mpk2018}
Madhuparna Karmakar.
\newblock Thermal transitions, pseudogap behavior, and
  $\mathrm{BCS}$$\ensuremath{-}$$\mathrm{BEC}$ crossover in
  $\mathrm{F}$ermi-$\mathrm{F}$ermi mixtures.
\newblock {\em Phys. Rev. A}, 97:033617, 2018.

\bibitem{mpk_epjd}
Madhuparna Karmakar and Pinaki Majumdar.
\newblock Anomalous pseudogap in population imbalanced $\mathrm{F}$ermi
  superfluids.
\newblock {\em The European Physical Journal D}, 70(10):220, Oct 2016.

\bibitem{mpk_jpcm2020}
Madhuparna Karmakar.
\newblock Pauli limited d-wave superconductors: quantum breached pair phase and
  thermal transitions.
\newblock {\em Journal of Physics: Condensed Matter}, 32:405604, 2020.

\bibitem{kim2011}
D.-H. Kim, J.~J. Kinnunen, J.-P. Martikainen, and P.~T\"orm\"a.
\newblock Exotic $\mathrm{S}$uperfluid $\mathrm{S}$tates of $\mathrm{L}$attice
  $\mathrm{F}$ermions in $\mathrm{E}$longated $\mathrm{T}$raps.
\newblock {\em Phys. Rev. Lett.}, 106:095301, 2011.

\bibitem{heikkinen2013}
Miikka O.~J. Heikkinen, Dong-Hee Kim, and P\"aivi T\"orm\"a.
\newblock Finite-temperature stability and dimensional crossover of exotic
  superfluidity in lattices.
\newblock {\em Phys. Rev. B}, 87:224513, 2013.

\bibitem{heikkinen2014}
M.~O.~J. Heikkinen, D.-H. Kim, M.~Troyer, and P.~T\"orm\"a.
\newblock Nonlocal $\mathrm{Q}$uantum $\mathrm{F}$luctuations and
  $\mathrm{F}$ermionic $\mathrm{S}$uperfluidity in the $\mathrm{I}$mbalanced
  $\mathrm{A}$ttractive $\mathrm{H}$ubbard $\mathrm{M}$odel.
\newblock {\em Phys. Rev. Lett.}, 113:185301, 2014.

\bibitem{gukelberger2016}
Jan Gukelberger, Sebastian Lienert, Evgeny Kozik, Lode Pollet, and Matthias
  Troyer.
\newblock Fulde-$\mathrm{F}$errell-$\mathrm{L}$arkin-$\mathrm{O}$vchinnikov
  pairing as leading instability on the square lattice.
\newblock {\em Phys. Rev. B}, 94:075157, 2016.

\bibitem{wolak2012}
M.~J. Wolak, B.~Gr\'emaud, R.~T. Scalettar, and G.~G. Batrouni.
\newblock Pairing in a two-dimensional $\mathrm{F}$ermi gas with population
  imbalance.
\newblock {\em Phys. Rev. A}, 86:023630, 2012.

\bibitem{bkt1}
V.L. Berezinsky.
\newblock {Destruction of long range order in one-dimensional and
  two-dimensional systems having a continuous symmetry group. I.
  $\mathrm{C}$lassical systems}.
\newblock {\em Sov. Phys. JETP}, 32:493--500, 1971.

\bibitem{bkt2}
V.~L. Berezinsky.
\newblock {Destruction of long-range order in one-dimensional and
  two-dimensional systems possessing a continuous symmetry group. II.
  $\mathrm{Q}$uantum systems}.
\newblock {\em Soviet Journal of Experimental and Theoretical Physics}, 34:610,
  1972.

\bibitem{bkt3}
J~M Kosterlitz and D~J Thouless.
\newblock Ordering, metastability and phase transitions in two-dimensional
  systems.
\newblock {\em Journal of Physics C: Solid State Physics}, 6:1181--1203, 1973.

\bibitem{koponen_njp2009}
T~K Koponen, T~Paananen, J-P Martikainen, M~R Bakhtiari, and P~T\"orm\"a.
\newblock {FFLO} state in 1-, 2- and 3-dimensional optical lattices combined
  with a non-uniform background potential.
\newblock {\em New Journal of Physics}, 10:045014, 2008.

\bibitem{koponen2007}
T.~K. Koponen, T.~Paananen, J.-P. Martikainen, and P.~T\"orm\"a.
\newblock Finite-$\mathrm{T}$emperature $\mathrm{P}$hase $\mathrm{D}$iagram of
  a $\mathrm{P}$olarized $\mathrm{F}$ermi $\mathrm{G}$as in an
  $\mathrm{O}$ptical $\mathrm{L}$attice.
\newblock {\em Phys. Rev. Lett.}, 99:120403, 2007.

\bibitem{loh2010}
Yen~Lee Loh and Nandini Trivedi.
\newblock Detecting the $\mathrm{E}$lusive
  $\mathrm{L}$arkin-$\mathrm{)}$vchinnikov $\mathrm{M}$odulated
  $\mathrm{S}$uperfluid $\mathrm{P}$hases for $\mathrm{I}$mbalanced
  $\mathrm{F}$ermi $\mathrm{G}$ases in $\mathrm{O}$ptical $\mathrm{L}$attices.
\newblock {\em Phys. Rev. Lett.}, 104:165302, 2010.

\bibitem{chiesa2013}
Simone Chiesa and Shiwei Zhang.
\newblock Phases of attractive spin-imbalanced fermions in square lattices.
\newblock {\em Phys. Rev. A}, 88:043624, 2013.

\bibitem{rosenberg2015}
Peter Rosenberg, Simone Chiesa, and Shiwei Zhang.
\newblock {FFLO} order in ultra-cold atoms in three-dimensional optical
  lattices.
\newblock {\em Journal of Physics: Condensed Matter}, 27:225601, 2015.

\bibitem{sheehy2007}
Daniel~E. Sheehy and Leo Radzihovsky.
\newblock $\mathrm{BEC}$$\ensuremath{-}$$\mathrm{BCS}$ crossover, phase
  transitions and phase separation in polarized resonantly-paired superfluids.
\newblock {\em Annals of Physics}, 322(8):1790 -- 1924, 2007.

\bibitem{xu2014}
Yong Xu, Chunlei Qu, Ming Gong, and Chuanwei Zhang.
\newblock Competing superfluid orders in spin-orbit-coupled fermionic cold-atom
  optical lattices.
\newblock {\em Phys. Rev. A}, 89:013607, 2014.

\bibitem{sun2013}
Kuei Sun and C.~J. Bolech.
\newblock Pair tunneling, phase separation, and dimensional crossover in
  imbalanced fermionic superfluids in a coupled array of tubes.
\newblock {\em Phys. Rev. A}, 87:053622, 2013.

\bibitem{parish2007_mft}
Meera~M. Parish, Stefan~K. Baur, Erich~J. Mueller, and David~A. Huse.
\newblock Quasi-$\mathrm{O}$ne-$\mathrm{D}$imensional $\mathrm{P}$olarized
  $\mathrm{F}$ermi $\mathrm{S}$uperfluids.
\newblock {\em Phys. Rev. Lett.}, 99:250403, 2007.

\bibitem{zhao_liu_ft2008}
Erhai Zhao and W.~Vincent Liu.
\newblock Theory of quasi-one-dimensional imbalanced $\mathrm{F}$ermi gases.
\newblock {\em Phys. Rev. A}, 78:063605, 2008.

\bibitem{vorontsov2005}
A.~B. Vorontsov, J.~A. Sauls, and M.~J. Graf.
\newblock Phase diagram and spectroscopy of
  $\mathrm{F}$ulde-$\mathrm{F}$errell-$\mathrm{L}$arkin- $\mathrm{O}$vchinnikov
  states of two-dimensional $d$-wave superconductors.
\newblock {\em Phys. Rev. B}, 72:184501, 2005.

\bibitem{vorontsov2006}
Anton~B. Vorontsov and Matthias~J. Graf.
\newblock Fermi-liquid effects in the
  $\mathrm{F}$ulde-$\mathrm{F}$errell-$\mathrm{L}$arkin-$\mathrm{O}$vchinnikov
  state of two-dimensional $d$-wave superconductors.
\newblock {\em Phys. Rev. B}, 74:172504, 2006.

\bibitem{beaird2010}
Robert Beaird, Anton~B. Vorontsov, and Ilya Vekhter.
\newblock Pauli-limited superconductivity with classical magnetic fluctuations.
\newblock {\em Phys. Rev. B}, 81:224501, 2010.

\bibitem{vorontsov2008}
A.~B. Vorontsov, I.~Vekhter, and M.~J. Graf.
\newblock Pauli-limited upper critical field in dirty $d$-wave superconductors.
\newblock {\em Phys. Rev. B}, 78:180505, 2008.

\bibitem{yang1998}
Kun Yang and S.~L. Sondhi.
\newblock Response of a ${d}_{{x}^{2}\ensuremath{-}{y}^{2}}$ superconductor to
  a $\mathrm{Z}$eeman magnetic field.
\newblock {\em Phys. Rev. B}, 57:8566--8570, 1998.

\bibitem{zhou2009}
Tao Zhou and C.~S. Ting.
\newblock Phase diagram and local tunneling spectroscopy of the
  $\mathrm{F}$ulde-$\mathrm{F}$errell-$\mathrm{L}$arkin- $\mathrm{O}$vchinnikov
  states of a two-dimensional square-lattice $d$-wave superconductor.
\newblock {\em Phys. Rev. B}, 80:224515, 2009.

\bibitem{ikeda2014}
Yuhki Hatakeyama and Ryusuke Ikeda.
\newblock Strong-$\mathrm{C}$oupling $\mathrm{A}$pproach to
  $\mathrm{A}$ntiferromagnetic $\mathrm{O}$rdering $\mathrm{D}$riven by
  $\mathrm{P}$aramagnetic $\mathrm{P}$air-$\mathrm{B}$reaking in
  $d$-$\mathrm{W}$ave $\mathrm{S}$uperconducting $\mathrm{P}$hase.
\newblock {\em Journal of the Physical Society of Japan}, 83(2):024713, 2014.

\bibitem{ikeda2015}
Yuhki Hatakeyama and Ryusuke Ikeda.
\newblock Antiferromagnetic order oriented by
  $\mathrm{F}$ulde-$\mathrm{F}$errell-$\mathrm{L}$arkin- $\mathrm{O}$vchinnikov
  superconducting order.
\newblock {\em Phys. Rev. B}, 91:094504, 2015.

\bibitem{ikeda2013}
Ken-ichi Hosoya and Ryusuke Ikeda.
\newblock Angular dependence of antiferromagnetic order induced by
  paramagnetism in a $d$-wave superconductor.
\newblock {\em Phys. Rev. B}, 88:094513, 2013.

\bibitem{ikeda2011}
Yuhki Hatakeyama and Ryusuke Ikeda.
\newblock Emergent antiferromagnetism in a $d$-wave superconductor with strong
  paramagnetic pair-breaking.
\newblock {\em Phys. Rev. B}, 83:224518, 2011.

\bibitem{ikeda2017}
Ken-ichi Hosoya and Ryusuke Ikeda.
\newblock Possible triplet superconducting order in a magnetic superconducting
  phase induced by paramagnetic pair breaking.
\newblock {\em Phys. Rev. B}, 95:224513, 2017.

\bibitem{adachi2015}
Kyosuke Adachi and Ryusuke Ikeda.
\newblock Possible $\mathrm{F}$ield$\ensuremath{-}$$\mathrm{T}$emperature
  $\mathrm{P}$hase $\mathrm{D}$iagrams of $\mathrm{T}$wo-$\mathrm{B}$ and
  $\mathrm{S}$uperconductors with $\mathrm{P}$aramagnetic $\mathrm{P}$air-
  $\mathrm{B}$reaking.
\newblock {\em Journal of the Physical Society of Japan}, 84(6):064712, 2015.

\bibitem{takemori2012}
Nayuta Takemori and Akihisa Koga.
\newblock Low-$\mathrm{T}$emperature $\mathrm{P}$roperties of the
  $\mathrm{F}$ermionic $\mathrm{M}$ixtures with $\mathrm{M}$ass
  $\mathrm{I}$mbalance in $\mathrm{O}$ptical $\mathrm{L}$attice.
\newblock {\em Journal of the Physical Society of Japan}, 81(6):063002, 2012.

\bibitem{ohashi2013}
Ryo Hanai, Takashi Kashimura, Ryota Watanabe, Daisuke Inotani, and Yoji Ohashi.
\newblock Excitation properties and effects of mass imbalance in the
  $\mathrm{BCS}$$\ensuremath{-}$$\mathrm{BEC}$ crossover regime of an ultracold
  $\mathrm{F}$ermi gas.
\newblock {\em Phys. Rev. A}, 88:053621, 2013.

\bibitem{roscher2015}
Dietrich Roscher, Jens Braun, and Joaquin~E. Drut.
\newblock Phase structure of mass- and spin-imbalanced unitary $\mathrm{F}$ermi
  gases.
\newblock {\em Phys. Rev. A}, 91:053611, 2015.

\bibitem{braun2014}
Jens Braun, Joaquin~E. Drut, Thomas Jahn, Martin Pospiech, and Dietrich
  Roscher.
\newblock Phases of spin- and mass-imbalanced ultracold $\mathrm{F}$ermi gases
  in harmonic traps.
\newblock {\em Phys. Rev. A}, 89:053613, 2014.

\bibitem{ohashi2014}
Ryo Hanai and Yoji Ohashi.
\newblock Heteropairing and component-dependent pseudogap phenomena in an
  ultracold $\mathrm{F}$ermi gas with different species with different masses.
\newblock {\em Phys. Rev. A}, 90:043622, 2014.

\bibitem{guo2009}
Hao Guo, Chih-Chun Chien, Qijin Chen, Yan He, and K.~Levin.
\newblock Finite-temperature behavior of an interspecies fermionic superfluid
  with population imbalance.
\newblock {\em Phys. Rev. A}, 80:011601, 2009.

\bibitem{gubbels2009}
K.~B. Gubbels, J.~E. Baarsma, and H.~T.~C. Stoof.
\newblock Lifshitz $\mathrm{P}$oint in the $\mathrm{P}$hase $\mathrm{D}$iagram
  of $\mathrm{R}$esonantly $\mathrm{I}$nteracting
  $^{6}\mathrm{Li}$$\ensuremath{-}$${}^{40}\mathrm{K}$ $\mathrm{M}$ixtures.
\newblock {\em Phys. Rev. Lett.}, 103:195301, 2009.

\bibitem{wang2009}
B.~Wang, Han-Dong Chen, and S.~Das~Sarma.
\newblock Quantum phase diagram of fermion mixtures with population imbalance
  in one-dimensional optical lattices.
\newblock {\em Phys. Rev. A}, 79:051604, 2009.

\bibitem{dalmonte2012}
M.~Dalmonte, K.~Dieckmann, T.~Roscilde, C.~Hartl, A.~E. Feiguin,
  U.~Schollw\"ock, and F.~Heidrich-Meisner.
\newblock Dimer, trimer, and
  $\mathrm{F}$ulde-$\mathrm{F}$errell-$\mathrm{L}$arkin-$\mathrm{O}$vchinnikov
  liquids in mass- and spin-imbalanced trapped binary mixtures in one
  dimension.
\newblock {\em Phys. Rev. A}, 85:063608, 2012.

\bibitem{pahl2014}
Shanna Pahl and Zlatko Koinov.
\newblock Phase $\mathrm{D}$iagram of a
  ${}^{6}\mathrm{Li}$$\ensuremath{-}$${}^{40}\mathrm{K}$ $\mathrm{M}$ixture in
  a $\mathrm{S}$quare $\mathrm{L}$attice.
\newblock {\em Journal of Low Temperature Physics}, 176:113, 2014.

\bibitem{braun2015}
Jens Braun, Joaqu\'{\i}n~E. Drut, and Dietrich Roscher.
\newblock Zero-$\mathrm{T}$emperature $\mathrm{E}$quation of $\mathrm{S}$tate
  of $\mathrm{M}$ass-$\mathrm{I}$mbalanced $\mathrm{R}$esonant $\mathrm{F}$ermi
  $\mathrm{G}$ases.
\newblock {\em Phys. Rev. Lett.}, 114:050404, 2015.

\bibitem{galitski2013}
Victor Galitski and Ian~B. Spielman.
\newblock Spin–orbit coupling in quantum gases.
\newblock {\em Nature}, 494:49, 2013.

\bibitem{dalibard2011}
Jean Dalibard, Fabrice Gerbier, Gediminas Juzeli\ifmmode~\bar{u}\else
  \={u}\fi{}nas, and Patrik \"Ohberg.
\newblock Colloquium: $\mathrm{A}$rtificial gauge potentials for neutral atoms.
\newblock {\em Rev. Mod. Phys.}, 83:1523--1543, 2011.

\bibitem{torma_rpp_review2018}
Jami~J Kinnunen, Jildou~E Baarsma, Jani-Petri Martikainen, and P{\"a}ivi
  T{\"o}rm{\"a}.
\newblock The
  fulde{\textendash}ferrell{\textendash}larkin{\textendash}ovchinnikov state
  for ultracold fermions in lattice and harmonic potentials: a review.
\newblock {\em Reports on Progress in Physics}, 81:046401, 2018.

\bibitem{higbie2002}
J.~Higbie and D.~M. Stamper-Kurn.
\newblock Periodically dressed $\mathrm{B}$ose{\textendash}$\mathrm{E}$instein
  condensate: A superfluid with an anisotropic and variable critical velocity.
\newblock {\em Phys. Rev. Lett.}, 88:090401, 2002.

\bibitem{lin_nature2011}
Y.~J. Lin, K.~Jim{\'e}nez-Garc{\'i}a, and I.~B. Spielman.
\newblock Spin{\textendash}orbit coupled
  $\mathrm{B}$ose{\textendash}$\mathrm{E}$instein condensates.
\newblock {\em Nature}, 471:83, 2011.

\bibitem{wang_prl2012}
Pengjun Wang, Zeng-Qiang Yu, Zhengkun Fu, Jiao Miao, Lianghui Huang, Shijie
  Chai, Hui Zhai, and Jing Zhang.
\newblock Spin{\textendash}orbit coupled degenerate $\mathrm{F}$ermi gases.
\newblock {\em Phys. Rev. Lett.}, 109:095301, 2012.

\bibitem{cheuk_prl2012}
Lawrence~W. Cheuk, Ariel~T. Sommer, Zoran Hadzibabic, Tarik Yefsah, Waseem~S.
  Bakr, and Martin~W. Zwierlein.
\newblock Spin{\textendash}injection spectroscopy of a spin{\textendash}orbit
  coupled $\mathrm{F}$ermi gas.
\newblock {\em Phys. Rev. Lett.}, 109:095302, 2012.

\bibitem{ruseckas2005}
J.~Ruseckas, G.~Juzeli\ifmmode~\bar{u}\else \={u}\fi{}nas, P.~\"Ohberg, and
  M.~Fleischhauer.
\newblock Non-$\mathrm{A}$belian $\mathrm{G}$auge $\mathrm{P}$otentials for
  $\mathrm{U}$ltracold $\mathrm{A}$toms with $\mathrm{D}$egenerate
  $\mathrm{D}$ark $\mathrm{S}$tates.
\newblock {\em Phys. Rev. Lett.}, 95:010404, 2005.

\bibitem{osterloh2005}
K.~Osterloh, M.~Baig, L.~Santos, P.~Zoller, and M.~Lewenstein.
\newblock Cold $\mathrm{A}$toms in $\mathrm{N}$on-$\mathrm{A}$belian
  $\mathrm{G}$auge $\mathrm{P}$otentials: $\mathrm{F}$rom the
  $\mathrm{H}$ofstadter $"\mathrm{M}oth"$ to $\mathrm{L}$attice
  $\mathrm{G}$auge $\mathrm{T}$heory.
\newblock {\em Phys. Rev. Lett.}, 95:010403, 2005.

\bibitem{campbell2011}
D.~L. Campbell, G.~Juzeli\ifmmode~\bar{u}\else \={u}\fi{}nas, and I.~B.
  Spielman.
\newblock Realistic $\mathrm{R}$ashba and $\mathrm{D}$resselhaus spin-orbit
  coupling for neutral atoms.
\newblock {\em Phys. Rev. A}, 84:025602, 2011.

\bibitem{zheng_pra2013}
Zhen Zheng, Ming Gong, Xubo Zou, Chuanwei Zhang, and Guangcan Guo.
\newblock Route to observable
  $\mathrm{F}$ulde{\textendash}$\mathrm{F}$errell{\textendash}$\mathrm{L}$arkin{\textendash}$\mathrm{O}$vchinnikov
  phases in three-dimensional spin-orbit-coupled degenerate $\mathrm{F}$ermi
  gases.
\newblock {\em Phys. Rev. A}, 87:031602, 2013.

\bibitem{xu_2014}
Yong Xu, Chunlei Qu, Ming Gong, and Chuanwei Zhang.
\newblock Competing superfluid orders in spin-orbit-coupled fermionic cold-atom
  optical lattices.
\newblock {\em Phys. Rev. A}, 89:013607, 2014.

\bibitem{iskin2012}
M.~Iskin.
\newblock Trapped $\mathrm{F}$ermi gases with $\mathrm{R}$ashba spin-orbit
  coupling in two dimensions.
\newblock {\em Phys. Rev. A}, 86:065601, 2012.

\bibitem{iskin2013}
M.~Iskin and A.~L. Suba\ifmmode \mbox{\c{s}}\else \c{s}\fi{}\ifmmode \imath
  \else~\i \fi{}.
\newblock Topological superfluid phases of an atomic $\mathrm{F}$ermi gas with
  in-and out-of-plane $\mathrm{Z}$eeman fields and equal
  $\mathrm{R}$ashba-$\mathrm{D}$resselhaus spin-orbit coupling.
\newblock {\em Phys. Rev. A}, 87:063627, 2013.

\bibitem{seo2013}
Kangjun Seo, Chuanwei Zhang, and Sumanta Tewari.
\newblock Topological uniform superfluid and
  $\mathrm{F}$ulde-$\mathrm{F}$errell-$\mathrm{L}$arkin- $\mathrm{O}$vchinnikov
  phases in three-dimensional to one-dimensional crossover of
  spin-orbit-coupled $\mathrm{F}$ermi gases.
\newblock {\em Phys. Rev. A}, 88:063601, 2013.

\bibitem{hu_njp2013}
Hui Hu and Xia-Ji Liu.
\newblock Fulde$\ensuremath{–}$$\mathrm{F}$errell superfluidity in ultracold
  $\mathrm{F}$ermi gases with $\mathrm{R}$ashba spin–orbit coupling.
\newblock {\em New Journal of Physics}, 15:093037, 2013.

\bibitem{liu_hu2013}
Xia-Ji Liu and Hui Hu.
\newblock Inhomogeneous $\mathrm{F}$ulde-$\mathrm{F}$errell superfluidity in
  spin-orbit-coupled atomic $\mathrm{F}$ermi gases.
\newblock {\em Phys. Rev. A}, 87:051608, 2013.

\bibitem{liu2013}
Xia-Ji Liu.
\newblock Fulde-$\mathrm{F}$errell pairing instability of a $\mathrm{R}$ashba
  spin-orbit-coupled $\mathrm{F}$ermi gas.
\newblock {\em Phys. Rev. A}, 88:043607, 2013.

\bibitem{dong2013}
Lin Dong, Lei Jiang, and Han Pu.
\newblock Fulde$\ensuremath{-}$$\mathrm{F}$errell pairing instability in
  spin{\textendash}orbit coupled $\mathrm{F}$ermi gas.
\newblock {\em New Journal of Physics}, 15(7):075014, 2013.

\bibitem{zhou2013}
Xiang-Fa Zhou, Guang-Can Guo, Wei Zhang, and Wei Yi.
\newblock Exotic pairing states in a $\mathrm{F}$ermi gas with
  three-dimensional spin-orbit coupling.
\newblock {\em Phys. Rev. A}, 87:063606, 2013.

\bibitem{iskin2_2013}
M.~Iskin.
\newblock Spin-orbit-coupling-induced
  $\mathrm{F}$ulde-$\mathrm{F}$errell-$\mathrm{L}$arkin-$\mathrm{O}$vchinnikov-
  like $\mathrm{C}$ooper pairing and skyrmion-like polarization textures in
  optical lattices.
\newblock {\em Phys. Rev. A}, 88:013631, 2013.

\bibitem{tewari2011}
Sumanta Tewari, Tudor~D Stanescu, Jay~D Sau, and S~Das Sarma.
\newblock Topologically non-trivial superconductivity in
  spin{\textendash}orbit-coupled systems: bulk phases and quantum phase
  transitions.
\newblock {\em New Journal of Physics}, 13:065004, 2011.

\bibitem{zhang2013}
Wei Zhang and Wei Yi.
\newblock Topological
  $\mathrm{F}$ulde–$\mathrm{F}$errell–$\mathrm{L}$arkin–$\mathrm{O}$vchinnikov
  states in spin–orbit-coupled $\mathrm{F}$ermi gases.
\newblock {\em Nature Communications}, 4:2711, 2013.

\bibitem{qu2013}
Chunlei Qu, Zhen Zheng, Ming Gong, Yong Xu, Li~Mao, Xubo Zou, Guangcan Guo, and
  Chuanwei Zhang.
\newblock Topological superfluids with finite-momentum pairing and
  $\mathrm{M}$ajorana fermions.
\newblock {\em Nature Communications}, 4:2710, 2013.

\bibitem{cao2014}
Ye~Cao, Shu-Hao Zou, Xia-Ji Liu, Su~Yi, Gui-Lu Long, and Hui Hu.
\newblock Gapless $\mathrm{T}$opological $\mathrm{F}$ulde-$\mathrm{F}$errell
  $\mathrm{S}$uperfluidity in $\mathrm{S}$pin-$\mathrm{O}$rbit
  $\mathrm{C}$oupled $\mathrm{F}$ermi $\mathrm{G}$ases.
\newblock {\em Phys. Rev. Lett.}, 113:115302, 2014.

\bibitem{liu2012}
Xia-Ji Liu and Hui Hu.
\newblock Topological superfluid in one-dimensional spin-orbit-coupled atomic
  $\mathrm{F}$ermi gases.
\newblock {\em Phys. Rev. A}, 85:033622, 2012.

\bibitem{xu2015}
Yong Xu and Chuanwei Zhang.
\newblock Berezinskii-$\mathrm{K}$osterlitz-$\mathrm{T}$houless
  $\mathrm{P}$hase $\mathrm{T}$ransition in
  2$\mathrm{D}$$\mathrm{S}$pin-$\mathrm{O}$rbit-coupled
  $\mathrm{F}$ulde-$\mathrm{F}$errell $\mathrm{S}$uperfluids.
\newblock {\em Phys. Rev. Lett.}, 114:110401, 2015.

\bibitem{cao2015}
Ye~Cao, Xia-Ji Liu, Lianyi He, Gui-Lu Long, and Hui Hu.
\newblock Superfluid density and
  $\mathrm{B}$erezinskii-$\mathrm{K}$osterlitz-$\mathrm{T}$houless transition
  of a spin-orbit-coupled $\mathrm{F}$ulde-$\mathrm{F}$errell superfluid.
\newblock {\em Phys. Rev. A}, 91:023609, 2015.

\bibitem{wang2018}
Jibiao Wang, Yanming Che, Leifeng Zhang, and Qijin Chen.
\newblock Instability of
  $\mathrm{F}$ulde-$\mathrm{F}$errell-$\mathrm{L}$arkin-$\mathrm{O}$vchinnikov
  states in atomic $\mathrm{F}$ermi gases in three and two dimensions.
\newblock {\em Phys. Rev. B}, 97:134513, 2018.

\bibitem{rosenberg2017}
Peter Rosenberg, Hao Shi, and Shiwei Zhang.
\newblock Ultracold $\mathrm{A}$toms in a $\mathrm{S}$quare $\mathrm{L}$attice
  with $\mathrm{S}$pin-$\mathrm{O}$rbit $\mathrm{C}$oupling: $\mathrm{C}$harge
  $\mathrm{O}$rder, $\mathrm{S}$uperfluidity, and $\mathrm{T}$opological
  $\mathrm{S}$ignatures.
\newblock {\em Phys. Rev. Lett.}, 119:265301, 2017.

\end{thebibliography}

\end{document}